\documentclass[12pt]{article}
\usepackage{amssymb}
\usepackage{amsmath}
\usepackage{amsthm}
\usepackage{cancel}
\usepackage{slashed}
\usepackage{fullpage}
\usepackage{color}
\usepackage{braket}
\usepackage{graphicx}
\usepackage{multirow}
\usepackage{verbatim}
\usepackage{cite}
\usepackage{bigints}
\usepackage{simplewick}
\usepackage{authblk}

\usepackage{caption}
\usepackage{subcaption}

\usepackage{feynmp}

\usepackage{hyperref}
\hypersetup{
    linktocpage,
     colorlinks,
     citecolor=darkgreen,
     linkcolor= darkgreen,
     urlcolor=darkgreen
}

\usepackage{pdfsync}
\usepackage{tikz}
\usetikzlibrary{shapes,backgrounds,calc}
\usepackage{verbatim}
\usetikzlibrary{decorations.pathmorphing, decorations.shapes}
\usepackage{xcolor}
\usetikzlibrary{arrows.meta}

\makeatletter

\pgfdeclaredecoration{gluon}{coil}
{
  \state{coil}[switch if less than=%
    0.5\pgfdecorationsegmentlength+%>
    \pgfdecorationsegmentaspect\pgfdecorationsegmentamplitude+%
    \pgfdecorationsegmentaspect\pgfdecorationsegmentamplitude to last,
               width=+\pgfdecorationsegmentlength]
  {
    \pgfpathcurveto
    {\pgfpoint@oncoil{0    }{ 0.555}{1}}
    {\pgfpoint@oncoil{0.445}{ 1    }{2}}
    {\pgfpoint@oncoil{1    }{ 1    }{3}}
    \pgfpathcurveto
    {\pgfpoint@oncoil{1.555}{ 1    }{4}}
    {\pgfpoint@oncoil{2    }{ 0.555}{5}}
    {\pgfpoint@oncoil{2    }{ 0    }{6}}
    \pgfpathcurveto
    {\pgfpoint@oncoil{2    }{-0.555}{7}}
    {\pgfpoint@oncoil{1.555}{-1    }{8}}
    {\pgfpoint@oncoil{1    }{-1    }{9}}
    \pgfpathcurveto
    {\pgfpoint@oncoil{0.445}{-1    }{10}}
    {\pgfpoint@oncoil{0    }{-0.555}{11}}
    {\pgfpoint@oncoil{0    }{ 0    }{12}}
  }
  \state{last}[next state=final]
  {
    \pgfpathcurveto
    {\pgfpoint@oncoil{0    }{ 0.555}{1}}
    {\pgfpoint@oncoil{0.445}{ 1    }{2}}
    {\pgfpoint@oncoil{1    }{ 1    }{3}}
    \pgfpathcurveto
    {\pgfpoint@oncoil{1.555}{ 1    }{4}}
    {\pgfpoint@oncoil{2    }{ 0.555}{5}}
    {\pgfpoint@oncoil{2    }{ 0    }{6}}
  }
  \state{final}{}
}

\def\pgfpoint@oncoil#1#2#3{%
  \pgf@x=#1\pgfdecorationsegmentamplitude%
  \pgf@x=\pgfdecorationsegmentaspect\pgf@x%
  \pgf@y=#2\pgfdecorationsegmentamplitude%
  \pgf@xa=0.083333333333\pgfdecorationsegmentlength%
  \advance\pgf@x by#3\pgf@xa%
}

\makeatother

\def\blobcolor{gray}

\usetikzlibrary{decorations.shapes}
\tikzset{myarrow/.style 2 args={
        decoration={markings,
            mark= at position #1 with {\arrow{#2}} ,
        },
        postaction={decorate}
    }
}

\tikzset{decorate glaubr/.style= {
 decorate, decoration={shape backgrounds,shape=circle,shape size=1pt,shape sep=2pt}, line width=0.3pt }}

\def\nn{\nonumber}

\def\cC{\mathcal{C}}

\def\cF{\mathcal{F}}
\def\cG{\mathcal{G}}

\def\cI{\mathcal{I}}

\def\cL{\mathcal{L}}
\def\cM{\mathcal{M}}

\def\cO{\mathcal{O}}

\def\cS{\mathcal{S}}

\def\tr{{\rm tr}}

\def\dg{\dagger}

\def\slash{\!\!\!/}

\def\be{\begin{equation}}
\def\ee{\end{equation}}

\def\l{\langle}
\def\r{\rangle}

\newcommand{\Eq}[1]{Eq.~\eqref{#1}}

\allowdisplaybreaks

\def\LPeq{\cong}

\definecolor{darkred}{rgb}{0.8,0.0,0.0}
\definecolor{darkblue}{rgb}{0.0,0.0,0.9}
\definecolor{darkerblue}{rgb}{0.0,0.0,0.5}
\definecolor{darkgreen}{rgb}{0.0,0.5,0.0}
\definecolor{black}{rgb}{0.0,0.0,0.0}
\definecolor{brown}{rgb}{0.6,0.4,0.2}

\newcommand{\black}{\color{black}}

\newcommand{\ccol}{\color{darkblue}}

\newcommand{\softcol}{\color{darkred}}

\newcommand{\scs}{ {{\softcol s}} }

\newcommand{\ccF}{ {{\ccol 4}} }
\newcommand{\ccTH}{ {{\ccol 3}} }

\newcommand{\ccO}{ {{\ccol 1}} }
\newcommand{\ccT}{ {{\ccol 2}} }
\newcommand{\cci}{ {{\ccol i}} }
\newcommand{\ccj}{ {{\ccol j}} }

\newcommand{\rN}{ {{\ccol N}} }
\newcommand{\rM}{ {{\ccol M}} }

\newcommand{\ccn}{ {{\ccol n}} }
\newcommand{\ccN}{ {{\ccol N}} }
\newcommand{\ccM}{ {{\ccol M}} }

\makeatletter
\newcommand*{\hyperlinkcite}[1]{\hyper@link{cite}{cite.#1\@extra@b@citeb}}
\makeatother

\newcommand{\Sp}{ \mathbf{Sp} }

\newcommand{\TT}{ \mathbf{T} }

\newcommand{\Sij}{ S_{\cci \ccj}}

\def\zero{0}
\def\one{1}
\def\two{2}

\def\fme{\varepsilon}
\def\pol{\varepsilon}
\def\ep{\epsilon}

\def\cMb{\overline{\mathcal{M}}}

\def\zero{0}
\def\one{1}
\def\two{2}
\def\n{n}
\def\nmo{n-1}
\def\m{m} 
\def\nmm{n-m}
\def\nmom{n-1-m}
\def\mpo{m+1}
\def\cMb{\overline{\mathcal{M}}}

\newcommand{\LGij}{\cL_{\cci \ccj}^{\text{G}}}

\title{
Factorization Violation and Scale Invariance
}

\author{Matthew D. Schwartz${}^1$}
\author{Kai Yan${}^1$}
\author{Hua Xing Zhu${}^2$}

\affil{${}^1${\small \emph{Department of Physics,
Harvard University, Cambridge, MA 02138, USA}}}
\affil{${}^2${\small \emph{Zhejiang Institute of Modern Physics, Department of Physics, Zhejiang University, Hangzhou 310027, China}}
}

\begin{document} 

\begin{fmffile}{feyngraph}
\unitlength = 1mm

\maketitle

\begin{abstract}
Factorization violating effects in hadron scattering are due mainly to spectator-spectator interactions.
While it is known that these interactions cancel in inclusive cross sections, like for the Drell-Yan process, not much is known about for what classes of observables factorization is violated. We show
that  for pure Glauber ladder graphs, all  amplitude-level factorization
violating effects completely cancel at  cross section level for any single-scale observable (such as
hadronic transverse energy or beam thrust). 
This result disproves previous claims that these pure Glauber graphs are factorization-violating.
Our proof exploits scale invariance
of two-to-two scattering amplitudes in an essential way. 
 The leading factorization-violating effects therefore
come from graphs with at least one soft gluon, involving the Lipatov vertex off of the Glauber ladders. This implies that real soft radiation must be involved in factorization-violation, shedding light on the connection between factorization-violation and the underlying event. 

\end{abstract}

\newpage
%\tableofcontents

\section{Introduction}

Factorization is essential to the predictive power of perturbative QCD at hadron colliders. The essential point of factorization is that it lets us separate the dynamics of the proton from the dynamics of the scattering that produces hard radiation. Unfortunately, factorization is known not to hold universally. There is both theoretical evidence for non-factorization through effects like super-leading logarithms, and experimental evidence, as calculations performed assuming factorization can have significant deviations from experiment. In order to continue to push the precision of predictions for the Large Hadron Collider and other machines, it will be essential to understand more about how and when factorization violation occurs. The hope is that with better understanding we might be able to either choose observables for which factorization violation is minimal or find some universality in factorization-violating effects.

Factorization has been shown to hold rigorously in perturbation theory in certain circumstances. For example, it holds at the amplitude level for processes with only final-state radiation. It also holds at the amplitude level for processes with only one colored particle in the initial state (like deep-inelastic scattering at the parton level). In addition, it holds for any process with a fixed number of external particles.
These results were reviewed and clarified recently in~\cite{FS1,FS2, Feige:2015rea}.   Factorization is known to be violated both at the amplitude and cross section level for processes with  two colored initial state particles and at least one final-state particle collinear to an initial-state one~\cite{ Catani:2011st,Forshaw:2012bi,Schwartz:2017nmr}. 
In such situations, while the amplitude can still be written as a splitting amplitude times the amplitude with the collinear pair replaced by their mother particle, the splitting amplitude necessarily depends on the quantum numbers of non-collinear particles.~\cite{Catani:2011st}.

Even if factorization is violated at the amplitude level, it may still cancel when amplitudes are squared and integrated over phase space. The celebrated example of this phenomenon is the Collins-Soper-Sterman (CSS) proof of factorization for Drell-Yan~\cite{Collins:1988ig}. There, when all relevant processes describing the hadro-production of a lepton pair are included, the cross section still can be written as the convolution of parton-distribution functions and a hard scattering kernel.

In recent years, there has been renewed interest in extending the CSS argument to other processes, and understanding its failure. Partly this has been motivated by the advent of jet substructure techniques~\cite{ PhysRevLett.39.1436,CATANI1991368,CATANI19933,Seymour:1993mx, Butterworth:2008iy}, where predictions of observables like jet mass or beam thrust are apparently more sensitive to factorization-violating effects than traditional kinematic observables, like the jet $p_T$ spectrum. Partly it has been driven by theory developments  that give new handles on 
factorization-violating effects. In particular, Soft-Collinear Effective Theory (SCET)~\cite{Bauer:2000ew,Bauer:2000yr,Bauer:2001ct,Bauer:2001yt,Bauer:2002nz,Beneke:2002ni} has allowed for higher precision jet substructure calculations. 
Recently, Rothstein and Stewart~\cite{Rothstein:2016bsq} have explained
how to account for Glauber effects within the SCET framework (see also~\cite{Liu:2008cc,Donoghue:2009cq,Bauer:2010cc,Fleming:2014rea}).   

In this current paper, we attempt to shed some light on when factorization occurs by combining the SCET Glauber picture with observations by CSS and others. In particular,
it has been argued in Ref.~\cite{Gaunt:2014ska,Rothstein:2016bsq} that factorization can be violated by diagrams with pure Glauber gluon exchange, but not involving any real gluon radiation. 
 We revisit these arguments and 
show that in fact factorization is {\it not} violated by the pure Glauber graphs. 
Our calculation amounts to showing that after summing over all possible cuts, contribution from pure Glauber graphs vanish,
\be
\int d \Pi_{\text{LIPS}}
 %\int  d^2 q_T d z d \bar{z} 
 \sum_{
%\text{cuts}} \delta(X- f_X (q_T, z, \bar z )) \begin{gathered}
\text{cuts}} \delta(X- f_X (p_3^\mu,p_4^\mu )) \begin{gathered}
\begin{tikzpicture}
 \node at (0,0) {
\resizebox{60mm}{!}{
     \fmfframe(0,0)(0,0){
    \begin{fmfgraph*}(90,40)
\fmfstraight
    \fmfleft{L1,L2}
    \fmfright{R1,R2}
    \fmf{photon}{L1,v1}
    \fmf{fermion}{v1,v}
    \fmf{fermion}{v,v2}
    \fmf{photon}{v2,L2}
    \fmf{photon}{R1,w1}
    \fmf{fermion}{w2,w}
    \fmf{fermion}{w,w1}
    \fmf{photon}{w2,R2}
    \fmf{quark,tension=0.5}{w1,t1,t2,t3,t4,v1}
    \fmf{quark,tension=0.5}{v2,b4,b3,b2,b1,w2}  
    \fmf{phantom,tension=0.2}{v,w}   
\fmffreeze 
    \fmf{dbl_dots,fore=red}{t1,b1} 
    \fmf{dbl_dots,fore=red}{t2,b2} 
    \fmf{dbl_dots,fore=red}{t3,b3} 
    \fmf{dbl_dots,fore=red}{t4,b4} 
     \fmfv{d.sh=circle, d.f=30,d.si=0.05w}{v}
     \fmfv{d.sh=circle, d.f=30,d.si=0.05w}{w}
     \end{fmfgraph*}
}
}
};
\draw[dashed,blue] (-1.5,1.2) -- (-1.5,-1.2);
\draw[dashed,blue] (-0.8,1.2) -- (-0.8,-1.2);
\draw[dashed,blue] (-0.05,1.2) -- (-0.05,-1.2);
\draw[dashed,blue] (0.75,1.2) -- (0.75,-1.2);
\draw[dashed,blue] (1.45,1.2) -- (1.45,-1.2);
\end{tikzpicture}
\end{gathered}
= 0
\ee
Here,  the vertical dashed lines denote possible cuts and $p_3^\mu$ and $p_4^\mu$ are the momenta of the two outgoing spectators generated by the cuts.

The cancellation we find implies that the differential cross-section for any single-scale observable, like hadronic transverse energy, or beam thrust, gets no contribution from pure-Glauber graphs.
The leading factorization-violating effects therefore must involve diagrams with the Lipatov vertex~\cite{Kuraev:1976ge}, that is, diagrams where soft gluons are exchanged between Glauber rungs. Our argument does not apply to doubly-differential observables, like those studied in~\cite{Zeng:2015iba}. 

We begin in Section~\ref{sec:fact} with a review of some results about factorization at the amplitude and cross section level. This section can be skipped by an informed reader, but may be useful to a reader who finds the literature on factorization marginally impenetrable. 
 Section~\ref{sec:ladders} presents our main new result, that the contribution of pure Glauber ladder graphs to cross section exactly vanishes for any single-scale observable. 
To not disrupt the logic,  we include in this section only summaries of the calculations, with details relegated to Appendices~\ref{app:exp} through~\ref{app:ContourPhase}.
While elements of our calculation are similar to the summation of Glauber ladder graphs into a phase in position space for forward scattering calculations, we work instead in momentum space where
a choice of variables respecting scale invariance can be made.
Section~\ref{sec:fv} discusses graphs beyond the Glauber ladders, necessarily involving soft gluons and the Lipatov vertex. For these graphs, scale-invariance is violated by quantum effects and so factorization can be violated. A brief summary and conclusions are in  Section~\ref{sec:conc}.

\section{Factorization and Factorization Violation \label{sec:fact}}
We begin with a review of some known results about factorization.  There are no new results in this section. The goal of the section is merely to clarify what we mean by factorization, and what is known, using relatively clear and precise language. 

Almost all the literature about factorization refers to statements about soft and collinear divergences in perturbative QCD. A clean way to describe perturbative factorization was developed in~\cite{FS1,FS2,Feige:2015rea,Schwartz:2017nmr}. Consider an initial state $\ket{Z}$ and a final state $\bra{X}$ each of which is made up of some quarks and gluons with various momenta. We can group those momenta into sectors either collinear to a set of directions $n_\ccj^\mu$ or soft. Each sector 
has an associated scale. For example, the scale $\lambda_\ccj$ associated with the $n_\ccj$-collinear
sector may be defined so that 
each momentum $p_\ccj$ in the $n_\ccj$ collinear sector has $n_\ccj \cdot p_\ccj \le \lambda_\ccj p_\ccj^0$. For the soft sector a parameter $\lambda_\scs$ can be defined so that the energy of each soft momentum $k^\mu$ has $k^0 \le \lambda_\scs Q$ where $Q$ is a hard scale, such as the center-of-mass energy, or the energy of some jet. Thus we can write  $\bra{X}= \bra{X_\scs}\bra{X_\ccO}\cdots\bra{X_\ccN}$ and $\ket{Z} = \ket{Z_{\scs}} \ket{Z_{\ccN+1} \cdots Z_{\ccM}}$. For any such
decomposition in which no initial-state direction is collinear to a final-state direction, amplitudes factorize:
 \be
\bra{X} \phi^\star \cdots \phi \ket{Z}
 \;\LPeq\; 
\cC(\Sij) \, \frac{\bra{X_\ccO} \phi^\star W_\ccO \ket{0}}{\bra{0} Y_\ccO^\dg W_\ccO \ket{0}} \,\cdots\,
 \frac{\bra{0} W_\rM^\dg\phi \ket{Z_\ccM}}{\bra{0} W_\rM^\dg Y_\rM \ket{0}}
\;\bra{X_\scs} Y_\ccO^\dg \cdots Y_\rM \ket{Z_\scs}
\label{sQEDmainXY}
\ee
In this relation $\LPeq$ means the two sides agree at leading power in all of the $\lambda_\ccj$ and $\lambda_\scs$, including all infrared divergences~(soft and collinear divergences). Another way to say this is that the Wilson coefficient $\cC(\Sij)$, defined as the ratio of the left-side amplitude to the factorized ampliutde on the right, is finite as all the $\lambda$ go to zero. Here, the matrix elements are written as amplitudes of operators in scalar QED, for simplicity; the same factorization formula holds in QCD but the color and spin notation is more cumbersome. The $W_\ccj$ and $Y_\ccj$ are Wilson lines. All operators in this factorization formula are written in terms of the fields of an ordinary quantum field theory (scalar QED or QCD) -- no effective field theory interactions are required. This factorization is closely related to factorization in SCET where there are soft fields and collinear fields for each direction with leading-power interactions. 

 One important implication of factorization is that it relates amplitudes with different external states. For example, the Wilson coefficient $\cC(\Sij)$ depends only on hard scales $\Sij=(P_\cci^\mu + P_\ccj^\mu)^2$ , with $P_\ccj^\mu$ the net momenta in the $n_\ccj$-collinear sector. Thus the factorization formula relates processes with different soft and collinear particles. 
 
 An application of factorization is to prove the universality of collinear splittings. Treating
 amplitudes as vectors in color space, a splitting amplitude is defined as the ratio
 between an amplitude  $\ket{\cMb}$ with $n-m$ particles to an amplitude $\ket{\cM}$ with $n$ particles. For $m=1$ the relationship can be written as
 \be
\ket{\cM(p_\ccO,\cdots, p_\ccn)} \LPeq \Sp(p_\ccO,p_\ccT ; p_\ccTH,\cdots, p_\ccn) \cdot \ket{\cMb ( P,p_\ccTH, \cdots p_\ccn) }
\label{gensplit}
\ee
Here  $\cM$ and $\cMb$ are amplitudes like in Eq.~\eqref{sQEDmainXY}. The momentum $P^\mu$ on the right is a single particle momentum that splits into $p_\ccO^\mu$ and $p_\ccT^\mu$: $P^\mu \LPeq p_\ccO^\mu  + p_\ccT^\mu$.  Factorization in Eq.~\eqref{sQEDmainXY} implies that
 \be
\ket{ \cMb} \LPeq
\dfrac
{\bra{P} \bar\psi \, W_\ccO \ket{0}
}
{ \tr\bra{0} Y^\dg_\ccO W_\ccO \ket{0} } \cdot 
\ket{   \cM_{{\black {\text{rest}}}}}
\;,
\qquad
\ket{ \cM }\LPeq
\dfrac
{\bra{p_\ccO,p_\ccT} \bar\psi \, W_\ccO \ket{0}
}
{ \tr\bra{0} Y^\dg_\ccO W_\ccO \ket{0} }
\cdot 
\ket{   \cM_{{\black {\text{rest}}}}}
\;
\ee
and therefore
\be
\Sp = \frac{
\bra{p_\ccO,p_\ccT} \bar\psi \, W_\ccO \ket{0}
}
{
\bra{P} \bar\psi \, W_\ccO \ket{0}
}
\label{Spscet}
\ee
The key point is that the splitting amplitude is {\it universal} -- it only depends on the the fields in the direction collinear to the splitting.
 
Now consider a situation where some outgoing particles are collinear to some incoming ones, such as in forward scattering. Then factorization at the amplitude level does not hold and Eq.~\eqref{sQEDmainXY} is invalid. Instead, we can write something similar by combining all the operators into one:
\be
\bra{X} \phi^\star \cdots \phi \ket{Z}
 \;\LPeq\; 
\cC(\Sij)  \bra{X} \phi^\star W_\ccO Y_\ccO^\dg  \cdots Y_\rN W_\rN^\dg\phi 
 \ket{Z}
\label{sQEDmainXY2}
\ee
To evaluate the right-hand side one needs to use the SCET Lagrangian which has collinear fields, soft fields, and Glauber interactions
\be
\cL_{\text{SCET}} = \sum_{\ccj} \cL_{\ccj} + \cL_{\scs} + \sum_{\cci \ccj} \LGij
\label{LSCET}
\ee
The labels $\ccj$ and $\scs$ have become quantum numbers labelling the sectors. Thus
interactions between different collinear sectors are forbidden by $\ccj$ and $\scs$ superselection rules 
(the terms in each $\cL_{\ccj}$ or $\cL_{\scs}$ only involve fields with the same quantum numbers). The denominator factors on the right-hand-side of Eq.~\eqref{sQEDmainXY} are replaced by a diagram-level zero-bin subtraction procedure to remove the soft/collinear overlap. With this understanding, when $\LGij$ can be ignored then Eq.~\eqref{sQEDmainXY} and Eq.~\eqref{sQEDmainXY2} are equivalent. 

The terms $\LGij$ in the SCET Lagrangian are the ``Glauber interactions", introduced in~\cite{Rothstein:2016bsq} (see also~\cite{Bauer:2010cc,Fleming:2014rea}). They contain interactions among collinear particles in the $\cci$ and $\ccj$ sectors, as well between these collinear directions and soft particles. The Glauber
interactions are non-local, involving explicit factors of $\frac{1}{\vec{k}_\perp^2}$ where $\vec{k}_\perp$ is the transverse-momentum transfer. The diagrams involving Glauber exchange to lowest order in $g_s$ are the same as those arising by taking the Glauber-scaling limit of diagrams in full QCD. For higher-order diagrams there is not a simple method-of-regions correspondence with QCD. Instead the all-orders form of these terms is fixed by symmetry arguments (reparameterization invariance) and direct matching calculations. In addition to adding these new interactions, SCET with $\LGij$ requires a new zero-bin subtraction: the Glauber limit of soft and collinear graphs must be subtracted diagram-by-diagram so as not to double count. Finally, there is an implicit non-analytic rapidity regulator required to define the theory. Some understanding of why the regulator must be non-analytic was discussed in~\cite{Schwartz:2017nmr}, but understanding the regulator in more detail (e.g. what properties it must have, what is regulator-independent, etc.) is an open area of research.

As the Glauber interactions involve different collinear sectors, they violate factorization. It is nevertheless a non-trivial check that the Glauber interactions can reproduce known factorization-violating effects in full QCD. One such check was performed in~\cite{Schwartz:2017nmr} where 
the generalized splitting function (i.e. one that depends on multiple directions) was calculated 
using the SCET formalism, finding a result in agreement with full QCD. Other checks, such as reproducing the Glauber phase in forward scattering or the BFKL equation, were discussed in~\cite{Rothstein:2016bsq}.

For factorization to hold in hadronic collisions, we would like to be able to separate the dynamics of the proton from the hard scattering. Let's start with a situation where factorization does hold, deep-inelastic scattering (DIS). In pictures, factorization implies
%
%%%%%%%%%%%%.  DIS  %%%%%%%%%%%%%
%
\be
\begin{gathered}
\begin{tikzpicture}
 \node at (0,0) {
% \parbox{20mm} {
\resizebox{40mm}{!}{
     \fmfframe(0,0)(0,0){
    \begin{fmfgraph*}(70,30)
    \fmfstraight
    \fmfleft{L1,L2,L3,L4,L5}
    \fmftop{T1,T2,T3,T4,T5}
    \fmfright{R1,R2,R3,R4,R5,R6,R7,R8,R9}
    \fmf{fermion,tension=6}{L1,v1}
    \fmf{fermion}{v1,v2}
    \fmf{fermion}{v2,R8}
    \fmf{photon,tension=4}{T2,v2}
    \fmf{gluon}{v1,R1}
    \fmf{gluon}{v1,R2}
    \fmf{gluon}{v1,R3}
    \fmf{gluon}{v2,R6}
    \fmf{gluon}{v2,R7}
    \fmfv{d.sh=circle, d.f=30,d.si=0.1w}{v2}
    \fmffreeze
     \fmf{phantom,tension=10}{v1,vs1,v2}
     \fmf{phantom,tension=10}{v2,vs2,R7}
     \fmf{phantom,tension=10}{v1,vs3,R2}
     \fmf{photon,right=0.5,fore=red}{vs3,vx,vs2}
     \fmf{photon,right=0.5,fore=red}{vs1,vs2}
     \fmf{photon,fore=red,tension=0}{vx,R5}
\end{fmfgraph*}
}
}
};
\node at (2.3,0.3) {$n_\ccT$};
\node at (2.3,-0.6) {$n_\ccO$};
\node at (-1.2,-0.4) {$n_\ccO$};
\node at (-0.5,0.1) {$n_\ccO$};
%\draw[black,fill=gray!20] (-1,-1.2) ellipse (1.6 and 0.5);
\end{tikzpicture}
\end{gathered}
\LPeq
\underbrace{
\begin{gathered}
\resizebox{40mm}{!}{
     \fmfframe(0,5)(0,-5){
\begin{fmfgraph*}(70,30)
\fmfstraight
    \fmfleft{L1,L2,L3,L4,L5}
    \fmftop{T1,T2,T3,T4,T5}
    \fmfright{R1,R2,R3,R4,R5,R6,R7,R8,R9}
    \fmf{phantom,tension=6}{L1,v1}
    \fmf{fermion}{v1,v2}
    \fmf{fermion}{v2,R8}
    \fmf{photon,tension=4}{T2,v2}
    \fmf{phantom}{v1,R1}
    \fmf{phantom}{v1,R2}
    \fmf{phantom}{v1,R3}
    \fmf{gluon}{v2,R6}
    \fmf{gluon}{v2,R7}
     \fmfv{d.sh=circle, d.f=30,d.si=0.1w}{v2}
    \fmffreeze
     \fmf{phantom,tension=10}{v1,vs1,v2}
     \fmf{phantom,tension=10}{v2,vs2,R7}
     \fmf{phantom,tension=10}{v1,vs3,R2}
     \fmf{photon,right=0.5,fore=red}{vs1,vx,vs2}
%    \fmffreeze
%     \fmf{phantom,right=0.8}{vs1,vx,vs2}
     \fmf{photon,right=0.2,fore=red}{vs1,vs2}
     \fmf{photon,right=0.2,fore=red,tension=0.8}{vx,R5}
\end{fmfgraph*}
}
}
\end{gathered}}_{\text{hard scattering}}
\times
\underbrace{
\begin{gathered}
\resizebox{40mm}{!}{
     \fmfframe(0,-5)(0,5){
\begin{fmfgraph*}(70,30)
\fmfstraight
    \fmfleft{L1,L2,L3,L4,L5}
    \fmftop{T1,T2,T3,T4,T5}
    \fmfright{R1,R2,R3,R4,R5,R6,R7,R8,R9}
    \fmf{fermion,tension=6}{L1,v1}
    \fmf{phantom}{v1,v2}
    \fmf{phantom}{v2,R8}
    \fmf{fermion,tension=0}{v1,R5}
    \fmf{phantom,tension=4}{T2,v2}
    \fmf{gluon}{v1,R1}
    \fmf{gluon}{v1,R2}
    \fmf{gluon}{v1,R3}
    \fmf{phantom}{v2,R6}
    \fmf{phantom}{v2,R7}
\end{fmfgraph*}
}
}
\end{gathered}}
_{\Sp}
\ee
The first diagram on the right-hand side further factorizes, into soft and collinear (not shown). The $n_\ccO$ and $n_\ccT$ labels indicate to which directions the various lines are collinear. The red shallow curled lines are meant to indicate soft gluons interacting between the two collinear sectors. In words, the dynamics of the initial state, collinear to the $n_\ccO^\mu$ direction, factorizes from the hard scattering.   In equations
\be
\bra{ X_\ccO; X_\ccT; X_\scs} 
\bar{\psi} \gamma^\mu \psi \ket{\gamma, p_\ccO}
\LPeq
\bra{X_\ccT, X_\scs} \bar{\psi} \gamma^\mu \psi \ket{\gamma, p_\ccO'}
\times \Sp(p_1 \to  p_1', X_\ccO)
% \bra{X_\ccO} \bar{\psi} W_\ccO \ket{0}
%\bra{X_\ccT\gamma^\mu \psi \ket{\gamma, p_\ccO}
\ee
where $p_\ccO'{}^\mu = \xi p_\ccO^\mu$ is the momentum participating in the hard partonic scattering. 

Does this imply that factorization holds in terms of parton distribution functions for the proton? Not necessarily. Indeed, factorization of DIS at the amplitude level is neither a necessary condition (factorization could hold at the cross section level) or a sufficient condition (we are just working in perturbation theory here, so we cannot say anything about non-perturbative physics). Nevertheless it is suggestive of factorization and a good start.

For Drell-Yan, the analogous factorization does not hold
%
%%%%%%%%%%%%.  DRELL YAN  %%%%%%%%%%%%%
%
\be
\begin{gathered}
\begin{tikzpicture}
 \node at (0,0) {
\resizebox{40mm}{!}{
     \fmfframe(0,0)(0,0){
    \begin{fmfgraph*}(70,35)
    \fmfkeep{dyfull}
    \fmfstraight
    \fmfleft{L1,L2,L3,L4,L5}
    \fmftop{T1,T2,T3,T4,T5}
    \fmfright{R1,R2,R3,R4,R5,Rmid,Rm5,Rm4,Rm3,Rm2,Rm1}
    \fmf{fermion,tension=6}{L1,v1}
    \fmf{fermion,tension=6}{v5,L5}
    \fmf{fermion}{v1,v}
    \fmf{fermion}{v,v5}
    \fmf{dashes,tension=4}{v,Rmid}
    \fmf{gluon}{v1,R1}
    \fmf{gluon}{v1,R2}
    \fmf{gluon}{v1,R3}
    \fmf{gluon}{v5,Rm1}
    \fmf{gluon}{v5,Rm2}
    \fmf{gluon}{v5,Rm3}
     \fmfv{d.sh=circle, d.f=30,d.si=0.1w}{v}
\fmffreeze
     \fmf{phantom,tension=10}{v1,vs1,v}
     \fmf{phantom,tension=10}{v5,vs5,v}
     \fmf{phantom,tension=10}{v1,vs1b,R1}
     \fmf{phantom,tension=10}{v5,vs5b,Rm1}
     \fmf{photon,left=0.5,fore=red}{vs1,vs5}
     \fmf{photon,left=0.5,fore=red}{vs1b,vx,vs5b}
     \fmf{photon,fore=red,tension=0,right=0.5}{vx,R5}
    \end{fmfgraph*}
}
}
};
\node at (-1.2, 0.6) {$n_\ccO$};
\node at (-0.2, 0.4) {$n_\ccO$};
\node at (2.2,0.6) {$n_\ccO$};
\node at (-1.2,-0.5) {$n_\ccT$};
\node at (-0.2,-0.3) {$n_\ccT$};
\node at (2.2,-0.7) {$n_\ccT$};
%\draw[black,fill=gray!20] (-1,-1.2) ellipse (1.6 and 0.5);
\end{tikzpicture}
\end{gathered}
\slashed{\LPeq}
\hspace{-10mm}
\begin{gathered}
\resizebox{40mm}{!}{
     \fmfframe(0,0)(0,0){
    \begin{fmfgraph*}(70,35)
  \fmfkeep{dyhard}
    \fmfstraight
    \fmfleft{L1,L2,L3,L4,L5}
    \fmftop{T1,T2,T3,T4,T5}
    \fmfright{R1,R2,R3,R4,R5,Rmid,Rm5,Rm4,Rm3,Rm2,Rm1}
    \fmf{phantom,tension=6}{L1,v1}
    \fmf{phantom,tension=6}{v5,L5}
    \fmf{fermion}{v1,v}
    \fmf{fermion}{v,v5}
    \fmf{dashes,tension=4}{v,Rmid}
    \fmf{phantom}{v1,R1}
    \fmf{phantom}{v1,R2}
    \fmf{phantom}{v1,R3}
    \fmf{phantom}{v5,Rm1}
    \fmf{phantom}{v5,Rm2}
    \fmf{phantom}{v5,Rm3}
     \fmfv{d.sh=circle, d.f=30,d.si=0.1w}{v}
\fmffreeze
     \fmf{phantom,tension=10}{v1,vs1,vs2,v}
     \fmf{phantom,tension=10}{v5,vs5,vs6,v}
     \fmf{photon,left=0.6,fore=red}{vs1,vs5}
     \fmf{photon,left=0.4,fore=red}{vs2,vx,vs6}
     \fmf{photon,right=0,fore=red,tension=0}{vx,R3}
    \end{fmfgraph*}
}
}
\end{gathered}
\times
\begin{gathered}
\resizebox{30mm}{!}{
     \fmfframe(0,-5)(0,5){
\begin{fmfgraph*}(70,30)
\fmfstraight
    \fmfleft{L1,L2,L3,L4,L5}
    \fmftop{T1,T2,T3,T4,T5}
    \fmfright{R1,R2,R3,R4,R5,R6,R7,R8,R9}
    \fmf{fermion,tension=6}{L1,v1}
    \fmf{phantom}{v1,v2}
    \fmf{phantom}{v2,R8}
    \fmf{fermion,tension=0}{v1,R5}
    \fmf{phantom,tension=4}{T2,v2}
    \fmf{gluon}{v1,R1}
    \fmf{gluon}{v1,R2}
    \fmf{gluon}{v1,R3}
    \fmf{phantom}{v2,R6}
    \fmf{phantom}{v2,R7}
\end{fmfgraph*}
}
}
\end{gathered}
\times
\begin{gathered}
\resizebox{30mm}{!}{
     \fmfframe(0,-5)(0,5){
\begin{fmfgraph*}(70,30)
\fmfstraight
    \fmfleft{L1,L2,L3,L4,L5}
    \fmfbottom{T1,T2,T3,T4,T5}
    \fmfright{R1,R2,R3,R4,R5,R6,R7,R8,R9}
    \fmf{fermion,tension=6}{L5,v1}
    \fmf{phantom}{v1,v2}
    \fmf{phantom}{v2,R2}
    \fmf{fermion,tension=0}{v1,R5}
    \fmf{phantom,tension=4}{T2,v2}
    \fmf{gluon}{v1,R9}
    \fmf{gluon}{v1,R8}
    \fmf{gluon}{v1,R7}
    \fmf{phantom}{v2,R4}
    \fmf{phantom}{v2,R3}
\end{fmfgraph*}
}
}
\end{gathered}
\label{DYnofact}
\ee 
The problem originates with the soft sector, but affects purely collinear emissions too through virtual effects.

There is a quick way to see what the problem is with soft radiation. Soft radiation factorizes when it is sensitive to only the net color charge going in a particular direction. This is similar to Gauss's law in electromagnetism -- at large distances only the net charge in a region matters to the leading approximation. But what is the net charge when there are incoming {\it and} outgoing particles in the same sector? For DIS, we can use a trick and move all the color of the $n_1$-collinear sector to a Wilson line in the scattering operator. By color conservation, we know the net color that the outgoing $n_\ccT$ jet sees is negative of the net $n_\ccO$ color. For Drell-Yan, this trick does not work. Indeed, the reason the minimal number of colored particles needed for factorization breaking is four is that with three or fewer, color conservation can be used to ensure that the scattering is only sensitive to the net color. This argument can be found in~\cite{Catani:2011st} and is explained in depth in~\cite{Schwartz:2017nmr}.

The obstruction to factorization can be traced to the invalidity of the eikonal approximation in describing soft radiation. 
Soft radiation refers to regions of real or virtual phase space in which all the components of a gluon's momentum are small compared to the energy scale $Q$ of the particles emitting the soft radiation, $|k^\mu| \ll Q$. 
For two momenta $p^\mu$ and $q^\mu$ we can always write
\be
\frac{1}{(p+k)^2 + i\fme } = { \frac{1}{2p\cdot k + i\fme}}  - { \frac{k^2}{((p+k)^2 +i\fme)(2 p\cdot k+ i\fme)} }\label{softsum}
\ee
The eikonal or Grammer-Yennie approximation amounts to dropping the second term on the right with respect to the first.  When the eikonal approximation can be used, soft-collinear factorization holds~\cite{FS1,FS2}. 
For $p^\mu$ hard ($p^0 \sim Q$) and $k^\mu$ soft ($k^0 \ll Q$), it seems like the second term can always be dropped since $\frac{k^2}{p\cdot k} \sim \frac{k^0}{Q} \ll 1$.
Unfortunately, it is not enough for $k^0 \ll Q$. The problem is that it is possible for $k^\mu \ll Q$ with
$\frac{k^2}{p\cdot k} \sim 1$. For example, $k^\mu \to 0$ holding $\frac{k_x^2}{k_0 Q}$ fixed. In light-cone coordinate the scaling of such mode can be written as $k\sim Q(\lambda^2, \lambda^2, \lambda)$. This kind of scaling toward $k^\mu = 0$ is known as Glauber scaling and can foil factorization. 

An intuitive way to understand the obstruction from Glauber scaling is to contrast the soft limit $k^\mu \to 0$ with the limit $Q\to \infty$. If $Q$ decouples, so we can  take $Q \to \infty$ to describe soft radiation, then the sources emitting soft radiation can be treated as scale-invariant Wilson lines $Y_\ccn$. Soft-collinear factorization is based on being able to use this scale-invariant limit. Glauber scaling involves $Q$ in an essential way, so it obstructs scale-invariance.

To prove that the eikonal limit can be used, one must show that the region of soft momenta described by Glauber scaling is contained in scaleless integrations over soft momenta. In CSS language, this happens when there is no pinch in the Glauber region. In the language of SCET, it is when contributions form the Glauber Lagrangian ($\LGij$ in Eq.\eqref{LSCET}) are not exactly canceled
by the Glauber-soft and Glauber-collinear zero-bin subtractions~\cite{Rothstein:2016bsq}. The relation between these two ideas was further explored in~\cite{Schwartz:2017nmr}. 

So factorization does not hold in Drell-Yan at the amplitude level, as in Eq.~\eqref{DYnofact}, because the Glauber region is not contained in the soft region. Graphs for Drell-Yan have a pinch in the Glauber region. 
What CSS showed was that despite the non-factorization at the amplitude level, factorization still holds for Drell-Yan as long as the observable is inclusive over all QCD radiation. In pictures
%
%%%%%%%%%%%%.  CSS  %%%%%%%%%%%%%
%
\be
\sum_X \int d \Pi_X \left|
\begin{gathered}
\resizebox{40mm}{!}{
     \fmfframe(0,0)(0,0){
% \fmfreuse{dyfull2}
    \begin{fmfgraph*}(70,35)
    \fmfkeep{dyfull}
    \fmfstraight
    \fmfleft{L1,L2,L3,L4,L5}
    \fmftop{T1,T2,T3,T4,T5}
    \fmfright{R1,R2,R3,R4,R5,Rmid,Rm5,Rm4,Rm3,Rm2,Rm1}
    \fmf{fermion,tension=6}{L1,v1}
    \fmf{fermion,tension=6}{v5,L5}
    \fmf{fermion}{v1,v}
    \fmf{fermion}{v,v5}
    \fmf{dashes,tension=4}{v,Rmid}
    \fmf{gluon}{v1,R1}
    \fmf{gluon}{v1,R2}
    \fmf{gluon}{v1,R3}
    \fmf{gluon}{v5,Rm1}
    \fmf{gluon}{v5,Rm2}
    \fmf{gluon}{v5,Rm3}
     \fmfv{d.sh=circle, d.f=30,d.si=0.1w}{v}
\fmffreeze
     \fmf{phantom,tension=10}{v1,vs1,v}
     \fmf{phantom,tension=10}{v5,vs5,v}
     \fmf{phantom,tension=10}{v1,vs1b,R1}
     \fmf{phantom,tension=10}{v5,vs5b,Rm1}
     \fmf{photon,left=0.5,fore=red}{vs1,vs5}
     \fmf{photon,left=0.5,fore=red}{vs1b,vx,vs5b}
     \fmf{photon,fore=red,tension=0,right=0.5}{vx,R5}
    \end{fmfgraph*}
}}
\end{gathered}
\right|^2
\LPeq
\sum_X \int d \Pi_X \left|
\begin{gathered}
\resizebox{40mm}{!}{
     \fmfframe(-25,0)(25,0){
    \begin{fmfgraph*}(70,35)
%   \fmfkeep{dyhard}
    \fmfstraight
    \fmfleft{L1,L2,L3,L4,L5}
    \fmftop{T1,T2,T3,T4,T5}
    \fmfright{R1,R2,R3,R4,R5,Rmid,Rm5,Rm4,Rm3,Rm2,Rm1}
    \fmf{phantom,tension=6}{L1,v1}
    \fmf{phantom,tension=6}{v5,L5}
    \fmf{fermion}{v1,v}
    \fmf{fermion}{v,v5}
    \fmf{dashes,tension=4}{v,Rmid}
    \fmf{phantom}{v1,R1}
    \fmf{phantom}{v1,R2}
    \fmf{phantom}{v1,R3}
    \fmf{phantom}{v5,Rm1}
    \fmf{phantom}{v5,Rm2}
    \fmf{phantom}{v5,Rm3}
     \fmfv{d.sh=circle, d.f=30,d.si=0.1w}{v}
\fmffreeze
     \fmf{phantom,tension=10}{v1,vs1,vs2,v}
     \fmf{phantom,tension=10}{v5,vs5,vs6,v}
     \fmf{photon,left=0.6,fore=red}{vs1,vs5}
     \fmf{photon,left=0.4,fore=red}{vs2,vx,vs6}
     \fmf{photon,right=0,fore=red,tension=0}{vx,R3}
    \end{fmfgraph*}
% \fmfreuse{dyhard2}
}}
\end{gathered}
\hspace{-15mm}
\right|^2
\times
\int  |\Sp|^2
\times
\int  |\Sp|^2
\ee 

To prove this, CSS showed that for soft gluons with momenta $k^\mu_i$ that interact with collinear momenta $p^\mu_j$, the transverse components of the soft
momenta can be neglected. That is, we can replace $k^\mu = (k^+,k^-,\vec k_\perp)$ by $k^\mu = (k^-,0,0)$ where $k^- $ is the component backwards to the jet direction (i.e. $k \cdot p \LPeq  k^+ p^-$). Once we know the result is unaffected by dropping transverse components then there is only one way to scale $k^\mu \to 0$; there is no Glauber region and the eikonal approximation  $\frac{k^2}{k\cdotp p} \to 0$ is always justified around $k^2=0$. 
In more detail, CSS's argument used old-fashioned perturbation theory.  They showed that the same cross section results with and without neglecting $k_\perp$ and $k_+$ by summing over all possible cuts in spectator-spectator graphs
%
%%%%%%%%%%%%%% CSS Glauber %%%%%%%%%%%
%
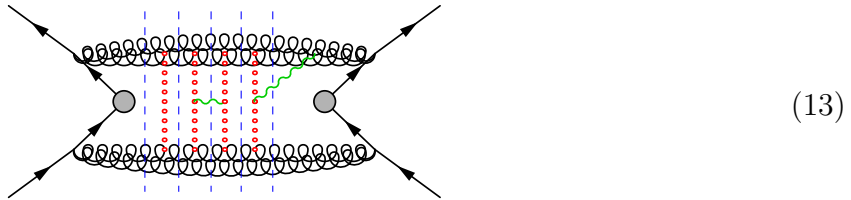
\be
\begin{gathered}
\begin{tikzpicture}
 \node at (0,0) {
\resizebox{60mm}{!}{
     \fmfframe(0,0)(0,0){
    \begin{fmfgraph*}(90,40)
\fmfstraight
    \fmfleft{L1,L2}
    \fmfright{R1,R2}
    \fmf{fermion}{L1,v1}
    \fmf{fermion}{v1,v}
    \fmf{fermion}{v,v2}
    \fmf{fermion}{v2,L2}
    \fmf{fermion}{R1,w1}
    \fmf{fermion}{w1,w}
    \fmf{fermion}{w,w2}
    \fmf{fermion}{w2,R2}
    \fmf{phantom,tension=0.5}{v1,t1,t2,t3,t4,t5,t6,t7,t8,t9,w1}
    \fmf{phantom,tension=0.5}{v2,b1,b2,b3,b4,b5,b6,b7,b8,b9,w2}  
    \fmf{phantom,tension=0.5}{v,w}   
\fmffreeze 
    \fmf{dbl_dots,fore=red}{t3,b3} 
    \fmf{dbl_dots,fore=red}{t4,x1,b4}
    \fmf{dbl_dots,fore=red}{t5,x2,b5} 
    \fmf{photon,fore=(0,,0.8,,0) ,tension=0}{x1,x2} 
    \fmf{dbl_dots,fore=red}{t6,x3,b6} 
    \fmf{photon,fore=(0,,0.8,,0),tension=0}{x3,b8}    
     \fmfv{d.sh=circle, d.f=30,d.si=0.05w}{v}
     \fmfv{d.sh=circle, d.f=30,d.si=0.05w}{w}
     \fmffreeze
\fmf{gluon,right=0.1}{v1,w1}
\fmf{gluon,right=0}{v1,w1}
\fmf{gluon,left=0.1}{v2,w2}
\fmf{gluon,left=0}{v2,w2}
     \end{fmfgraph*}
}
}
};
\draw[dashed,blue] (-1.1,1.2) -- (-1.1,-1.2);
\draw[dashed,blue] (-0.65,1.2) -- (-0.65,-1.2);
\draw[dashed,blue] (-0.22,1.2) -- (-0.22,-1.2);
\draw[dashed,blue] (0.18,1.2) -- (0.18,-1.2);
\draw[dashed,blue] (0.6,1.2) -- (0.6,-1.2);
\end{tikzpicture}
\end{gathered}
\ee
as well as in active-spectator graphs. In addition, their argument exploited observations about the polarizations of the gluons coupling to the collinear sector.  
While there is no doubt that the CSS proof is correct, it gives little guidance as to what we might learn about situations where factorization is violated. 

\subsection{Factorization violation}
Factorization holds for the Drell-Yan process where only the lepton momenta are measured. 
A typical Drell-Yan observable is the
\begin{itemize}
\item {\bf Lepton transverse momentum} ${\mathbf q_T}$: defined in the Drell-Yan process as the transverse momentum of the lepton pair $q_T = |\vec{p}_{1,\perp} + \vec{p}_{2,\perp}|$ where $p_1^\mu$ and $p_2^\mu$ are the lepton momenta.
\end{itemize}
By momentum conservation, $q_T$ is also equal to the net transverse momentum of all the hadronic final state particles. 

For other observables, results about factorization are murkier. Two other variables we will discuss are
\begin{itemize}
\item {\bf Hadronic transverse energy ${\mathbf E_T}$}. Assuming all measured particles are massless, transverse energy is the scalar sum of the particles' transverse momenta $E_T = \sum_j | \vec{p}_{j T}|$.
\item {\bf Beam thrust}. This is a hadronic event shape observable. For a process involving vector boson production ($pp \to V + X$), beam thrust defined in hadronic center-of-mass frame is 
$\tau_B = \frac{1}{Q} \sum_j | \vec{p}_{j T}| \exp (-|Y_j| - Y_V)$
where $Q$ and $Y_V$ are the vector-boson mass and rapidity and $\vec{p}_{j T}$ and $Y_j$ are the transverse
momenta and rapidities of the other  final state particles~\cite{Stewart:2009yx,Stewart:2010pd}.
\end{itemize}
$q_T$ was shown to factorize by CSS. For the other observables no rigorous results are known. 
 The general lore is that factorization violation shows up in event generator simulations as sensitivity of an observable to the underlying event.~\cite{Gaunt:2014ska,Zeng:2015iba,Rothstein:2016bsq,Papaefstathiou:2010bw,Grazzini:2014uha,AlcarazMaestre:2012vp}.
 
 Consider for example beam thrust. In the original papers~\cite{Banfi:2004nk,Stewart:2009yx,Stewart:2010pd}, beam thrust was thought to 
 factorize. A rough argument along the lines of CSS about why the Glauber contributions should cancel was presented in~\cite{Stewart:2009yx}. Unfortunately, beam thrust is extremely sensitive to models of the underlying event; turning underlying event on or off in simulations has an order-one effect on the beam thrust distribution, completely destroying predictivity of the theoretical calculation~\cite{Alioli:2016wqt}. Beam thrust was revisited by Gaunt~\cite{Gaunt:2014ska} who argued that the pure Glauber contribution is factorization-violating.
 Similar conclusions were reached in~\cite{Rothstein:2016bsq}.

Let us briefly review Gaunt's argument. He first considered summing cuts of a diagram with a single Glauber exchange. There are two possible cuts.
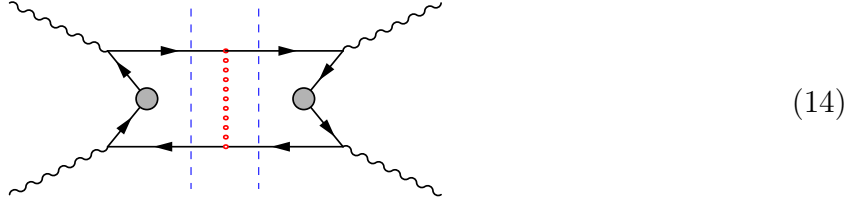
\be
\begin{gathered}
\begin{tikzpicture}
 \node at (0,0) {
\resizebox{60mm}{!}{
     \fmfframe(0,0)(0,0){
    \begin{fmfgraph*}(90,40)
\fmfstraight
    \fmfleft{L1,L2}
    \fmfright{R1,R2}
    \fmf{photon}{L1,v1}
    \fmf{fermion}{v1,v}
    \fmf{fermion}{v,v2}
    \fmf{photon}{v2,L2}
    \fmf{photon}{R1,w1}
    \fmf{fermion}{w2,w}
    \fmf{fermion}{w,w1}
    \fmf{photon}{w2,R2}
    \fmf{quark,tension=0.5}{w1,t1,v1}
    \fmf{quark,tension=0.5}{v2,b1,w2}  
    \fmf{phantom,tension=0.5}{v,w}   
\fmffreeze 
    \fmf{dbl_dots,fore=red}{t1,b1} 
     \fmfv{d.sh=circle, d.f=30,d.si=0.05w}{v}
     \fmfv{d.sh=circle, d.f=30,d.si=0.05w}{w}
     \end{fmfgraph*}
}
}
};
\draw[dashed,blue] (-0.5,1.2) -- (-0.5,-1.2);
\draw[dashed,blue] (0.4,1.2) -- (0.4,-1.2);
\end{tikzpicture}
\end{gathered}
\ee
The two cuts contribute to the cross section with opposite signs and the sum of them vanishes, independent of the observable. Next, he looked at diagrams with two Glauber gluons exchanged. There are three cuts through such graphs:
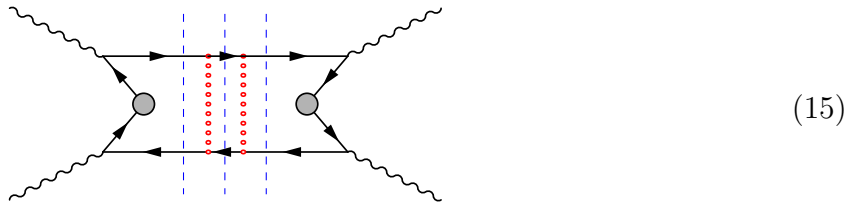
\be
\begin{gathered}
\begin{tikzpicture}
 \node at (0,0) {
\resizebox{60mm}{!}{
     \fmfframe(0,0)(0,0){
    \begin{fmfgraph*}(90,40)
\fmfstraight
    \fmfleft{L1,L2}
    \fmfright{R1,R2}
    \fmf{photon}{L1,v1}
    \fmf{fermion}{v1,v}
    \fmf{fermion}{v,v2}
    \fmf{photon}{v2,L2}
    \fmf{photon}{R1,w1}
    \fmf{fermion}{w2,w}
    \fmf{fermion}{w,w1}
    \fmf{photon}{w2,R2}
    \fmf{quark,tension=0.5}{w1,t1}
    \fmf{quark,tension=1.5}{t1,t2}
    \fmf{quark,tension=0.5}{t2,v1}
    \fmf{quark,tension=0.5}{v2,b2}  
    \fmf{quark,tension=1.5}{b2,b1}  
    \fmf{quark,tension=0.5}{b1,w2}  
    \fmf{phantom,tension=0.5}{v,w}   
\fmffreeze 
    \fmf{dbl_dots,fore=red}{t1,b1} 
    \fmf{dbl_dots,fore=red}{t2,b2} 
     \fmfv{d.sh=circle, d.f=30,d.si=0.05w}{v}
     \fmfv{d.sh=circle, d.f=30,d.si=0.05w}{w}
     \end{fmfgraph*}
}
}
};
\draw[dashed,blue] (-0.6,1.2) -- (-0.6,-1.2);
\draw[dashed,blue] (-0.05,1.2) -- (-0.05,-1.2);
\draw[dashed,blue] (0.5,1.2) -- (0.5,-1.2);
\end{tikzpicture}
\end{gathered}
\ee
Gaunt argued that for the observables $E_T$ and beam thrust, the Glauber effects do not cancel when summing over these cuts and there is a factorization-violating effect. Gaunt showed that for a single Glauber exchange diagram, one can achieve the cancellation by relabelling the transverse components of integrated momentum. However, for double Glauber exchange, Gaunt argued that no change of variables should exist to achieve the cancellation.

A similar argument to Gaunt's can be found in Stewart and Rothstein's treatise on Glauber gluons in SCET. These authors argued that when all the ladder graphs are summed, the result is a Glauber phase $\exp(i \phi (\vec b_\perp))$ with $\vec b_\perp$ the impact parameter conjugate to the relative transverse momentum of the two spectators $\Delta \vec p_\perp$. When one calculates the cross section there is an integral
over this $\vec b_\perp$ as well as the $\vec b_\perp'$ for the complex-conjugate amplitude. The argument is that these phases do not cancel unless the observable
 is independent of $\Delta \vec p_\perp$. Therefore factorization should be violated by Glauber ladder diagrams for any observable other than $q_T$. 

While both Gaunt and Rothstein/Stewart provide strong arguments, their arguments rely on assuming no unusual cancellations can happen. Their choices of variables are certainly suggestive 
that it would take a miracle for cancellation to happen. 
But miracles do happen when there are symmetries. In this paper, we show that a choice of variables that respects the scale invariance of the problem
makes a general cancelation manifest. 

Our main result is that the sum of all the ladder graphs is zero when integrated over the kinematic variables other than a single infrared-safe observable:
\be
 \int d z d \bar{z} \sum_{
\text{cuts}} \begin{gathered}
\begin{tikzpicture}
 \node at (0,0) {
\resizebox{60mm}{!}{
     \fmfframe(0,0)(0,0){
    \begin{fmfgraph*}(90,40)
\fmfstraight
    \fmfleft{L1,L2}
    \fmfright{R1,R2}
    \fmf{photon}{L1,v1}
    \fmf{fermion}{v1,v}
    \fmf{fermion}{v,v2}
    \fmf{photon}{v2,L2}
    \fmf{photon}{R1,w1}
    \fmf{fermion}{w2,w}
    \fmf{fermion}{w,w1}
    \fmf{photon}{w2,R2}
    \fmf{quark,tension=0.5}{w1,t1,t2,t3,t4,v1}
    \fmf{quark,tension=0.5}{v2,b4,b3,b2,b1,w2}  
    \fmf{phantom,tension=0.2}{v,w}   
\fmffreeze 
    \fmf{dbl_dots,fore=red}{t1,b1} 
    \fmf{dbl_dots,fore=red}{t2,b2} 
    \fmf{dbl_dots,fore=red}{t3,b3} 
    \fmf{dbl_dots,fore=red}{t4,b4} 
     \fmfv{d.sh=circle, d.f=30,d.si=0.05w}{v}
     \fmfv{d.sh=circle, d.f=30,d.si=0.05w}{w}
     \end{fmfgraph*}
}
}
};
\draw[dashed,blue] (-1.5,1.2) -- (-1.5,-1.2);
\draw[dashed,blue] (-0.8,1.2) -- (-0.8,-1.2);
\draw[dashed,blue] (-0.05,1.2) -- (-0.05,-1.2);
\draw[dashed,blue] (0.75,1.2) -- (0.75,-1.2);
\draw[dashed,blue] (1.45,1.2) -- (1.45,-1.2);
\end{tikzpicture}
\end{gathered}
= 0
\ee
The variables $z$ and $\bar{z}$ that respect the symmetry of the scattering will be discussed in the next section.

The leading factorization violating effect comes from graphs with soft gluons exchanged between the Glauber rungs:
\be
\begin{gathered}
\begin{tikzpicture}
 \node at (0,0) {
\resizebox{60mm}{!}{
     \fmfframe(0,0)(0,0){
    \begin{fmfgraph*}(90,40)
\fmfstraight
    \fmfleft{L1,L2}
    \fmfright{R1,R2}
    \fmf{photon}{L1,v1}
    \fmf{fermion}{v1,v}
    \fmf{fermion}{v,v2}
    \fmf{photon}{v2,L2}
    \fmf{photon}{R1,w1}
    \fmf{fermion}{w2,w}
    \fmf{fermion}{w,w1}
    \fmf{photon}{w2,R2}
    \fmf{quark,tension=0.5}{w1,t1}
    \fmf{quark,tension=1.5}{t1,t2}
    \fmf{quark,tension=0.5}{t2,v1}
    \fmf{quark,tension=0.5}{v2,b2}  
    \fmf{quark,tension=1.5}{b2,b1}  
    \fmf{quark,tension=0.5}{b1,w2}  
    \fmf{phantom,tension=0.5}{v,w}   
\fmffreeze 
    \fmf{dbl_dots,fore=red}{t1,x1,x2,b1} 
    \fmf{dbl_dots,fore=red}{t2,y1,y2,b2}
 \fmffreeze
    \fmf{gluon,fore=(0,,0.8,,0)}{x1,y1} 
    \fmf{gluon,fore=(0,,0.8,,0)}{x2,y2} 
     \fmfv{d.sh=circle, d.f=30,d.si=0.05w}{v}
     \fmfv{d.sh=circle, d.f=30,d.si=0.05w}{w}
     \end{fmfgraph*}
}
}
};
\draw[dashed,blue] (-0.6,1.2) -- (-0.6,-1.2);
\draw[dashed,blue] (-0.05,1.2) -- (-0.05,-1.2);
\draw[dashed,blue] (0.5,1.2) -- (0.5,-1.2);
\end{tikzpicture}
\end{gathered}
\ee
These graphs involve the {\it Lipatov vertex} -- the 3-gluon vertex connected 2 Glauber gluons to a soft gluons~\cite{Kuraev:1976ge}. The Lipatov vertex is currently known to one loop in QCD~\cite{Fadin:1993wh,DelDuca:1998cx}. The Lipatov vertex is embedded in the Glauber operator in SCET in Eq.~\eqref{LSCET}.

\section{Glauber ladder graphs \label{sec:ladders}}
In this section, we prove that summing over all Glauber ladder graphs, factorization is preserved.
We start with the 2-loop result which demonstrates all the essential features of our argument. We then discuss the all-orders result.

\subsection{2-loop Glauber cancellation \label{sec:2loop}}
Consider the amplitude for Drell-Yan production from quark initial states with two outgoing gluons:
\be  
\cM  \Big[ q (p_\ccO)  +  \bar{q } ( p_\ccT) \rightarrow   g (p_\ccTH) +  g( p_\ccF)  +    V \Big]
\ee
At up to 2-loop order order, the diagrams that can produce factorization violating effects are
\begin{align} 
\cM_\zero
=
 \begin{tikzpicture}[baseline={([yshift=-.5ex]current bounding box.center)},scale=1.1]
\draw (-1,-0.8) -- (0,0);
\draw (-1,0.8) -- (0,0);
\draw [fill=\blobcolor] (0,0) circle [radius = 0.15];
\draw[color=black,decorate,decoration={gluon, amplitude=1.2pt,    segment length=1.8pt, aspect=0.6}] (-0.8,0.6) -- (0.6,0.6);
    \draw (-0.8,0.6) -- (0.6,0.6);
\draw[color=black,decorate,decoration={gluon, amplitude=1.2pt,
    segment length=1.8pt, aspect=0.6}] (-0.8,-0.6) -- (0.6,-0.6);
    \draw (-0.8,-0.6) -- (0.6,-0.6);    
 \draw [-{Latex[length=3pt]}] (-0.9,0.9) node[above,  scale=0.5]{$\ccO$}--(-0.7,0.7);
 \draw [-{Latex[length=3pt]}] (-0.9,-0.9) node[below,  scale=0.5]{$\ccT$}--(-0.7,-0.7);
 \draw [-{Latex[length=3pt]}] (0.3,0.7)--(0.6,0.7) node[above, scale=0.5]{$\ccTH$};
  \draw [-{Latex[length=3pt]}] (0.3,-0.7)--(0.6,- 0.7) node[below, scale=0.5]{$\ccF$};
 \draw (0.75,0.6) node[scale=0.5]{$a$};
 \draw (0.75,- 0.6) node[ scale=0.5]{$b$}; 
\end{tikzpicture},
\qquad  
\cM_{G \one} = \begin{tikzpicture}[baseline={([yshift=-.5ex]current bounding box.center)},scale=1.1]
\draw (-1,-0.8) -- (0,0);
\draw (-1,0.8) -- (0,0);
\draw [fill=\blobcolor] (0,0) circle [radius = 0.15];
\draw[color=black,decorate,decoration={gluon, amplitude=1.2pt,
    segment length=1.8pt, aspect=0.6}] (-0.8,0.6) -- (0.8,0.6);
    \draw (-0.8,0.6) -- (0.8,0.6);
\draw[color=black,decorate,decoration={gluon, amplitude=1.2pt,
    segment length=1.8pt, aspect=0.6}] (-0.8,-0.6) -- (0.8,-0.6);
    \draw (-0.8,-0.6) -- (0.8,-0.6);    
\draw [red,fill=red] (0.4,0.6) circle [radius=0.8pt];
\draw [red,fill=red] (0.4,-0.6) circle [radius=0.8pt];
\draw [red,decorate glaubr] (0.4,0.6) to [bend left=0] (0.4,-0.6);
 \draw [-{Latex[length=3pt]}] (0.5,0.2)--(0.5,0.0)node[below, scale=0.4]{$\tiny{ \ell}$};
 \draw [-{Latex[length=3pt]}] (-0.9,0.9) node[above,  scale=0.5]{$\ccO$}--(-0.7,0.7);
 \draw [-{Latex[length=3pt]}] (-0.9,-0.9) node[below,  scale=0.5]{$\ccT$}--(-0.7,-0.7);
 \draw [-{Latex[length=3pt]}] (0.5,0.7)--(0.8,0.7) node[above, scale=0.5]{$\ccTH$};
  \draw [-{Latex[length=3pt]}] (0.5,-0.7)--(0.8,- 0.7) node[below, scale=0.5]{$\ccF$};
\end{tikzpicture}, 
\qquad 
\cM_{G\two} = 
\begin{tikzpicture}[baseline={([yshift=-.5ex]current bounding box.center)},scale=1.1]
\draw (-1,-0.8) -- (0,0);
\draw (-1,0.8) -- (0,0);
\draw [fill=\blobcolor] (0,0) circle [radius = 0.15];
\draw[color=black,decorate,decoration={gluon, amplitude=1.2pt,
    segment length=1.8pt, aspect=0.6}] (-0.8,0.6) -- (0.9,0.6);
    \draw (-0.8,0.6) -- (0.9,0.6);
\draw[color=black,decorate,decoration={gluon, amplitude=1.2pt,
    segment length=1.8pt, aspect=0.6}] (-0.8,-0.6) -- (0.9,-0.6);
    \draw (-0.8,-0.6) -- (0.9,-0.6);    
\draw [red,fill=red] (0.3,0.6) circle [radius=0.8pt];
\draw [red,fill=red] (0.33,-0.6) circle [radius=0.8pt];
\draw [red,decorate glaubr] (0.3,0.6) to [bend left=0] (0.3,-0.6);
\draw [red,fill=red] (0.6,0.6) circle [radius=0.8pt];
\draw [red,fill=red] (0.6,-0.6) circle [radius=0.8pt];
\draw [red,decorate glaubr] (0.6,0.6) to [bend left=0] (0.6,-0.6);
\draw [-{Latex[length=3pt]}] (0.65,0.2)--(0.65,0.0)node[below, scale=0.4]{$\tiny{ k}$};
 \draw [-{Latex[length=3pt]}] (0.35,0.2)--(0.35,0.0)node[below, scale=0.4]{$\tiny{ \ell-k}$};
 \draw [-{Latex[length=3pt]}] (-0.9,0.9) node[above,  scale=0.5]{$\ccO$}--(-0.7,0.7);
 \draw [-{Latex[length=3pt]}] (-0.9,-0.9) node[below,  scale=0.5]{$\ccT$}--(-0.7,-0.7);
 \draw [-{Latex[length=3pt]}] (0.6,0.7)--(0.9,0.7) node[above, scale=0.5]{$\ccTH$};
  \draw [-{Latex[length=3pt]}] (0.6,-0.7)--(0.9,- 0.7) node[below, scale=0.5]{$\ccF$};
\end{tikzpicture}
\end{align}
In these diagrams, the red dotted exchanges are Glauber gluons. The Feynman rules for these gluons can be found in~\cite{Bauer:2000yr} using SCET, or more simply by power expanding the amplitude in QCD using Glauber scaling: $|\vec k_\perp| \gg k^+, k^-$. Glauber scaling makes the propagators for a momentum $k^\mu$ depend only on $\vec k_\perp$, i.e. $k^2 = k^+k^- - \vec k_\perp^2 \LPeq - \vec k_\perp^2$. Since the propagators only involve transverse momentum, they are more like contact interactions than propagators but we draw them as extended lines to exhibit their origin from diagrams in QCD. 

The leading order matrix element is
\be
 \cM_\zero =  
  \frac{1}{s_{\ccO\ccTH} s_{\ccT \ccF}}  N_{\mu\nu} p_{\ccTH,\perp}^{\mu} p_{\ccF,\perp}^\nu
  \label{M0N}
\ee
We assume the outgoing gluons are collinear to the incoming quarks, so the tree-level
matrix element is given at leading power by the product of splitting amplitudes:
\be
 \cM_\zero \LPeq    \Sp^\zero (\bar{ \ccO}_{q}, \ccTH_g )   \, \Sp^\zero ( \bar{\ccT}_{\bar q }, \ccF_g ) \, \bar{v}_{\bar n } \Gamma u_n 
\ee
This leads to a simplified form for the matrix $N_{\mu\nu}$ in Eq.~\eqref{M0N}:
\be
 N_{\mu \nu}
  = \bar{v}_{\bar n } ( p_\ccT )    \frac{ n\slash}{2}  \frac{\bar{n}\slash}{ 2 }    \left( \frac{2 \,  \pol_{\ccF, \nu}  }{ n \cdot p_\ccF  }   
  +\frac{   \pol\slash_{\ccF,\perp}  \gamma_{\perp,\nu}       }{  n \cdot  ( p_\ccT - p_\ccF )   }    \right)
   \Gamma  
 \left(  \frac{2\,  \pol_{\ccTH, \mu } }{ \bar n \cdot   p_\ccTH     }  +  \frac{  \gamma_{\perp, \mu }     \pol\slash_{\ccTH,\perp}   }{ \bar n \cdot  ( p_\ccO - p_\ccTH )   }    \right)   \frac{n\slash}{2}  \frac{\bar n\slash}{2}    u_n (p_\ccO )    
\ee

By inserting  the Glauber potential  operators and integrating out the light-cone components of the Glauber momenta, 
the one-loop and two-loop Glauber ladder graphs reduce to the following integrals in the $d=2- 2\ep$ transverse plane,
\begin{multline} 
 \cM_{G\one}   =   \left( -\frac{i g_s^2  }{2}\right)  (\TT_\ccTH \cdot \TT_\ccF)   \frac{1}{s_{\ccO\ccTH} s_{\ccT \ccF}}   N_{\mu\nu}     \\
  \times \vec{p}^2_{\ccTH,\perp}   \vec{p}^2_{\ccF,\perp}   \int \frac{d^{d-2} \ell}{(2 \pi)^{d-2} }   
 \frac{1}{(\vec{p}_{\ccTH,\perp }+ \vec{\ell}_\perp)^2}  
 \frac{1}{(\vec{p}_{\ccF,\perp} - \vec{\ell}_\perp)^2 }
  \frac{1}{\vec{\ell}_\perp^2}  (p^\mu_{\ccTH,\perp}  + \ell^\mu_\perp)(p^\nu_{\ccF,\perp} - \ell^\nu_\perp)   
 \end{multline}
where $s_{\cci\ccj} = (p_\cci + p_\ccj)^2$  and
\begin{multline}
\cM_{G\two} =    \left( -\frac{ g_s^4  }{8}\right)  (\TT_\ccTH \cdot \TT_\ccF)^2   \frac{1}{s_{\ccO\ccTH} s_{\ccT \ccF}}    N_{\mu\nu}     \\
 \times \vec{p}^2_{\ccTH,\perp}   \vec{p}^2_{\ccF,\perp}     \int \frac{d^{d-2} \ell}{(2 \pi)^{d-2} }   \frac{d^{d-2} k}{(2 \pi)^{d-2} }   
 \frac{1}{(\vec{p}_{\ccTH,\perp }+ \vec{\ell}_\perp)^2}  
 \frac{1}{(\vec{p}_{\ccF,\perp} - \vec{\ell}_\perp)^2 }
  \frac{1}{(\vec{\ell}_\perp - \vec{k }_\perp    )^2}  
  \frac{1 }{\vec{k}_\perp^2}(p^\mu_{\ccTH,\perp}  + \ell^\mu_\perp)(p^\nu_{\ccF,\perp} - \ell^\nu_\perp)   
\end{multline}  
Here $s_{ij} = (p_i + p_j)^2$, and we use the convention that all particle momenta are incoming; outgoing momenta are obtained by crossing $p \to -p$. 
We have computed the amplitudes $\cM_0, \cM_{G1}$ and $\cM_{G2}$ in $d$ dimensions as a series in $\ep = \frac{4-d}{2}$,
using some master integrals from~\cite{Chavez:2012kn}. Details are given in the appendices and will be summarized here.

In computing the matrix-element squared, we find that up to 2-loop order the IR divergences exactly cancel. This cancellation is non-trivial, but also expected from general results. 
Of the 8 possible helicity combinations, only 2 are independent. 
 If both collinear quark/gluon pairs have the same helicity (or if both pairs have opposite helicity), such that $h_\ccTH = \pm h_\ccO, h_\ccF= \pm h_\ccT$,  then we find (see Appendix~\ref{app:2loopreg})
 \begin{align}
| \cM |^2_{++} & \equiv  \Big| \cM_\zero + \cM_{G\one} + \cM_{G\two} + \cdots \Big|^2_{h_\ccTH= \pm h_\ccO, h_\ccF= \pm h_\ccT}   \nn \\
& = \Big| \cM_\zero (+,+,\pm,\pm) \Big|^2 \left[
  1+
 \frac{   {\rm c}_{\rm 2G} }{ {\rm c}_\zero } \,  \frac{ \alpha_s^2}{8}(u+v+1) \ln^2 \frac{u}{v}   + \cO(\alpha_s^3)\right]
 \label{hel1}
\end{align} 
If one collinear quark/gluon pair has the same helicity and the other pair has the opposite helicity, such that $h_\ccTH = \pm h_\ccO, h_\ccF= \mp h_\ccT$, then
\begin{align}  
| \cM |^2_{+-} &\equiv
\Big| \cM_\zero + \cM_{G\one} + \cM_{G\two} + \cdots \Big|^2_{h_\ccTH= \pm h_\ccO, h_\ccF= \mp h_\ccT}   \nn \\
& = 
 \Big|  \cM_\zero (+,-,+,+) \Big|^2 \left[
1+
\frac{   {\rm c}_{\rm 2G} }{ {\rm c}_\zero }  \, \frac{ \alpha_s^2}{4}   \ln u \ln v + \cO(\alpha_s^3)\right]
 \label{hel2}
\end{align}
where the color sum is implicit in $|\cM|^2$. The 2-loop and tree-level color factors are $    {\rm c}_{\rm 2G} = \frac18 C_A^2 (C_A^2 +2) C_F, \,   {\rm c}_\zero  =  C_A C_F^2$, and  
\be
\label{M0def}
\big| \cM_\zero  (h_\ccO, h_\ccT,  h_\ccTH,  h_\ccF  ) \big|^2  
=
{\rm c}_\zero \,  \frac{1}{q_T^4 \, u v } \,
 \frac{ s_{ \ccT \ccTH } s_{\ccO \ccF } }{ s_{\ccO \ccT}^2  } \, \Big| \text{Split}_{-h_\ccO, h_\ccTH} \Big( \frac{1}{ z_{\ccO \ccTH } }  \Big) \Big|^2  \Big|\text{Split}_{-h_\ccT, h_\ccF} \Big( \frac{1}{  z_{ \ccT\ccF }}\Big)  \Big|^2 , 
\ee
 with
 \be
   z_{\ccO \ccTH }  \equiv  \frac{ \bar n \cdot (p_\ccO + p_\ccTH ) }{ \bar n \cdot p_\ccO }< 1  , 
  \quad \text{and} \quad   z_{\ccT \ccF}  \equiv  \frac{  n \cdot (p_\ccT+ p_\ccF) }{  n \cdot p_\ccT }< 1  .
  \label{z12}
\ee
and
\be 
\Big|\text{Split}_{++} ( z_{\cci \ccj} ) \Big|^2=  g_s^2 \frac{1}{1-z_{\cci \ccj}}, \quad  \Big|\text{Split}_{+-} ( z_{\cci \ccj} ) \Big|^2=  g_s^2\frac{z_{\cci \ccj}^2}{1-z_{\cci \ccj}},
\label{splitdef}
\ee  
These results are expressed in terms of the dimensionful variable $q_T=|\vec{q}_\perp|$ and two dimensionless variables $u$ and $v$. These are defined as
\begin{align} 
\vec{q}_\perp \equiv \vec{p}_{\ccTH,\perp}+ \vec{p}_{\ccF,\perp}  , \quad   u \equiv \frac{\vec{p}^2_{\ccTH,\perp}}{q_T^2}, \quad v \equiv \frac{\vec{p}^2_{\ccF,\perp}}{q_T^2}.
\label{uvdef} 
\end{align} 
At fixed $q_T$, the collinear limits are $u\to0, v\to 1$ or $v \to 0, u \to 1$.

To proceed, let's consider how to integrate over $d^2 p_{\ccTH,\perp}$, or equivalently $u$ and $v$.
Since
$\vec{q}_\perp \equiv \vec{p}_{\ccTH,\perp}+ \vec{p}_{\ccF,\perp}$,
the three vectors form a triangle in the transverse plane. We can rotate this triangle
so that $\vec q_\perp$ is conveniently oriented along the real axis, and rescale out $q_T$, leading to
a simpler triangle defined by one point $z$ in the complex plane
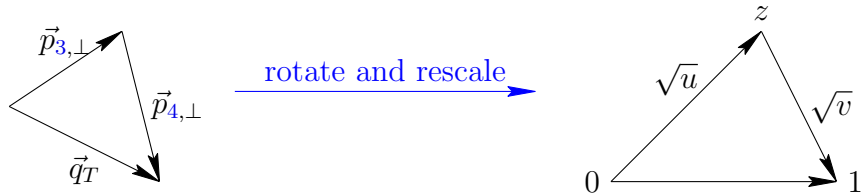
\be
\tikzset{>={Stealth[inset=2pt,length=12pt,angle'=20,round]}}
  \begin{tikzpicture}
    \coordinate (A) at (1,1);
    \coordinate (B) at ($(A)+(1.5,1)$);
    \coordinate (C) at ($(A)+(2,-1)$);
    \draw [->] (A)  -- (B)node[midway,above]{$\vec{p}_{\ccTH,\perp}$};
    \draw [->] (A)  -- (C)node[midway,below]{$\vec{q}_T$};
    \draw [->] (B)  -- (C)node[midway,right]{$\vec{p}_{\ccF,\perp}$};
    \draw[->,blue] (4,1.2) -- (8,1.2)node[midway,above]{$\text{rotate and rescale}$};
     \coordinate (A2) at (9,0);
    \coordinate (B2) at ($(A2)+(3,0)$);
    \coordinate (C2) at ($(A2)+(2, 2)$);
    \draw [->] (A2)node[left]{$0$}  -- (B2)node[right]{$1$}node[midway,below]{$$};
    \draw [->] (A2)  -- (C2)node[above]{$z$}node[midway,above]{$\sqrt{u}~~$};
    \draw [<-] (B2)  -- (C2)node[midway,right]{$\sqrt{v}$};
\end{tikzpicture}
\nonumber
\ee
\vspace{-2mm}
The relation between $u,v$ and $z$ is then
\be
u = z\bar{z}, \qquad v = (1-z)(1-\bar{z}) \label{conformvar}
\ee
Using $z$ and $\bar{z}$ facilitates integrating over $d^2 p_{\ccTH,\perp}$. Explicitly,
\be
d^2 p_{\ccTH,\perp} = d p_{\ccTH,x}d p_{\ccTH,y}= q_T^2 d \text{Re}(z) d \text{Im}(z) = q_T^2 d^2 z
\ee
The phase-space integrals then become regular two-dimensional integrals over conformal coordinates. 
\begin{align}
q_T^2\int \frac{d^{2}  p_{\ccTH, \perp} }{(2\pi)^{2}} \, |\cM|^2_{h_\ccO,h_\ccT} &\equiv    {\rm c}_{\rm 2G}   \, \frac{ s_{ \ccT \ccTH } s_{\ccO \ccF } }{ s_{\ccO \ccT}^2  } \,
 \Big| \text{Split} ( z_{\ccO \ccTH }^{-1 } ) \Big|^2  \Big|\text{Split}  ( z_{ \ccT\ccF }^{-1} )  \Big|^2 \,  
 \Big( \frac{\alpha_s}{ 4\pi} \Big)^2   I_\text{reg}^{h_\ccO,h_\ccT}  
\label{intregular}
\end{align} 
Explicitly, for $ | \cM |^2_{++}$ , the relevant integral is (see Appendix~\ref{app:ContourPhase})
\be 
I_\text{reg}^{++}  \equiv \int d^2 z \, \frac{1}{uv}   \frac{u+v-1}{ 2} \ln^2  \left( \frac{u}{v} \right)     = 4\pi \zeta_3
\label{intpp}
\ee
For $ | \cM |^2_{+-} $, the relevant integral is 
\be 
I_\text{reg}^{+-}  \equiv \int d^2 z \, \frac{1}{uv}  \ln u \ln v    = 4\pi \zeta_3
\label{intpm}
\ee
Intriguingly both integrals give the same result. This result is non-zero, and not power suppressed as $q_T\to0$ so it would seem to indicate factorization violation. 

If this were the end of the story, we would have found a non-zero factorization-violating contribution to any observable, even $q_T$ for which factorization is proven to hold.
In fact, there is another piece contributing to the cross section. 
The term of order $\ep^1$ in $|\cM|^2$ has the form
\be
\label{m2e2}
|\cM|^{2}_{\ep}  = \frac{   {\rm c}_{\rm 2G} }{ {\rm c}_\zero } \, \frac{\alpha_s^2}{8}  \left( \frac{ \tilde{\mu}^2 }{ q_T^2 } \right)^{2\ep}  e^{2\ep\gamma_E}    c_\Gamma^2  \,  \ep \,  \Big\{  
 |\cM_\zero|^2  (uv)^{-2\ep}   \left[ 6 \zeta_3  
+ \cO(\ep) \right]  + \text{integrable}  \Big\}
+\cdots
\ee
where $ \tilde{\mu}^2 \equiv  4\pi \mu^2 $, and $c_\Gamma = \frac{\Gamma(1-\ep)^2 \Gamma(1+\ep)}{\Gamma(1-2 \ep)}$. 
Due to the factor $\frac{1}{uv}$ in $|\cM_0|^2$ in Eq.~\eqref{M0def}, the first term in braces is singular when integrated over 4-dimensional phase space. Thus
in $d$ dimensions it will give a $\ep^{-1}$ factor which cancels the $\ep^1$ prefactor giving an $\cO(\ep^0)$ contribution. The remaining terms
in the braces are integrable over phase space and thus do not contribute as $\ep \to 0$. The singular term is independent of helicity, so we leave the helicity
labels implicit.

Performing the integral over the singular piece in $d$ dimensions, we find
\begin{multline}
\mu^{2\ep} q_T^2 \int \frac{d^{d-2}  p_{\ccTH, \perp} }{(2\pi)^{d-2}} \, |\cM|^2_\ep  =    {\rm c}_{\rm 2G}  \, \frac{ s_{ \ccT \ccTH } s_{\ccO \ccF } }{ s_{\ccO \ccT}^2  } \, \Big| \text{Split} ( z_{\ccO \ccTH }^{-1 } ) \Big|^2  \Big|\text{Split}  ( z_{ \ccT\ccF }^{-1} )  \Big|^2
 \\
 \times  \Big( \frac{\alpha_s}{ 4\pi} \Big)^2 \,  \ep \, \Big\{ 6 \zeta_3 I_\text{sing} (\ep )  + \cO(\ep^0) 
\Big\} 
  \label{intsing}
\end{multline} 
where
\begin{align} 
I_\text{sing} (\ep) & \equiv  \pi e^{2\ep\gamma_E}    c_\Gamma^2   (q_T^2 )^{1+2\ep}  (\tilde{\mu}^2)^{3\ep}  \int  \frac{d^{2-2\ep}    p_{\ccTH, \perp}}{\pi^{1-\ep}} \frac{1}{[ \vec{p}_{\ccTH, \perp}^2]^{1+2\ep} \, [(\vec{q}_T- \vec{p}_{\ccTH, \perp})^2]^{1+2\ep} } \\
& = \pi e^{2\ep\gamma_E}    c_\Gamma^2  \left(  \frac{ \tilde{\mu}^2}{ q_T^2} \right)^{3\ep} \frac{\Gamma^2 (-3\ep) \, \Gamma(1+4\ep)}{\Gamma(-6\ep) \, \Gamma^2(1+2\ep)}\\ 
& =    - \frac{2 \pi }{3 \ep} + \cO(\ep^0) 
\end{align} 
When added to the contributions from $\cO(\ep^0)$ in $|\cM|^2$, Eqs.~\eqref{intregular} and \eqref{intpp}-\eqref{intpm},
we see that the net contribution is zero:
\be
q_T^2 \frac{d \sigma}{d q^2_T}\Big |_{\text{2-loop Glauber}}= \sigma_0
 \mu^{2 \ep} q_T^2 \int \frac{d^{d-2}  p_{\ccTH, \perp} }{(2\pi)^{d-2}} \, \Big(  |\cM|^2_{h_\ccO,h_\ccT}+|\cM|^2_\ep \Big) =0 +  \cO(\ep)
\label{qTsum}
\ee
Thus at 2 loops, the Glauber ladder graphs do not generate a factorization-violating effect
for $q_T$.\footnote{Although we extend the integration region outside of the small-$p_T$ region to establish the cancellation, this is exactly what is required
by the effective field theory. The region where $p_T$ is not small has no Glauber pinch and is correctly described by a factorized expression.}

What changes if we use a different observable, like $E_T$ or beam thrust?
We know by dimensional analysis  that  $q_T^{4+4\ep} |\cM|^2_{\text{2-loop}}$ is dimensionless, depending only on $(z, \bar z)$, thus we define
\be 
|\widetilde{\cM}|^2_{\text{2-loop}} (z, \bar z ) \equiv q_T^{4} \left( \frac{q_T}{\mu} \right)^{4\ep} |\cM|^2_{\text{2-loop}}
\ee
and a  rescaled transverse momentum $\widetilde{p}_{\ccTH,\perp}^\mu$
so that
\be 
%\widetilde{p}_{\ccTH,\perp}^\mu  \equiv \frac{ p_{\ccTH,\perp}^\mu }{|q_T|}, \quad %  \tilde{q}_{T}^\mu \equiv \frac{ q_{T}^\mu}{|q_T|}, \quad 
%\text{such that} \quad 
|\vec{\widetilde{p}}_{\ccTH,\perp}|^2 = | z|^2, \quad  |\vec{1} - \vec{\widetilde{p}}_{\ccTH,\perp}|^2 = |1-z|^2
\ee
Working with $q_T$ and $\vec{\widetilde{p}}_{\ccTH, \perp}$ as the independent variables, \Eq{qTsum} can be written as the sum of the following two integrals  canceling each other at $\cO(\ep^0)$, 
\begin{align} 
\frac12 \frac{d \sigma}{d \ln q_T}\Big |_{\text{2-loop Glauber}}= \sigma_0 \int \frac{d^2 z}{(2\pi)^2}   |\widetilde{\cM}|^2_{h_\ccO,h_\ccT} (z, \bar z) +
\sigma_0 \int \frac{d^{d-2}  \widetilde{p}_{\ccTH, \perp} }{(2\pi)^{d-2}} \, \left( \frac{\mu}{q_T} \right)^{6\ep} \, |\widetilde{\cM}|^2_\ep (z, \bar z) + \cO(\ep)
\end{align}
Then we can change variables from $q_T$ to $E_T$ easily. For $2 \to 2$ scattering
\be
E_T=|\vec{p}_{\ccTH,\perp}| +|\vec{p}_{\ccF,\perp}| =( \sqrt{u} + \sqrt{v}) q_T= ( |z| + |1-z| ) q_T \label{qttoet}
\ee 
Thus, noting that that the Jacobian $\frac{\partial \ln q_T}{ \partial \ln E_T} = 1$,
 %working with $E_T, \widetilde{p}_{\ccTH, \perp}^\mu$ as the independent variables, we have 
 \begin{multline} 
 \frac12 \frac{d \sigma}{d \ln E_T}\Big |_{\text{2-loop Glauber}}=  \\
 \sigma_0 \int \frac{d^2 z}{(2\pi)^2}   |\widetilde{\cM}|^2_{h_\ccO,h_\ccT} (z, \bar z) +
 \sigma_0  \left( \frac{\mu}{E_T} \right)^{6\ep} \int \frac{d^{d-2}  \widetilde{p}_{\ccTH, \perp} }{(2\pi)^{d-2}} \, \left(  |z| + |1-z|  \right)^{6\ep} \, |\widetilde{\cM}|^2_\ep (z, \bar z) + \cO(\ep)
 \label{zform}
\end{multline}
The integration over $z$ and $\bar{z}$ in the first integral is unaffected by the change of variables.
 The second integral is changed by a factor of $ \left(  |z| + |1-z|  \right)^{6\ep} $. However, the behavior of the integrand in the 
 singular regions $z\to0$ and $z\to1$ is unchanged. This new factor only contributes at order $\cO(\ep)$.  
 Therefore the cancellation between the two integrals still holds at $\cO(\ep^0)$. The key point is that the inclusive integration over $z$ and $\bar{z}$ is observable independent. 
A more detailed discussion of the change of variables is given in Appendix~\ref{sec:variable}.

For another example, consider beam thrust. For $2\to 2$ scattering, 
\be
\ln \tau_B = 2\ln q_T + \ln \left( \frac{ |z|^2 }{  (z_{\ccO\ccTH}^{-1} -1) q^+ } +  \frac{  |1-z |^2}{ (z_{\ccT\ccF}^{-1} -1) q^- }  \right) 
\ee
where $q^\mu = (q^+, q^-,  q_T)$ is the total momentum of the lepton pair. 
Changing from $q_T$ to $\tau_B$ results in an expression similar to Eq.~\eqref{zform}. The integration around the singular region again
only contributes new terms that start at $\cO(\ep)$, and thus there is no factorization-violating effect from the 2-loop Glauber ladder graphs. 

The same argument holds for any infrared-safe single-scale transverse observable $X$. Any such observable must be expressible as
\be
\ln X =  a \ln q_T + g(z,\bar{z}) \label{Xrel}
\ee
for some $a>0$ and some function $g(z, \bar z )$ that is regular as $z\to0$ and $z\to1$.  If $g(z,\bar{z})$ were not regular in the collinear limits, 
of if $a \le 0$,  then the observable cannot be infrared safe. Because $g(z,\bar{z})$ is regular in collinear limits, the same inclusive integrals over $z$ can be done without 
generating singularities, and thus an observable $X$ does not exhibit factorization-violating effects at this order.

\subsection{All-orders Glauber cancellation}
We showed that at 2-loop order,  the Glauber ladder diagrams alone do not violate factorization for any single scale observable. Now we will show that the cancellation we found persists to all orders.

By direct calculation, we find that to  all orders in perturbation theory, the sum of Glauber ladder diagrams
has the form
\be
 \cM_G  ( \{ h_\cci \}  )
  =   e^{ i \phi_{\bf g}  ( \widehat{\alpha}_s (\TT_\ccTH \cdot \TT_\ccF)  ) } 
 \bigg[ 1+  f_{\text{reg}}^{ \{ h_\cci \}   } \big( z, \bar{z}   ; \frac{\alpha_s}{2} (\TT_\ccTH \cdot \TT_\ccF) \big) + \ep  f_{ \text{sing}}  \big( \widehat{\alpha}_s (\TT_\ccTH \cdot \TT_\ccF) \big)   + \cdots  \bigg]   \cM^\zero( \{ h_\cci \}  )  
  \label{AllLoopIR}
\ee
where the $\cdots$ are terms that do not contribute to the cross section. The combination
\be
\widehat{\alpha}_s(\mu) \equiv  \frac{ \alpha_s (\mu)}{2}   c_\Gamma e^{\ep \gamma_E}  ( q_T^2 u v )^{-\ep}
\ee
with $c_\Gamma= \frac{\Gamma(1-\ep)^2\Gamma(1+\ep) }{\Gamma (1-2 \ep) }$ appears naturally
and the phase is
\be
{\bf{ \phi}_g} ( \alpha )  \equiv  \alpha  \left(  \frac{1}{ \ep} +  2 \gamma_E \right)  
 -i \ln \frac{ \Gamma (1+ i  \alpha )   }{ \Gamma(1- i   \alpha  )}
\label{eq:IRphase}   \\ 
 \ee
A key property of the expression in Eq.~\eqref{AllLoopIR} is that all of the infrared divergences are contained in the phase. This is  shown in Appendix~\ref{app:exp}. Another
property is that once this phase is factored out, the order $\ep^0$ term denoted $f_{\text{reg}}$ is integrable over the collinear regions. This decomposition and the calculation of ${\bf{ \phi}_g} (\widehat{\alpha_s} )$ are given in Appendix~\ref{app:pssing}.
The expression for $f_{\text{reg}}$ at 2-loops, which appeared in Eqs.~\eqref{hel1} and \eqref{hel2}, is computed in Appendix~\ref{app:2loopreg}.

The leading non-integrable piece we call $f_{ \text{sing}}$. It can be extracted from a general formula we derive for the singular part of $\cM_G$. At order $\alpha_s^n$, this singular part is
\be
 \cM_{G,\text{sing}}^{n}  =  \frac{1}{n!} \left( \frac{  -i \alpha_s}{ 2 } \right)^n  (\TT_\ccTH \cdot \TT_\ccF)^n  \,
 e^{  n  \ep   \gamma_E} (q_T^2)^{- n\ep} (uv)^{-n\ep} [\Gamma (- \ep ) ]^n   \frac{ \Gamma(1- \ep) \Gamma ( 1 + n\ep)    }{ \Gamma (1- (n+1) \ep ) }  \,   \cM_\zero 
\ee
Expanding $ e^{-i \phi_{\bf g}(\widehat{\alpha}_s)}  \cM_\text{sing}^{n}$ gives
\be
 \ep f_\text{sing}(\alpha; \ep) =   ( 3 \zeta_3 \, \ep  +\frac92 \zeta_4 \, \ep^2 
 + \cdots  ) \alpha^2+  (  \zeta_4\,  \ep -16 \zeta_5 \, \ep^2 +  \cdots )  i \alpha^3 + (-5 \zeta_5 \, \ep + \cdots )  \alpha^4 + \cO(\alpha^5)
 \ee
The first term in this expansion was used in Eq.~\eqref{m2e2}.

Since all of the IR divergences cancel in $|\cM|^2$, the cross section has the form
\be
\frac{d \sigma}{d \ln q_T}\Big|_{\text{Glauber ladders}} = \int dz d \bar{z} F_\text{reg}(z,\bar{z})
+ \ep \frac{1}{q_T^2} \Big(\frac{\mu}{q_T}\Big)^{2  (n+1) \ep} \int d^{d-2} p_{\ccTH,\perp}   F_{\text{sing}}(z,\bar{z})
=0
\label{dsigqt}
\ee
In Section \ref{sec:2loop} we showed that at two loops both of these terms are separately non-zero and only their sum vanishes. 
For observable $q_T$, this must also be true to all orders as a result of CSS poof of factorization for $q_T$.

Now say we change variables from $\ln q_T$ to $\ln E_T$, using Eq. \eqref{qttoet}.
 The first integral in Eq. \eqref{dsigqt} is unaffected since the integrand depends only on the conformal coordinates. 
 Changing $q_T$ into $E_T$ also has no effect on the singular behavior of the second integral, since $E_T \to q_T$ in the collinear limits.
Thus both integrals give the same result at $\cO(\ep^0)$,
and there is no factorization-violating effect from Glauber ladders for $E_T$, to all orders. 
The same argument holds for any infrared-safe single-scale observable $X$. 

% 
%%%%%%%%%%%%%%%%%%%%%%%%%%%%%%%%%
% Factorization violation
%%%%%%%%%%%%%%%%%%%%%%%%%%%%%%%%%

\section{Factorization-violating effects \label{sec:fv}}
Does $q_T$ factorization imply  factorization for other observables to all orders? The answer is, not surprisingly, no. The scaling arguments we have been using only go so far. They do however give some indications of how factorization violation can show up. 

Non-ladder potentially-factorization-violating contributions to the cross section involve soft emissions from the Glauber legs.  For example, virtual or real contributions could look like
\begin{align} 
\cM_{V} = 
\begin{tikzpicture}[baseline={([yshift=-.5ex]current bounding box.center)},scale=1.1]
\draw (-1,-0.8) -- (0,0);
\draw (-1,0.8) -- (0,0);
\draw [fill=\blobcolor] (0,0) circle [radius = 0.1];
\draw [fill=green!80!black, opacity=0.75] (0.55,0) ellipse (0.35 and 0.2);
\draw[color=black,decorate,decoration={gluon, amplitude=1.2pt,
    segment length=1.8pt, aspect=0.6}] (-0.8,0.6) -- (0.9,0.6);
    \draw (-0.8,0.6) -- (0.9,0.6);
\draw[color=black,decorate,decoration={gluon, amplitude=1.2pt,
    segment length=1.8pt, aspect=0.6}] (-0.8,-0.6) -- (0.9,-0.6);
    \draw (-0.8,-0.6) -- (0.9,-0.6);    
\draw [red,fill=red] (0.3,0.6) circle [radius=1pt];
\draw [red,fill=red] (0.3,-0.6) circle [radius=1pt];
\draw [red,decorate glaubr] (0.3,0.6) to [bend left=0] (0.3,0.15);
\draw [red,decorate glaubr] (0.3,-0.15) to [bend left=0] (0.3,-0.6);
\draw [red,fill=red] (0.8,0.6) circle [radius=0.8pt];
\draw [red,fill=red] (0.8,-0.6) circle [radius=0.8pt];
\draw [red,decorate glaubr] (0.8,0.6) to [bend left=0] (0.8,0.15);
\draw [red,decorate glaubr] (0.8,-0.15) to [bend left=0] (0.8,-0.6);
\draw [-{Latex[length=3pt]}] (0.7,-0.2)--(0.7,-0.38)node[below, scale=0.4]{$\qquad \tiny{  \ell_2 + k   }$};
 \draw [-{Latex[length=3pt]}] (0.35,-0.2)--(0.35,-0.38)node[below, scale=0.4]{$\tiny{ \ell_1 -k}$};
 \draw [-{Latex[length=3pt]}] (0.7,0.52)--(0.7,0.34)node[below, scale=0.4]{$\; \tiny{ \ell_2 }$};
 \draw [-{Latex[length=3pt]}] (0.35,0.52)--(0.35,0.34)node[below, scale=0.4]{$\; \tiny{ \ell_1 }$};
 \draw [-{Latex[length=3pt]}] (-0.9,0.9) node[above,  scale=0.5]{$\ccO$}--(-0.7,0.7);
 \draw [-{Latex[length=3pt]}] (-0.9,-0.9) node[below,  scale=0.5]{$\ccT$}--(-0.7,-0.7);
 \draw [-{Latex[length=3pt]}] (0.6,0.7)--(0.9,0.7) node[above, scale=0.5]{$\ccTH$};
  \draw [-{Latex[length=3pt]}] (0.6,-0.7)--(0.9,- 0.7) node[below, scale=0.5]{$\ccF$};
  \draw (0.55,0.0) node[scale=0.6]{$S$};
\end{tikzpicture}
 , 
\quad 
\cM_{R} = 
\begin{tikzpicture}[baseline={([yshift=-.5ex]current bounding box.center)},scale=1.1]
\draw (-1,-0.8) -- (0,0);
\draw (-1,0.8) -- (0,0);
\draw [fill=\blobcolor] (0,0) circle [radius = 0.1];
\draw [fill=green!80!black] (0.45,0) ellipse (0.2 and 0.2);
\draw[color=black,decorate,decoration={gluon, amplitude=1.2pt,
    segment length=1.8pt, aspect=0.6}] (-0.8,0.6) -- (0.75,0.6);
    \draw (-0.8,0.6) -- (0.75,0.6);
\draw[color=black,decorate,decoration={gluon, amplitude=1.2pt,
    segment length=1.8pt, aspect=0.6}] (-0.8,-0.6) -- (0.75,-0.6);
    \draw (-0.8,-0.6) -- (0.75,-0.6);    
\draw[color=green!80!black,decorate,decoration={gluon, amplitude=1.2pt,
   segment length=1.8pt, aspect=0.6}] (0.65,-0.25) to [bend left= 45] (0.65,0.25);  
\draw [red,fill=red] (0.3,0.6) circle [radius=1pt];
\draw [red,fill=red] (0.3,-0.6) circle [radius=1pt];
\draw [red,decorate glaubr] (0.3,0.6) to [bend left=0] (0.3,0.15);
\draw [red,decorate glaubr] (0.3,-0.15) to [bend left=0] (0.3,-0.6);
 \draw [-{Latex[length=3pt]}] (0.35,-0.2)--(0.35,-0.38)node[below, scale=0.4]{$\tiny{ \ell_1 -k}$};
 \draw [-{Latex[length=3pt]}] (0.35,0.52)--(0.35,0.34)node[below, scale=0.4]{$\; \tiny{ \ell_1 }$};
 \draw [-{Latex[length=3pt]}] (-0.9,0.9) node[above,  scale=0.5]{$\ccO$}--(-0.7,0.7);
 \draw [-{Latex[length=3pt]}] (-0.9,-0.9) node[below,  scale=0.5]{$\ccT$}--(-0.7,-0.7);
 \draw [-{Latex[length=3pt]}] (0.45,0.7)--(0.75,0.7) node[above, scale=0.5]{$\ccTH$};
  \draw [-{Latex[length=3pt]}] (0.45,-0.7)--(0.75,- 0.7) node[below, scale=0.5]{$\ccF$};
 \draw (0.45,0.0)   node[scale=0.6]{$S$};
\end{tikzpicture} 
\end{align} 
In this section we will focus on the potentially-factorization-violating contributions from graphs with soft gluon exchanges between the Glauber ladders. Similar graphs have also been briefly discussed in Ref.~\cite{Gaunt:2014ska}.

For graphs like these, there is neither proof nor expectation that the IR divergences should resum into a phase. Instead, we expect there to be IR divergences in $|\cM|^2$ that cancel only when the phase space integral is done. 
Let us denote the cross section from all graphs with Glaubers as $\sigma_G$. 
Quite generally, we can write
\be
\frac{d \sigma_G}{d \ln q_T} = \int du dv\ldots  dw f(q_T,u,v,w,\ldots)
\ee
with $q_T$ the only dimensionful variable and the $u,v,w,\cdots$ variables dimensionless. 
For an IR or rapidity divergent contribution, another scale $\mu$ or $\nu$ can appear. This scale can only appear logarithmically, so the $q_T$ dependence must be logarithmic and we can write a series expansion
\be
\frac{d \sigma_G}{d \log q_T} = \int du dv \ldots  dw \left[f_0(u,v,\ldots,w)+ \ln\frac{q_T}{\mu} f_1(u,v,w,\ldots)
+ \ln^2\frac{q_T}{\mu} f_2(,u,v,\ldots,w) + \cdots\right]
\label{qtparam}
\ee
Some of these integrals may be IR divergent. We know however, that the final result must be IR finite, so the IR divergences must cancel among the various contributions. Moreover, we know by $q_T$-factorization that the sum of all of these contributions is exactly zero. Since each term multiplies a different power of $\ln q_T$, each one separately must integrate to zero.

Now consider the cross section for $E_T$ rather than $q_T$. For $2\to 2$ scattering,
 $E_T = q_T(\sqrt{u} + \sqrt{v})$. So $\ln q_T = \ln E_T + \ln(\sqrt{u}+\sqrt{v})$. 
 For graphs with more final-state gluons, we would have a more complicated function
 but still a linear relation between $\ln q_T$ and $\ln E_T$. So let us write
 \be
 \ln q_T = \ln E_T + g(u,v,\cdots w)
 \ee
 Other single scale variables, like beam thrust, will have a similarly linear relation with a different function $g$. 
Then
\begin{multline}
\frac{d \sigma_G}{d \log E_T} = \int du dv \ldots  dw \Big[f_0(u,v,\ldots,w)
+ \ln\frac{E_T}{\mu} f_1(u,v,w,\ldots) \\ 
+ g(u,v,\ldots,w) f_1(u,v,\ldots,w) +2 \ln \frac{E_T}{\mu} g(u,v,\ldots,w) f_2(u,v,\ldots,w)
+ \cdots\Big]
\label{genfv}
\end{multline}
The terms on the first line vanish by $q_T$ factorization, but the terms on the second may not.

\subsection{Observable dependence}
Let us now put a little more detail into Eq.~ \eqref{genfv}. 
We discuss two types of observables for which factorization may not hold
\begin{enumerate}
\item Non-global observables that do not include the collinear region, so that the real soft gluon phase space is not fully integrated over. 
\item  Global observables that are rapidity-independent in the collinear region.
\end{enumerate}

The first category includes non-global observables only sensitive to soft emissions within some region of the detector, but  inclusive over the collinear particles along the beam directions.  An example of such an observable is the mass of the hardest jet.
The second category includes global hadronic event shapes, like $E_T$ or transverse thrust

 To understand the factorization breaking effects due to the these two types of measurements, 
let us look at diagrams with one soft gluon as an example. 
The leading order real-emission diagram has a single soft gluon coming off of a Glauber line:
\be
\cM_R^1 = 
\begin{tikzpicture}[baseline={([yshift=-.5ex]current bounding box.center)},scale=1.1]
\draw (-1,-0.8) -- (0,0);
\draw (-1,0.8) -- (0,0);
\draw [fill=\blobcolor] (0,0) circle [radius = 0.15];
\draw[color=black,decorate,decoration={gluon, amplitude=1.2pt,
    segment length=1.8pt, aspect=0.6}] (-0.8,0.6) -- (0.8,0.6);
    \draw (-0.8,0.6) -- (0.8,0.6);
\draw[color=black,decorate,decoration={gluon, amplitude=1.2pt,
    segment length=1.8pt, aspect=0.6}] (-0.8,-0.6) -- (0.8,-0.6);
    \draw (-0.8,-0.6) -- (0.8,-0.6);    
\draw [red,fill=red] (0.4,0.6) circle [radius=0.8pt];
\draw [red,fill=red] (0.4,-0.6) circle [radius=0.8pt];
\draw [red,decorate glaubr] (0.4,0.6) to [bend left=0] (0.4,-0.6);
\draw [red,fill=red] (0.4,0) circle [radius=0.8pt];
\draw[color=green!80!black,decorate,decoration={gluon, amplitude=1.2pt,
    segment length=1.8pt, aspect=0.6}] (0.4,0) to [bend right=0] (0.8,0);
 \draw [-{Latex[length=3pt]}] (0.5,0.5)--(0.5,0.2)node[right, scale=0.4]{$\tiny{ \ell_1}$};
 \draw [-{Latex[length=3pt]}] (0.5,-0.1)--(0.8,-0.1)node[below, scale=0.4]{$k$};
 \draw [-{Latex[length=3pt]}] (-0.9,0.9) node[above,  scale=0.5]{$p_\ccO$}--(-0.7,0.7);
 \draw [-{Latex[length=3pt]}] (-0.9,-0.9) node[below,  scale=0.5]{$p_\ccT$}--(-0.7,-0.7);
 \draw [-{Latex[length=3pt]}] (0.5,0.7)--(0.8,0.7) node[above, scale=0.5]{$p_\ccTH$};
  \draw [-{Latex[length=3pt]}] (0.5,-0.7)--(0.8,- 0.7) node[below, scale=0.5]{$q-p_\ccTH-k$};
\end{tikzpicture}
\label{MR1}
\ee
This diagram contributes to the cross section at order $g_s^6 |\cM_0|^2$. Virtual diagrams
that contribute at this same order are the square of 1-loop graphs with a Glauber gluon and a single soft loop:
\be
\cM_V^1  = 
\begin{tikzpicture}[baseline={([yshift=-.5ex]current bounding box.center)},scale=1.1]
\draw (-1,-0.8) -- (0,0);
\draw (-1,0.8) -- (0,0);
\draw [fill=\blobcolor] (0,0) circle [radius = 0.15];
\draw[color=black,decorate,decoration={gluon, amplitude=1.2pt,
    segment length=1.8pt, aspect=0.6}] (-0.8,0.6) -- (0.8,0.6);
    \draw (-0.8,0.6) -- (0.8,0.6);
\draw[color=black,decorate,decoration={gluon, amplitude=1.2pt,
    segment length=1.8pt, aspect=0.6}] (-0.8,-0.6) -- (0.8,-0.6);
    \draw (-0.8,-0.6) -- (0.8,-0.6);    
\draw [red,fill=red] (0.4,0.6) circle [radius=0.8pt];
\draw [red,fill=red] (0.4,-0.6) circle [radius=0.8pt];
\draw [red,decorate glaubr] (0.4,0.6) to [bend left=0] (0.4,0.2);
\draw [red,decorate glaubr] (0.4,-0.2) to [bend left=0] (0.4,-0.6);
\draw[color=green!80!black,decorate,decoration={gluon, amplitude=1.2pt,
    segment length=1.8pt, aspect=0.6}] (0.4,0.2) to [bend left=50] (0.4,-0.2); 
  \draw[color=green!80!black,decorate,decoration={gluon, amplitude=1.2pt,
    segment length=1.8pt, aspect=0.6}] (0.4,-0.2) to [bend left=50] (0.4, 0.2);   
\draw [red,fill=red] (0.4,0.2) circle [radius=0.8pt];
\draw [red,fill=red] (0.4,-0.2) circle [radius=0.8pt];
\draw [-{Latex[length=3pt]}] (0.5,0.5)--(0.5,0.3)node[below, scale=0.4]{$\tiny{ \ell_1}$};
\draw [-{Latex[length=3pt]}] (-0.9,0.9) node[above,  scale=0.5]{$p_\ccO$}--(-0.7,0.7);
\draw [-{Latex[length=3pt]}] (-0.9,-0.9) node[below,  scale=0.5]{$p_\ccT$}--(-0.7,-0.7);
\draw [-{Latex[length=3pt]}] (0.5,0.7)--(0.8,0.7) node[above, scale=0.5]{$p_\ccTH$};
 \draw [-{Latex[length=3pt]}] (0.5,-0.7)--(0.8,- 0.7) node[below, scale=0.5]{$q-p_\ccTH$};
\end{tikzpicture} 
 \quad 
 +
 \quad 
 \begin{tikzpicture}[baseline={([yshift=-.5ex]current bounding box.center)},scale=1.1]
\draw (-1,-0.8) -- (0,0);
\draw (-1,0.8) -- (0,0);
\draw [fill=\blobcolor] (0,0) circle [radius = 0.15];
\draw[color=black,decorate,decoration={gluon, amplitude=1.2pt,
    segment length=1.8pt, aspect=0.6}] (-0.8,0.6) -- (0.8,0.6);
    \draw (-0.8,0.6) -- (0.8,0.6);
\draw[color=black,decorate,decoration={gluon, amplitude=1.2pt,
    segment length=1.8pt, aspect=0.6}] (-0.8,-0.6) -- (0.8,-0.6);
    \draw (-0.8,-0.6) -- (0.8,-0.6);    
\draw [red,fill=red] (0.4,0.6) circle [radius=0.8pt];
\draw [red,fill=red] (0.4,-0.6) circle [radius=0.8pt];
\draw [red,decorate glaubr] (0.4,0.6) to [bend left=0] (0.4,-0.6);
\draw[color=green!80!black,decorate,decoration={gluon, amplitude=1.2pt,
    segment length=1.8pt, aspect=0.6}] (0.4,0 ) to [bend left=50] (0.6, 0.1); 
\draw[color=green!80!black,decorate,decoration={gluon, amplitude=1.2pt,
    segment length=1.8pt, aspect=0.6}] (0.6, -0.1) to [bend left=50] (0.4,0);   
\draw[color=green!80!black,decorate,decoration={gluon, amplitude=1.2pt,
    segment length=1.8pt, aspect=0.6}] (0.6, 0.1) to [bend left=50] (0.6, -0.1);       
\draw [red,fill=red] (0.4,0) circle [radius=0.8pt];
% \draw [-{Latex[length=3pt]}] (-0.9,0.9) node[above,  scale=0.5]{$p_\ccO$}--(-0.7,0.7);
% \draw [-{Latex[length=3pt]}] (-0.9,-0.9) node[below,  scale=0.5]{$p_\ccT$}--(-0.7,-0.7);
% \draw [-{Latex[length=3pt]}] (0.5,0.7)--(0.8,0.7) node[above, scale=0.5]{$p_\ccTH$};
%  \draw [-{Latex[length=3pt]}] (0.5,-0.7)--(0.8,- 0.7) node[below, scale=0.5]{$q-p_\ccTH$};
\end{tikzpicture}  
\ee
There are also  2-loop graphs with two Glauber gluons relevant at this order:
\be
\cM_V^2 = 
\begin{tikzpicture}[baseline={([yshift=-.5ex]current bounding box.center)},scale=1.1]
\draw (-1,-0.8) -- (0,0);
\draw (-1,0.8) -- (0,0);
\draw [fill=\blobcolor] (0,0) circle [radius = 0.15];
\draw[color=black,decorate,decoration={gluon, amplitude=1.2pt,
    segment length=1.8pt, aspect=0.6}] (-0.8,0.6) -- (0.9,0.6);
    \draw (-0.8,0.6) -- (0.9,0.6);
\draw[color=black,decorate,decoration={gluon, amplitude=1.2pt,
    segment length=1.8pt, aspect=0.6}] (-0.8,-0.6) -- (0.9,-0.6);
    \draw (-0.8,-0.6) -- (0.9,-0.6);    
\draw [red,fill=red] (0.3,0.6) circle [radius=0.8pt];
\draw [red,fill=red] (0.33,-0.6) circle [radius=0.8pt];
\draw [red,decorate glaubr] (0.3,0.6) to [bend left=0] (0.3,-0.6);
\draw [red,fill=red] (0.6,0.6) circle [radius=0.8pt];
\draw [red,fill=red] (0.6,-0.6) circle [radius=0.8pt];
\draw [red,decorate glaubr] (0.6,0.6) to [bend left=0] (0.6,-0.6);
\draw [red,fill=red] (0.3,0) circle [radius=0.8pt];
\draw [red,fill=red] (0.6,0) circle [radius=0.8pt];
\draw[color=green!80!black,decorate,decoration={gluon, amplitude=1.2pt,
    segment length=1.8pt, aspect=0.6}] (0.3,0) to [bend right=0] (0.6,0);
\draw [-{Latex[length=3pt]}] (0.65,-0.1)--(0.65,-0.3)node[below, scale=0.4]{$\qquad \tiny{ \ell_2 + k}$};
 \draw [-{Latex[length=3pt]}] (0.35,-0.1)--(0.35,-0.3)node[below, scale=0.4]{$\tiny{ \ell_1 -k}$};
 \draw [-{Latex[length=3pt]}] (0.65,0.5)--(0.65,0.3)node[below, scale=0.4]{$\; \tiny{ \ell_2 }$};
 \draw [-{Latex[length=3pt]}] (0.35,0.5)--(0.35,0.3)node[below, scale=0.4]{$\; \tiny{ \ell_1 }$};
 \draw [-{Latex[length=3pt]}] (-0.9,0.9) node[above,  scale=0.5]{$p_\ccO$}--(-0.7,0.7);
 \draw [-{Latex[length=3pt]}] (-0.9,-0.9) node[below,  scale=0.5]{$p_\ccT$}--(-0.7,-0.7);
 \draw [-{Latex[length=3pt]}] (0.6,0.7)--(0.9,0.7) node[above, scale=0.5]{$p_\ccTH$};
  \draw [-{Latex[length=3pt]}] (0.6,-0.7)--(0.9,- 0.7) node[below, scale=0.5]{$q-p_\ccTH$};
\end{tikzpicture}  
\quad 
+ 
\begin{tikzpicture}[baseline={([yshift=-.5ex]current bounding box.center)},scale=1.1]
\draw (-1,-0.8) -- (0,0);
\draw (-1,0.8) -- (0,0);
\draw [fill=\blobcolor] (0,0) circle [radius = 0.15];
\draw[color=black,decorate,decoration={gluon, amplitude=1.2pt,
    segment length=1.8pt, aspect=0.6}] (-0.8,0.6) -- (0.9,0.6);
    \draw (-0.8,0.6) -- (0.9,0.6);
\draw[color=black,decorate,decoration={gluon, amplitude=1.2pt,
    segment length=1.8pt, aspect=0.6}] (-0.8,-0.6) -- (0.9,-0.6);
    \draw (-0.8,-0.6) -- (0.9,-0.6);    
\draw [red,fill=red] (0.3,0.6) circle [radius=0.8pt];
\draw [red,fill=red] (0.33,-0.6) circle [radius=0.8pt];
\draw [red,decorate glaubr] (0.3,0.6) to [bend left=0] (0.3, 0.2 );
\draw [red,decorate glaubr] (0.3,-0.6) to [bend left=0] (0.3,-0.2 );
\draw[color=green!80!black,decorate,decoration={gluon, amplitude=1.2pt,
    segment length=1.8pt, aspect=0.6}] (0.3,0.2) to [bend left=50] (0.3,-0.2); 
  \draw[color=green!80!black,decorate,decoration={gluon, amplitude=1.2pt,
    segment length=1.8pt, aspect=0.6}] (0.3,-0.2) to [bend left=50] (0.3, 0.2);   
\draw [red,fill=red] (0.3,0.2) circle [radius=0.8pt];
\draw [red,fill=red] (0.3,-0.2) circle [radius=0.8pt];
\draw [red,fill=red] (0.6,0.6) circle [radius=0.8pt];
\draw [red,fill=red] (0.6,-0.6) circle [radius=0.8pt];
\draw [red,decorate glaubr] (0.6,0.6) to [bend left=0] (0.6,-0.6);
% \draw [-{Latex[length=3pt]}] (-0.9,0.9) node[above,  scale=0.5]{$p_\ccO$}--(-0.7,0.7);
% \draw [-{Latex[length=3pt]}] (-0.9,-0.9) node[below,  scale=0.5]{$p_\ccT$}--(-0.7,-0.7);
% \draw [-{Latex[length=3pt]}] (0.6,0.7)--(0.9,0.7) node[above, scale=0.5]{$p_\ccTH$};
%  \draw [-{Latex[length=3pt]}] (0.6,-0.7)--(0.9,- 0.7) node[below, scale=0.5]{$q-p_\ccTH$}; 
\end{tikzpicture}  
\quad 
+
\begin{tikzpicture}[baseline={([yshift=-.5ex]current bounding box.center)},scale=1.1]
\draw (-1,-0.8) -- (0,0);
\draw (-1,0.8) -- (0,0);
\draw [fill=\blobcolor] (0,0) circle [radius = 0.15];
\draw[color=black,decorate,decoration={gluon, amplitude=1.2pt,
    segment length=1.8pt, aspect=0.6}] (-0.8,0.6) -- (0.9,0.6);
    \draw (-0.8,0.6) -- (0.9,0.6);
\draw[color=black,decorate,decoration={gluon, amplitude=1.2pt,
    segment length=1.8pt, aspect=0.6}] (-0.8,-0.6) -- (0.9,-0.6);
    \draw (-0.8,-0.6) -- (0.9,-0.6);    
\draw [red,fill=red] (0.3,0.6) circle [radius=0.8pt];
\draw [red,fill=red] (0.33,-0.6) circle [radius=0.8pt];
\draw [red,decorate glaubr] (0.3,0.6) to [bend left=0] (0.3,- 0.6 );
\draw[color=green!80!black,decorate,decoration={gluon, amplitude=1.2pt,
    segment length=1.8pt, aspect=0.6}] (0.3,0 ) to [bend left=50] (0.5 , 0.1 ); 
\draw[color=green!80!black,decorate,decoration={gluon, amplitude=1.2pt,
    segment length=1.8pt, aspect=0.6}] (0.5, -0.1) to [bend left=50] (0.3,0);   
\draw[color=green!80!black,decorate,decoration={gluon, amplitude=1.2pt,
    segment length=1.8pt, aspect=0.6}] (0.5, 0.1) to [bend left=50] (0.5, -0.1);           
\draw [red,fill=red] (0.6,0.6) circle [radius=0.8pt];
\draw [red,fill=red] (0.6,-0.6) circle [radius=0.8pt];
\draw [red,decorate glaubr] (0.6,0.6) to [bend left=0] (0.6,-0.6);
\draw [red,fill=red] (0.3,0) circle [radius=0.8pt];
% \draw [-{Latex[length=3pt]}] (-0.9,0.9) node[above,  scale=0.5]{$p_\ccO$}--(-0.7,0.7);
% \draw [-{Latex[length=3pt]}] (-0.9,-0.9) node[below,  scale=0.5]{$p_\ccT$}--(-0.7,-0.7);
% \draw [-{Latex[length=3pt]}] (0.6,0.7)--(0.9,0.7) node[above, scale=0.5]{$p_\ccTH$};
%  \draw [-{Latex[length=3pt]}] (0.6,-0.7)--(0.9,- 0.7) node[below, scale=0.5]{$q-p_\ccTH$}; 
\end{tikzpicture}  
\quad 
+
\begin{tikzpicture}[baseline={([yshift=-.5ex]current bounding box.center)},scale=1.1]
\draw (-1,-0.8) -- (0,0);
\draw (-1,0.8) -- (0,0);
\draw [fill=\blobcolor] (0,0) circle [radius = 0.15];
\draw[color=black,decorate,decoration={gluon, amplitude=1.2pt,
    segment length=1.8pt, aspect=0.6}] (-0.8,0.6) -- (0.9,0.6);
    \draw (-0.8,0.6) -- (0.9,0.6);
\draw[color=black,decorate,decoration={gluon, amplitude=1.2pt,
    segment length=1.8pt, aspect=0.6}] (-0.8,-0.6) -- (0.9,-0.6);
    \draw (-0.8,-0.6) -- (0.9,-0.6);    
\draw [red,fill=red] (0.3,0.6) circle [radius=0.8pt];
\draw [red,fill=red] (0.3,-0.6) circle [radius=0.8pt];
\draw [red,decorate glaubr] (0.6,0.6) to [bend left=0] (0.6, 0.2 );
\draw [red,decorate glaubr] (0.6,-0.6) to [bend left=0] (0.6,-0.2 );
\draw[color=green!80!black,decorate,decoration={gluon, amplitude=1.2pt,
    segment length=1.8pt, aspect=0.6}] (0.6,0.2) to [bend left=50] (0.6,-0.2); 
  \draw[color=green!80!black,decorate,decoration={gluon, amplitude=1.2pt,
    segment length=1.8pt, aspect=0.6}] (0.6,-0.2) to [bend left=50] (0.6, 0.2);   
 \draw [red,fill=red] (0.6,0.2) circle [radius=0.8pt];
\draw [red,fill=red] (0.6,-0.2) circle [radius=0.8pt];
\draw [red,fill=red] (0.6,0.6) circle [radius=0.8pt];
\draw [red,fill=red] (0.6,-0.6) circle [radius=0.8pt];
\draw [red,decorate glaubr] (0.3,0.6) to [bend left=0] (0.3,-0.6);
% \draw [-{Latex[length=3pt]}] (-0.9,0.9) node[above,  scale=0.5]{$p_\ccO$}--(-0.7,0.7);
% \draw [-{Latex[length=3pt]}] (-0.9,-0.9) node[below,  scale=0.5]{$p_\ccT$}--(-0.7,-0.7);
% \draw [-{Latex[length=3pt]}] (0.6,0.7)--(0.9,0.7) node[above, scale=0.5]{$p_\ccTH$};
%  \draw [-{Latex[length=3pt]}] (0.6,-0.7)--(0.9,- 0.7) node[below, scale=0.5]{$q-p_\ccTH$}; 
\end{tikzpicture}  
\quad 
+
\begin{tikzpicture}[baseline={([yshift=-.5ex]current bounding box.center)},scale=1.1]
\draw (-1,-0.8) -- (0,0);
\draw (-1,0.8) -- (0,0);
\draw [fill=\blobcolor] (0,0) circle [radius = 0.15];
\draw[color=black,decorate,decoration={gluon, amplitude=1.2pt,
    segment length=1.8pt, aspect=0.6}] (-0.8,0.6) -- (0.9,0.6);
    \draw (-0.8,0.6) -- (0.9,0.6);
\draw[color=black,decorate,decoration={gluon, amplitude=1.2pt,
    segment length=1.8pt, aspect=0.6}] (-0.8,-0.6) -- (0.9,-0.6);
    \draw (-0.8,-0.6) -- (0.9,-0.6);    
\draw [red,fill=red] (0.3,0.6) circle [radius=0.8pt];
\draw [red,fill=red] (0.3,-0.6) circle [radius=0.8pt];
\draw [red,decorate glaubr] (0.3,0.6) to [bend left=0] (0.3,- 0.6 );
\draw[color=green!80!black,decorate,decoration={gluon, amplitude=1.2pt,
    segment length=1.8pt, aspect=0.6}] (0.6,0 ) to [bend left=50] (0.8, 0.1 ); 
\draw[color=green!80!black,decorate,decoration={gluon, amplitude=1.2pt,
    segment length=1.8pt, aspect=0.6}] (0.8, -0.1) to [bend left=50] (0.6,0);   
\draw[color=green!80!black,decorate,decoration={gluon, amplitude=1.2pt,
    segment length=1.8pt, aspect=0.6}] (0.8, 0.1) to [bend left=50] (0.8, -0.1);           
\draw [red,fill=red] (0.6,0.6) circle [radius=0.8pt];
\draw [red,fill=red] (0.6,-0.6) circle [radius=0.8pt];
\draw [red,decorate glaubr] (0.6,0.6) to [bend left=0] (0.6,-0.6);
\draw [red,fill=red] (0.6,0) circle [radius=0.8pt];
% \draw [-{Latex[length=3pt]}] (-0.9,0.9) node[above,  scale=0.5]{$p_\ccO$}--(-0.7,0.7);
% \draw [-{Latex[length=3pt]}] (-0.9,-0.9) node[below,  scale=0.5]{$p_\ccT$}--(-0.7,-0.7);
% \draw [-{Latex[length=3pt]}] (0.6,0.7)--(0.9,0.7) node[above, scale=0.5]{$p_\ccTH$};
%  \draw [-{Latex[length=3pt]}] (0.6,-0.7)--(0.9,- 0.7) node[below, scale=0.5]{$q-p_\ccTH$}; 
\end{tikzpicture} 
\ee
These contribute at order $g_s^6 |\cM_0|^2$ as well as through interference with the tree-level graphs $\cM_0$. Note that there are no graphs where a soft gluon connects to a collinear line -- collinear fields only interact with ultrasoft gluons at leading power

First, let's consider observables in class 1, that are inclusive over the beam. 
Let's call the observable $X$ and the measurement function on the emitted radiation
$f_X(\vec{k})$.
For these observables we know that if we were inclusive over everything, all the Glauber graphs would exactly cancel, like for $q_T$. So the uncanceled part is only due to the real emission in the measured region of phase space. 
Thus we only need to consider Lipatov diagrams, like  $\cM_R^\one$ with real soft emissions. The leading factorization violating effect can then be computed by
\be
\frac{d \hat{\sigma}_G }{ d X}   = 
 \int d^{2} q_T d^{2} p_{\ccTH, \perp} \int_\Omega d^4k\,  2\pi \delta(k^2) \, |  \cM_R^\one |^2
 \delta(f_X(\vec k)- X)
\ee
where $\Omega$ is the area of phase space of the observable. 
In the limit that the angular region $A = \Delta y \Delta \phi$  in phase space  is small, the real emission amplitude will be independent of rapidity $y$ and azimuthal angle $\phi$. Then we expect the factorization violating effect to be proportional to the area and integrand to only depend on the transverse momentum of the emitted gluon.  Thus,
\be
\label{RealSoftInR}
\frac{d \hat{\sigma}_G }{ d X}   \sim A 
 \int d^{2} q_T d^{2} p_{\ccTH, \perp} d^2 k_\perp\,
 \gamma_\text{Lip}  (q_T, p_{\ccTH, \perp}, k_\perp)
 \delta(f_X(\vec k_\perp)- X)
\ee
where  $ \gamma_\text{Lip} $ is given by 
\begin{align}
 \gamma_\text{Lip}   (q_T, p_{\ccTH, \perp}, k_\perp)  & =   N_{\mu\nu}   N_{\rho\sigma}    
   \int \frac{d^{d-2} \ell_{1, \perp} }{(2 \pi)^{d-2} }   \frac{d^{d-2} \ell_{2, \perp} }{(2 \pi)^{d-2} }    \\
& \times 
 \frac{ (p^\mu_{\ccTH,\perp}  + \ell^\mu_{1, \perp} ) }{(\vec{p}_{\ccTH,\perp }+ \vec{\ell}_{1,\perp} )^2}  
 \frac{ (q_T^\nu  - p^\nu_{\ccTH,\perp} - \ell^\nu_{1, \perp})    }{( \vec{q}_T - \vec{p}_{\ccTH,\perp} - \vec{\ell}_{1, \perp} )^2 }   
 \frac{ (p^\rho_{\ccTH,\perp}  + \ell^\rho_{2, \perp} ) }{(\vec{p}_{\ccTH,\perp }+ \vec{\ell}_{2,\perp} )^2}  
 \frac{ (q_T^\sigma  - p^\sigma_{\ccTH,\perp} - \ell^\sigma_{2, \perp})    }{( \vec{q}_T - \vec{p}_{\ccTH,\perp} - \vec{\ell}_{2, \perp} )^2 }  
  \nn \\
& \times \frac{1}{ \vec{\ell}^2_{1, \perp}}  \frac{1}{ \vec{\ell}^2_{2, \perp}} 
\left[  
 \frac{ -2(\vec{\ell}_{1, \perp} +\vec{ \ell}_{2, \perp})^2 }{(\vec{\ell}_{1,\perp} - \vec{k }_\perp   )^2 (\vec{\ell}_{2,\perp} + \vec{k }_\perp   )^2}  + 
  \frac{2  \vec{\ell}_{1, \perp}^2 }{  (\vec{\ell}_{1,\perp} - \vec{k }_\perp   )^2  \vec{k}_\perp^2}   + 
  \frac{2  \vec{\ell}_{2, \perp}^2 }{  (\vec{\ell}_{2,\perp} + \vec{k }_\perp   )^2  \vec{k}_\perp^2}     
   \right] \nn
\end{align}
 Holding $k_\perp$ fixed, the phase-space integral $  \int  d^{2} q_T d^{2} p_{\ccTH, \perp} \, \gamma_\text{Lip}  (q_T, p_{\ccTH, \perp}, k_\perp) $ is IR finite. The $k_\perp \rightarrow 0$ limit is not allowed by the measurement function. Therefore \Eq{RealSoftInR} will give us a positive finite number proportional to the area $A$ and to $\alpha_s^4$.

Now let's proceed with the second type of observable, which measures particles emitted in all rapidity regimes. 
 In particular, we will assume that the  observable is  independent of the  rapidity of soft or collinear particles but 
 only sensitive to their transverse momenta. The measurement function acting on a two-body and three-body final state can be writen as 
 \begin{align} 
 f_X(  \vec p_{\ccTH, \perp}, \vec q_\perp- \vec p_{\ccTH, \perp}  )   & = |q_T|^a \, f_X( z, \bar z )    \\
 f_X( \vec p_{\ccTH, \perp}, \vec k_\perp, \vec q_T- \vec p_{\ccTH, \perp}- \vec k_\perp )  
 & =   |q_T|^a \, f_X( z, \bar z, w, \bar w ) 
 \end{align} 
 where $z, w$ are complex dimensionless variables defined by
 \begin{align} 
 z \bar z \equiv \frac{\vec p_{\ccTH, \perp}^2}{ q_T^2} , \quad  w \bar w \equiv \frac{ \vec k_\perp^2 }{q_T^2 }, \quad  (1-z-w)(1-\bar z- \bar w) \equiv \frac{\vec p_{\ccF, \perp}^2}{ q_T^2}
 \end{align}
By infrared safety,  $ f_X( z, \bar z ) $ and $ f_X( z, \bar z , w, \bar w) $ have the following properties,
\be
 \lim_{z \rightarrow 0 \text{ or }1}   f( z, \bar z )   =1, \quad  \lim_{w \rightarrow 0 }   f( z, \bar z , w, \bar w )   = f (z,\bar z),
\ee
 
Unlike the non-global observables, here all the diagrams contribute to  $\frac{d \sigma}{ d X}$.
 Since the diagrams  have entangled virtual and real IR divergences in different regimes, it is hard to see directly whether  $\frac{d\hat{\sigma}}{ d X}$ is non-vanishing.  The way  we will deal with it by adding and subtracting a subtraction term.
We do this by defining a new measurement function $f'$  that is not sensitive to soft radiation collinear to
the spectators so that $f'$ satisfies
\be
 f_X'( \vec p_{\ccTH, \perp}, \vec k_\perp, \vec q_T- \vec p_{\ccTH, \perp}- \vec k_\perp )  
 = f_X(\vec  p_{\ccTH,\perp} ,  \vec q_T- \vec p_{\ccTH, \perp} ) 
\ee
and for 2-body final states $f_X' = f_{X}$.

By adding and subtracting the cross section with $f'_X$, the cross section for $f_X$ can be written as the sum of two terms:
\begin{align}
\frac{d\hat{\sigma} }{ d \ln  X} = \left( \frac{d\hat{\sigma} }{ d \ln X} - \frac{d\hat{\sigma}' }{ d \ln X}  \right)  + \frac{d\hat{\sigma}' }{ d \ln X}  \label{MeasDiff}
\end{align}
where $\sigma'$ denotes the cross section computed with $f'_X$. 
The first term only acts non-trivially on wide-angle soft real emission diagrams, where the divergence as $k_\perp$ goes to zero is cured by the measurement function. To compute the second term, we can first integrate inclusively over soft momentum $k$, after which soft real and virtual divergences cancel. 
 Then the integrand contains only finite integrable functions in two-body phase space. 
In the following we will show in detail that either term in  \Eq{MeasDiff} could  be non-vanishing. 

Let's start with the first term. 
To better understand the behavior of this integral, especially how it depends on the measurement function, 
we need to study the scaling behavior of the  squared amplitude.  
Since the measurement is independent of the soft gluon rapidity,
 the integral generates a rapidity 
 divergence. Regulating this divergence with a scale $\nu$ gives a rapidity log term that
 breaks scale invariance.
 Thus, we 
we can no longer prove the cancellation using scale invariance as we did for Glauber ladder diagrams.,
even for a single-scale observable. 
In order to show the failure of cancellation, we only need to keep track of the rapidity logs.
Expanding the matrix element, we see
\begin{align} 
 |\cM|^2_{\text{Lip}}   (q_T, p_{\ccTH, \perp}, k_\perp )   = \int \frac{ d k^0 d k^z }{ ( 2 \pi)^2 }   |\cM^\one_R|^2  
 \overset{\text{regulated}}{=}   \left( \frac2\eta+ \ln \frac{\nu^2}{ k_\perp^2 } \right) \, \gamma_\text{Lip} (q_T, p_{\ccTH, \perp}, k_\perp )   + \eta\text{-finite.}
\label{eq:MR}
\end{align} 
Note that the $k_z$ integral in Eq.~\eqref{eq:MR} leads to rapidity divergence. We have used the rapidity regulator of Ref.~\cite{Chiu:2011qc,Chiu:2012ir} to regularize it, which leads to the divergent $2/\eta$ term in Eq.~\eqref{eq:MR}. The rapidity scale $\nu$ in Eq.~\eqref{eq:MR} is similar to the $\mu$ in dimensional regularization. 
This determines the scaling behavior of the squared amplitude 
\begin{align} 
 |\cM|^2_{\text{Lip}}   (q_T, p_{\ccTH, \perp}, k_\perp )   
 = \frac{1}{q_T^{8}}  
  \left( -\ln q_T^2 \, \widetilde{\gamma}_\text{Lip}  (z, \bar z, w, \bar w )  + \cO(\ep)  +  |\widetilde{\cM}|_\text{Lip}^2 (z, \bar z, w, \bar w )  \right)  \label{MLipScale} \,,
\end{align} 
where we use $\widetilde{\gamma}_{\text{Lip}}$ and $\widetilde{\cM}_{\text{Lip}}$ to denote the dimensionless version of $\gamma_{\text{Lip}}$ and $\cM_{\text{Lip}}$ by dividing with appropriate  power of $q_T$. 
The second term in \Eq{MLipScale} respects scale invariance and is therefore insensitive to the difference in measurement functions.  
The first term  in \Eq{MLipScale}  is of the form anticipated in Eq.~\eqref{qtparam}. It  will give us a non-vanishing integral 
\begin{align} 
 & \frac{d \sigma}{ d \ln X }  - \frac{d \sigma'}{ d \ln X }    \nn \\
  & = \int \frac{d^2 q_T d^2 p_{\ccTH , \perp} d^2 k_{\perp} }{(2\pi)^{6}}\,  |\cM|_\text{Lip}^2 (q_T, p_{\ccTH, \perp}, k_\perp) 
  \Big[ 
  \delta[  f_X(\vec q_T,\vec  p_{\ccTH,\perp},\vec k_\perp ) -X]
  -
  \delta[  f_X'(\vec q_T,\vec  p_{\ccTH,\perp}) -X]
  \Big]
  \\
& \overset{{\cO(\ep^0)}}{\sim} \int d^2 z  d^2 w \, \gamma_{\text{Lip}} (z, \bar z , w, \bar w ) \int d \ln q_T \, ( -\ln q_T^2) 
\Big[ \delta(f_X - X) - \delta(f_X' - X)\Big]
\nn \\
& =  \int d^2 z d^2 w \, \gamma_{\text{Lip}} (z, \bar z , w, \bar w ) \, \frac2a \ln \frac{f_X(z, \bar z, w, \bar w) }{ f_X(z, \bar z ) } 
\end{align} 

Now let's move onto the second term in \Eq{MeasDiff} . In order to compute $\frac{d \hat{\sigma}'}{ d X}$,
 let us take the Lipatov diagram and fully integrate over the real soft momentum $k$, then add it to the virtual diagrams.
Doing so allows us to write down a squared amplitude $|\cM_V|^2_\text{inc} $, which corresponds to the sum of all four-loop cut diagrams with fixed $q_T$ and $p_{\ccTH, \perp}$,   
 \begin{align}
 |\cM_V|^2_\text{inc} (q_T, p_{\ccTH, \perp} )
   & \equiv 
   \,  \cM_V^\two \big( \cM^\zero \big)^* +  \cM^\zero \big( \cM_V^\two \big)^*  \nn \\
 &  + \cM_V^\one \big( \cM_V^\one \big)^*   + \int \frac{d^{d} k }{ (2\pi)^d } 2\pi \delta(k^2)   \cM_R^\one  \big( \cM_R^\one \big)^*    
\end{align} 
Then the cross section becomes a two-body phase-space integral over $ |\cM_V|^2_\text{inc} $, 
\begin{align} 
  \frac{d \sigma'}{ d \ln X}   
  =  \int \frac{d^2 q_T d^2 p_{\ccTH , \perp} }{(2\pi)^{4}} \,    |\cM_V|^2_\text{inc} (q_T, p_{\ccTH, \perp} ) 
   \;  
   \delta[ f'(\vec{q}_T, \vec {p}_{\ccTH, \perp} )-X] 
\end{align}
Focusing on the  divergent terms, we are able to determine the scaling behaviour of $|\cM_V|^2_\text{inc}$, which takes the following form
\begin{align} 
 |\cM_V|^2_\text{inc} (q_T, p_{\ccTH, \perp} )  =  \frac{1}{q_T^{4}}
 & \left[ \Big( \frac12 \ln^2 q_T^2  - \ln \nu^2 \ln q_T^2  \Big)  \,  \Gamma_\text{inc}  (z, \bar z )   \right.   \nn  \\
   &    \left.  \quad  - \ln q_T^2 \, \gamma_\text{inc}  (z, \bar z )  + \cO(\ep) 
  +  |\widetilde{\cM}_V|^2_\text{inc} ( z, \bar z ) \right] 
\end{align}  
where $ \Gamma_\text{inc} $ and $ \gamma_\text{inc} $ are determined by integrals in  transverse dimensions  over the soft and Glauber momenta. 
Again, the cross section with measurement $f_X'$  can be written as its difference with measurement $q_T$, 
\begin{align} 
    \frac{d \sigma'}{ d \ln X } - \frac1a\frac{ d \sigma}{ d \ln q_T}
&=  \frac{2}{a^2}  \int d^2 z  \,   \Gamma_\text{inc} (z, \bar z) \,   \ln^2 f(z, \bar z )  \nn \\
& + \frac2a \int d^2 z \,
\left[ \ln \frac{\nu^2}{ X^2} \, \Gamma_\text{inc} (z, \bar z)    + \gamma_{\text{inc}} (z, \bar z )  \right] \, \ln f(z, \bar z ) 
\end{align} 

Putting the pieces together
\begin{multline}
 \frac{d \sigma}{ d \ln X}  \sim \frac2a  \ln \frac{\nu^2}{ X^2} \int d^2 z \, \Gamma_\text{inc} (z, \bar z) \,  \ln f(z, \bar z )  +   \frac{2}{a^2}  \int d^2 z  \,   \Gamma_\text{inc} (z, \bar z) \,   \ln^2 f(z, \bar z )    \\
 +  \frac2a \int d^2 z\,  \left[ 
\gamma_{\text{inc}} (z, \bar z ) \, \ln f(z, \bar z ) + \int d^2 w \, \widetilde{\gamma}_{\text{Lip}} (z, \bar z , w, \bar w ) \, \ln \frac{f (z, \bar z, w, \bar w) }{ f(z, \bar z ) }  \right] 
 \end{multline}
Generically we expect the integrals above not to vanish. The $\nu$ dependence here should cancel with contributions from Glauber diagrams with one additional collinear gluon, and $\nu^2$ should be replaced by a hard scale related to $q^+ q^-$. 
 It would certainly be interesting to see whether the explicit forms of these integrals can help determine
some observables for which the integrals do vanish, and factorization violation is then
postponed to higher order.

% active-spectator-spectator interations 
So far our discussion is restricted to diagrams with soft emissions from the Glauber line. 
At  leading order, real-emission diagrams can also have a single soft gluon coming off of an active collinear parton. Example
tree-level and 1-loop diagrams are
\be
\cM_{Ra}^{0} = 
\begin{tikzpicture}[baseline={([yshift=-.5ex]current bounding box.center)},scale=1.1]
\draw (-1,-0.8) -- (0,0);
\draw (-1,0.8) -- (0,0);
\draw [fill=\blobcolor] (0,0) circle [radius = 0.15];
\draw[color=black,decorate,decoration={gluon, amplitude=1.2pt,
    segment length=1.8pt, aspect=0.6}] (-0.8,0.6) -- (0.6,0.6);
    \draw (-0.8,0.6) -- (0.6,0.6);
\draw[color=black,decorate,decoration={gluon, amplitude=1.2pt,
    segment length=1.8pt, aspect=0.6}] (-0.8,-0.6) -- (0.6,-0.6);
    \draw (-0.8,-0.6) -- (0.6,-0.6);    
\draw[color=green!80!black,decorate,decoration={gluon, amplitude=1.2pt,
    segment length=1.8pt, aspect=0.6}] (-0.1,0.1) to [bend right=0] (0.4,0);
 \draw [-{Latex[length=3pt]}] (0.3,-0.1)--(0.6,-0.1)node[below, scale=0.4]{$k$};
 \draw [-{Latex[length=3pt]}] (-0.9,0.9) node[above,  scale=0.5]{$p_\ccO$}--(-0.7,0.7);
 \draw [-{Latex[length=3pt]}] (-0.9,-0.9) node[below,  scale=0.5]{$p_\ccT$}--(-0.7,-0.7);
 \draw [-{Latex[length=3pt]}] (0.3,0.7)--(0.6,0.7) node[above, scale=0.5]{$p_\ccTH$};
  \draw [-{Latex[length=3pt]}] (0.3,-0.7)--(0.6,- 0.7) node[below, scale=0.5]{$q-k-p_\ccTH$};
\end{tikzpicture}
, \quad 
\cM_{Ra}^{1} = 
\begin{tikzpicture}[baseline={([yshift=-.5ex]current bounding box.center)},scale=1.1]
\draw (-1,-0.8) -- (0,0);
\draw (-1,0.8) -- (0,0);
\draw [fill=\blobcolor] (0,0) circle [radius = 0.15];
\draw[color=black,decorate,decoration={gluon, amplitude=1.2pt,
    segment length=1.8pt, aspect=0.6}] (-0.8,0.6) -- (0.8,0.6);
    \draw (-0.8,0.6) -- (0.8,0.6);
\draw[color=black,decorate,decoration={gluon, amplitude=1.2pt,
    segment length=1.8pt, aspect=0.6}] (-0.8,-0.6) -- (0.8,-0.6);
    \draw (-0.8,-0.6) -- (0.8,-0.6);    
\draw [red,fill=red] (0.4,0.6) circle [radius=0.8pt];
\draw [red,fill=red] (0.4,-0.6) circle [radius=0.8pt];
\draw [red,decorate glaubr] (0.4,0.6) to [bend left=0] (0.4,-0.6);
\draw[color=green!80!black,decorate,decoration={gluon, amplitude=1.2pt,
    segment length=1.8pt, aspect=0.6}] (-0.1,0.1) to [bend right=0] (0.6,0);
 \draw [-{Latex[length=3pt]}] (0.5,0.5)--(0.5,0.2)node[right, scale=0.4]{$\tiny{ \ell_1}$};
 \draw [-{Latex[length=3pt]}] (0.5,-0.1)--(0.8,-0.1)node[below, scale=0.4]{$k$};
 \draw [-{Latex[length=3pt]}] (-0.9,0.9) node[above,  scale=0.5]{$p_\ccO$}--(-0.7,0.7);
 \draw [-{Latex[length=3pt]}] (-0.9,-0.9) node[below,  scale=0.5]{$p_\ccT$}--(-0.7,-0.7);
 \draw [-{Latex[length=3pt]}] (0.5,0.7)--(0.8,0.7) node[above, scale=0.5]{$p_\ccTH$};
  \draw [-{Latex[length=3pt]}] (0.5,-0.7)--(0.8,- 0.7) node[below, scale=0.5]{$q-k-p_\ccTH$};
\end{tikzpicture}
\ee
A factorization-violating effect can come from interference of $ \cM_{Ra}^0$ with two-loop real-emission diagrams $\cM_{R}^2$:
\be
\cM_R^2 = 
\begin{tikzpicture}[baseline={([yshift=-.5ex]current bounding box.center)},scale=1.1]
\draw (-1,-0.8) -- (0,0);
\draw (-1,0.8) -- (0,0);
\draw [fill=\blobcolor] (0,0) circle [radius = 0.15];
\draw[color=black,decorate,decoration={gluon, amplitude=1.2pt,
    segment length=1.8pt, aspect=0.6}] (-0.8,0.6) -- (0.9,0.6);
    \draw (-0.8,0.6) -- (0.9,0.6);
\draw[color=black,decorate,decoration={gluon, amplitude=1.2pt,
    segment length=1.8pt, aspect=0.6}] (-0.8,-0.6) -- (0.9,-0.6);
    \draw (-0.8,-0.6) -- (0.9,-0.6);    
\draw [red,fill=red] (0.3,0.6) circle [radius=0.8pt];
\draw [red,fill=red] (0.33,-0.6) circle [radius=0.8pt];
\draw [red,decorate glaubr] (0.3,0.6) to [bend left=0] (0.3,-0.6);
\draw [red,fill=red] (0.6,0.6) circle [radius=0.8pt];
\draw [red,fill=red] (0.6,-0.6) circle [radius=0.8pt];
\draw [red,decorate glaubr] (0.6,0.6) to [bend left=0] (0.6,-0.6);
\draw [red,fill=red] (0.3,0) circle [radius=0.8pt];
%\draw [red,fill=red] (0.6,0) circle [radius=0.8pt];
\draw[color=green!80!black,decorate,decoration={gluon, amplitude=1.2pt,
    segment length=1.8pt, aspect=0.6}] (0.3,0) to [bend right=0] (0.8,0);
 \draw [-{Latex[length=3pt]}] (0.65,-0.1)--(0.95,-0.1)node[below, scale=0.4]{$k$};    
%\draw [-{Latex[length=3pt]}] (0.65,-0.1)--(0.65,-0.3)node[below, scale=0.4]{$\qquad \tiny{ \ell_2 + k}$};
% \draw [-{Latex[length=3pt]}] (0.35,-0.1)--(0.35,-0.3)node[below, scale=0.4]{$\tiny{ \ell_1 -k}$};
 \draw [-{Latex[length=3pt]}] (0.65,0.5)--(0.65,0.3)node[below, scale=0.4]{$\; \tiny{ \ell_2 }$};
% \draw [-{Latex[length=3pt]}] (0.35,0.5)--(0.35,0.3)node[below, scale=0.4]{$\; \tiny{ \ell_1 }$};
 \draw [-{Latex[length=3pt]}] (-0.9,0.9) node[above,  scale=0.5]{$p_\ccO$}--(-0.7,0.7);
 \draw [-{Latex[length=3pt]}] (-0.9,-0.9) node[below,  scale=0.5]{$p_\ccT$}--(-0.7,-0.7);
 \draw [-{Latex[length=3pt]}] (0.6,0.7)--(0.9,0.7) node[above, scale=0.5]{$p_\ccTH$};
  \draw [-{Latex[length=3pt]}] (0.6,-0.7)--(0.9,- 0.7) node[below, scale=0.5]{$q-k-p_\ccTH$};
\end{tikzpicture}  
\quad
+ 
\quad
\begin{tikzpicture}[baseline={([yshift=-.5ex]current bounding box.center)},scale=1.1]
\draw (-1,-0.8) -- (0,0);
\draw (-1,0.8) -- (0,0);
\draw [fill=\blobcolor] (0,0) circle [radius = 0.15];
\draw[color=black,decorate,decoration={gluon, amplitude=1.2pt,
    segment length=1.8pt, aspect=0.6}] (-0.8,0.6) -- (0.9,0.6);
    \draw (-0.8,0.6) -- (0.9,0.6);
\draw[color=black,decorate,decoration={gluon, amplitude=1.2pt,
    segment length=1.8pt, aspect=0.6}] (-0.8,-0.6) -- (0.9,-0.6);
    \draw (-0.8,-0.6) -- (0.9,-0.6);    
\draw [red,fill=red] (0.3,0.6) circle [radius=0.8pt];
\draw [red,fill=red] (0.33,-0.6) circle [radius=0.8pt];
\draw [red,decorate glaubr] (0.3,0.6) to [bend left=0] (0.3,-0.6);
\draw [red,fill=red] (0.6,0.6) circle [radius=0.8pt];
\draw [red,fill=red] (0.6,-0.6) circle [radius=0.8pt];
\draw [red,decorate glaubr] (0.6,0.6) to [bend left=0] (0.6,-0.6);
%\draw [red,fill=red] (0.3,0) circle [radius=0.8pt];
\draw [red,fill=red] (0.6,0) circle [radius=0.8pt];
\draw[color=green!80!black,decorate,decoration={gluon, amplitude=1.2pt,
    segment length=1.8pt, aspect=0.6}] (0.6,0) to [bend right=0] (0.9,0);
 \draw [-{Latex[length=3pt]}] (0.65,-0.1)--(0.95,-0.1)node[below, scale=0.4]{$k$};    
%\draw [-{Latex[length=3pt]}] (0.65,-0.1)--(0.65,-0.3)node[below, scale=0.4]{$\qquad \tiny{ \ell_2 + k}$};
% \draw [-{Latex[length=3pt]}] (0.35,-0.1)--(0.35,-0.3)node[below, scale=0.4]{$\tiny{ \ell_1 -k}$};
% \draw [-{Latex[length=3pt]}] (0.65,0.5)--(0.65,0.3)node[below, scale=0.4]{$\; \tiny{ \ell_2 }$};
 \draw [-{Latex[length=3pt]}] (0.35,0.5)--(0.35,0.3)node[below, scale=0.4]{$\; \tiny{ \ell_1 }$};
 \draw [-{Latex[length=3pt]}] (-0.9,0.9) node[above,  scale=0.5]{$p_\ccO$}--(-0.7,0.7);
 \draw [-{Latex[length=3pt]}] (-0.9,-0.9) node[below,  scale=0.5]{$p_\ccT$}--(-0.7,-0.7);
 \draw [-{Latex[length=3pt]}] (0.6,0.7)--(0.9,0.7) node[above, scale=0.5]{$p_\ccTH$};
  \draw [-{Latex[length=3pt]}] (0.6,-0.7)--(0.9,- 0.7) node[below, scale=0.5]{$q-k-p_\ccTH$};
\end{tikzpicture}  
\ee
and from interference between $ \cM_{Ra}^1$ and the one-loop diagram $ \cM_R^1$ in Eq.~\eqref{MR1}.
These interference terms contribute to the cross section  at order $g_s^6 |\cM_0|^2$. 
Both $\cM_{Ra}^1 (q, p_{\ccTH}, k)$ and $\cM_R^2 (q, p_{\ccTH}, k)$ contain IR divergences from the Glauber loop labeled in the diagram. 
With the presence of Lipatov vertex,  the color generators of the two Glauber operator insertions in $\cM_R^2$ do not commute, therefore
IR divergences do not cancel at amplitude-squared level.  
The interference term $\cM_{Ra}^1 (\cM_R^1 )^* +\cM_{Ra}^0 (\cM_R^2 )^* $ could therefore contain $q_T$-dependent IR poles, generating factorization-violating  effects.

The same issue appears  in $\cM_R$ at higher loop order, for example, though diagarms
\be  
\cM_{R}^n = 
\begin{tikzpicture}[baseline={([yshift=-.5ex]current bounding box.center)},scale=1.1]
\draw (-1,-0.8) -- (0,0);
\draw (-1,0.8) -- (0,0);
\draw [fill=\blobcolor] (0,0) circle [radius = 0.15];
\draw[color=black,decorate,decoration={gluon, amplitude=1.2pt,
    segment length=1.8pt, aspect=0.6}] (-0.8,0.6) -- (1.0,0.6);
    \draw (-0.8,0.6) -- (0.9,0.6);
\draw[color=black,decorate,decoration={gluon, amplitude=1.2pt,
    segment length=1.8pt, aspect=0.6}] (-0.8,-0.6) -- (1.0,-0.6);
    \draw (-0.8,-0.6) -- (0.9,-0.6);    
\draw [red,fill=red] (0.25,0.6) circle [radius=0.8pt];
\draw [red,fill=red] (0.25,-0.6) circle [radius=0.8pt];
\draw [red,decorate glaubr] (0.25,0.6) to [bend left=0] (0.25,-0.6);
\draw [red,fill=red] (0.5,0.6) circle [radius=0.8pt];
\draw [red,fill=red] (0.5,0) circle [radius=0.8pt];
\draw [red,fill=red] (0.5,-0.6) circle [radius=0.8pt];
\draw [red,decorate glaubr] (0.5,0.6) to [bend left=0] (0.5,-0.6);
\draw [red,fill=red] (0.8,0.6) circle [radius=0.8pt];
\draw [red,fill=red] (0.8,-0.6) circle [radius=0.8pt];
\draw [red,decorate glaubr] (0.8,0.6) to [bend left=0] (0.8,-0.6);
\draw[color=green!80!black,decorate,decoration={gluon, amplitude=1.2pt,
    segment length=1.8pt, aspect=0.6}] (0.5,0) to [bend right=0] (0.9,0);
   \draw [-{Latex[length=3pt]}] (0.75,-0.1)--(0.95,-0.1)node[below, scale=0.4]{$k$}; 
% \node[scale=0.5] at (0.95,-0.1) {$k$}; 
\node[scale=0.5] at (0.7,0.1) {$\cdots$};
\node[scale=0.5] at (0.4,0.1) {$\cdots$};
%\draw [-{Latex[length=3pt]}] (1,0.2)--(1,0.0)node[right, scale=0.5]{$\ell$};
 \draw [-{Latex[length=3pt]}] (-0.9,0.9) node[above,  scale=0.5]{$p_\ccO$}--(-0.7,0.7);
 \draw [-{Latex[length=3pt]}] (-0.9,-0.9) node[below,  scale=0.5]{$p_\ccT$}--(-0.7,-0.7);
 \draw [-{Latex[length=3pt]}] (0.6,0.7)--(0.9,0.7) node[above, scale=0.5]{$p_\ccTH$};
  \draw [-{Latex[length=3pt]}] (0.6,-0.7)--(0.9,- 0.7) node[below, scale=0.5]{$q-k-p_\ccTH$};
\end{tikzpicture}
, \quad 
\cM_{Ra}^n = 
\begin{tikzpicture}[baseline={([yshift=-.5ex]current bounding box.center)},scale=1.1]
\draw (-1,-0.8) -- (0,0);
\draw (-1,0.8) -- (0,0);
\draw [fill=\blobcolor] (0,0) circle [radius = 0.15];
\draw[color=black,decorate,decoration={gluon, amplitude=1.2pt,
    segment length=1.8pt, aspect=0.6}] (-0.8,0.6) -- (1.0,0.6);
    \draw (-0.8,0.6) -- (0.9,0.6);
\draw[color=black,decorate,decoration={gluon, amplitude=1.2pt,
    segment length=1.8pt, aspect=0.6}] (-0.8,-0.6) -- (1.0,-0.6);
    \draw (-0.8,-0.6) -- (0.9,-0.6);    
\draw [red,fill=red] (0.3,0.6) circle [radius=0.8pt];
\draw [red,fill=red] (0.3,-0.6) circle [radius=0.8pt];
\draw [red,decorate glaubr] (0.3,0.6) to [bend left=0] (0.3,-0.6);
\draw [red,fill=red] (0.5,0.6) circle [radius=0.8pt];
\draw [red,fill=red] (0.5,-0.6) circle [radius=0.8pt];
\draw [red,decorate glaubr] (0.5,0.6) to [bend left=0] (0.5,-0.6);
\draw [red,fill=red] (0.85,0.6) circle [radius=0.8pt];
\draw [red,fill=red] (0.85,-0.6) circle [radius=0.8pt];
\draw [red,decorate glaubr] (0.85,0.6) to [bend left=0] (0.85,-0.6);
\draw[color=green!80!black,decorate,decoration={gluon, amplitude=1.2pt,
    segment length=1.8pt, aspect=0.6}] (-0.1,0.1) to [bend right=0] (0.65,0);
  \draw [-{Latex[length=3pt]}] (0.6,-0.1)--(0.8,-0.1)node[below, scale=0.4]{$k$}; 
 %\node[scale=0.5]  at (0.7,-0.1) {$k$}; 
\node[scale=0.5] at (0.7,0.1) {$\cdots$};
%\draw [-{Latex[length=3pt]}] (1,0.2)--(1,0.0)node[right, scale=0.5]{$\ell$};
 \draw [-{Latex[length=3pt]}] (-0.9,0.9) node[above,  scale=0.5]{$p_\ccO$}--(-0.7,0.7);
 \draw [-{Latex[length=3pt]}] (-0.9,-0.9) node[below,  scale=0.5]{$p_\ccT$}--(-0.7,-0.7);
 \draw [-{Latex[length=3pt]}] (0.6,0.7)--(0.9,0.7) node[above, scale=0.5]{$p_\ccTH$};
  \draw [-{Latex[length=3pt]}] (0.6,-0.7)--(0.9,- 0.7) node[below, scale=0.5]{$q-k-p_\ccTH$};
\end{tikzpicture}
\ee
 Since color generators to the left of the Lipatov vertex do not commute to the right, 
   IR divergences in  real-emission diagrams do not cancel in the squared amplitudes $|\cM_R|^2_\text{N-Glauber}$ or $ (\cM_R \cM_{Ra}^*)_\text{N-Glauber}$,  for $N>3$. Understanding the structure of the IR divergences of the active/spectator soft-emission diagrams terms could be important to understanding  
   factorization-violation in more detail and should be an interesting area for future research.

\section{Summary and Conclusion} \label{sec:conc}
The main result of this paper is that for any single scale observable, factorization is not violated due to processes involving only Glauber exchange between spectator quarks. 
 Previous studies, working in position space~\cite{Rothstein:2016bsq} or using more standard kinematic variables~\cite{Gaunt:2014ska}, suggested that the sum of all of these graphs would not vanish for a generic observable without some magical cancellation. 
 In our proof, we show that the magical cancellation does happen, and is clearest using  a natural set of conformal coordinates $z$ and $\bar{z}$. With these coordinates, 
 the integral over phase space for  spectator emission is completely inclusive over $z$ for all infrared-safe single-scale observables, and hence the Glauber cancellation for $q_T$ implies  the Glauber cancellation for all such observables. 

To gain more insight into the cancellation, we computed the complete contribution of Glauber ladder graphs at next-to-leading order. In order to see the cancellation explicitly, we needed the terms up to order $\ep^0$ at 2 loops
and the terms up to order $\ep$ at 1-loop. These order $\ep$ terms are needed because a $\frac{1}{\ep}$ phase space divergence appears when integrating over the spectator transverse momentum. 
We demonstrated a non-trivial cancellation between the $\cO(\ep^0)$ IR-finite 2-loop contribution and the $\frac{1}{\ep} \times \ep$ 1-loop contribution, confirming our general result.

To demonstrate the cancellation to all orders, we showed that all of the IR divergences in the ladder diagrams exponentiate into a phase. After the phase is pulled out, the $\cO(\ep^0)$ terms are integrable over phase space while the $\cO(\ep)$ terms are not. Thus at each order in perturbation theory, a non-trivial cancellation between finite contributions and phase-space singular contributions will occur. While we have focused on hadronic observables associated with a hard Drell-Yan scattering process in this paper, the method we use is general and can also be applied to more complicated processes. For example, it has been shown that dijet production at hadron colliders violate factorization if the dijets are back-to-back and have small total transverse momentum~\cite{Collins:2007nk}. It would be interesting to see whether or how factorization is violated through explicit calculation along the line of this work. Another interesting process to consider using our formalism is $t\bar{t}$ production at small transverse momentum~\cite{Zhu:2012ts,Li:2013mia,Catani:2014qha}, where it has been shown that final state interactions of top quarks and spectators leads to factorization violation~\cite{Mitov:2012gt}.

One corollary of our proof is that that the leading order factorization-violating effects must involve soft radiation, through the Lipatov vertex associated with the Glauber lines. The graphs involving such soft radiation can have IR divergences that do not exponentiate and/or rapidity divergences. Such divergences, when regulated, generate a scale in the amplitude which prohibits the application of our scale-invariance argument for observable independence. We briefly studied the forms that these higher order terms can have. Diagramatically these contributions are similar to the factorization violation in Regge factorization~\cite{DelDuca:2014cya,Rothstein:2016bsq}. It would be interesting to evaluate these diagrams explicitly, and see if there is any universality in the perturbative factorization violating terms. With the explicit expression for the factorization-violating contribution, it would also be interesting to search for hadronic observables that can delay factorization-violation to higher order or that avoid factorization-violating effects all together.

\section*{Acknowledgements}
We thank Iain Stewart for discussions, and Jonathan Gaunt, Iain Stewart, Ira Rothstein for valuable comments. MDS and KY are supported in part by the Department of Energy under contract DE-SC0013607. HXZ is supported in part by The Thousand Youth Plan of China under contract 588020-X01702/076. 

\appendix

\section{More on the change of variables}
\label{sec:variable}

The main ingredient in our proof of Glauber cancellation for $E_T$ observable in Sec.~\ref{sec:ladders} is the invariance of integration measure when changing from $q_T$ to $E_T$ in $d=4$ dimensions,
\begin{align}
\frac{dq_T}{q_T} du \, dv = \frac{dE_T}{E_T} du \, dv
\label{eq:qteqet}
\end{align}
where the definition of $u$ and $v$ is given in Eq.~\eqref{uvdef}. The above equality should be understood in the sense of integrating over an dimensionless squared amplitude. The algebra needed to arrive at this result is relatively simple, but we give it here for completeness. 

By dimensional analysis, the integration in the transverse momentum space can be parameterized, e.g., by
\begin{align}
  d\Phi_T = \frac{1}{q_T^3} d q_T \, dp_{\ccTH,T} \, dp_{\ccF,T} 
\end{align}
Here $p_{\ccTH,T} = |p_{\ccTH,\perp}|$, $p_{\ccF,T}=|p_{\ccF,T}|$. We have suppressed a factor related to azimuthal angle integral, which is irrelevant to our discussion. 
$q_T$, $p_{\ccTH,T}$, and $p_{\ccF,T}$ gives a complete parameterization of the transverse momentum space, by specifying the length of three edges of the triangle depicted in Sec.~\ref{sec:ladders}. Recalling that $p_{\ccTH,T} = \sqrt{u} q_T$, $p_{\ccF,T}=\sqrt{v} q_T$, we change variables to $q_T$, $u$, and $v$ by
\begin{align}
  d\Phi_T = & \frac{1}{q_T^3} dq_T\, du\, dv\,
\left|  \frac{\partial(q_T,p_{\ccTH,T},p_{\ccF,T})}{\partial(q_T,u,v)} \right|\,,
\\ 
=& \frac{1}{4 q_T \sqrt{u} \sqrt{v}} dq_T\, du\, dv 
\label{eq:qtuv}
\end{align}

We can also change variables to $E_T$, $u$, and $v$, using the same definition of $u$ and $v$ as in Eq.~\eqref{uvdef},
\begin{align}
  d\Phi_T = & \frac{1}{q_T^3}  \left|\frac{\partial(q_T, p_3,
  p_4)}{\partial(E_T, u, u )} \right| dE_T\, du\, dv
\\
= & 
\frac{1}{q_T^3}  \left|\frac{\partial \left(\frac{E_T}{\sqrt{u} + \sqrt{v}} , E_T \frac{\sqrt{u}}{\sqrt{u} + \sqrt{v}},
  E_T \frac{\sqrt{v}}{\sqrt{u} + \sqrt{v} }\right)}{\partial( E_T, u, v)} \right| dE_T\, du\, dv\, 
\\
= & \frac{1}{q_T^3} \frac{E_T^2}{4 \sqrt{u}\sqrt{v}
    (\sqrt{u}+\sqrt{v})^3} dE_T\, du\, dv\, 
\\
= &   \frac{1}{4 \sqrt{u}\sqrt{v}
    E_T} dE_T\, du\, dv\, 
\label{eq:etuv}
\end{align}
Comparing Eq.~\eqref{eq:qtuv} and Eq.~\eqref{eq:etuv}, we arrive at Eq.~\eqref{eq:qteqet}.
A similar equality holds if we change from $du\,dv$ to $dz\,d\bar{z}$. In  $d= 4 - 2 \ep$ dimensions, the integration measure becomes 
\begin{align}
 \frac{dE_T}{E_T} du \, dv =  \frac{dq_T}{q_T} du \, dv  \left( 1 + \alpha \ep \ln(\sqrt{u} + \sqrt{v}) + \cO(\ep^2) \right)
\end{align}
where $\alpha=2 + 2n$ with $n$ Glauber loops. The difference of the two parameterizations is $\cO(\ep)$ or above. Since after cancelling the infrared divergent phase factor in Eq.~\eqref{eq:IRphase}, individual pure Glauber graphs can give at most finite corrections, these $\cO(\ep)$ terms can be safely neglected. 

The equality in Eq.~\eqref{eq:qteqet} does not hold when the squared amplitude develops anomalous dimension, either from an incomplete cancellation of soft radiation, or from a rapidity divergence. These two cases are discussed in Sec.~\ref{sec:fv}.

\section{Exponentiation of IR divergences \label{app:exp}}
Consider diagram $\cM_{Gn}$ with $n$ Glauber gluons exchanged:
\begin{align} 
\cM_{G}^n( q_T, p_{\ccTH, \perp} ) = 
\begin{tikzpicture}[baseline={([yshift=-.5ex]current bounding box.center)},scale=1.1]
\draw (-1,-0.8) -- (0,0);
\draw (-1,0.8) -- (0,0);
\draw [fill=\blobcolor] (0,0) circle [radius = 0.15];
\draw[color=black,decorate,decoration={gluon, amplitude=1.2pt,
    segment length=1.8pt, aspect=0.6}] (-0.8,0.6) -- (1.0,0.6);
    \draw (-0.8,0.6) -- (0.9,0.6);
\draw[color=black,decorate,decoration={gluon, amplitude=1.2pt,
    segment length=1.8pt, aspect=0.6}] (-0.8,-0.6) -- (1.0,-0.6);
    \draw (-0.8,-0.6) -- (0.9,-0.6);    
\draw [red,fill=red] (0.3,0.6) circle [radius=0.8pt];
\draw [red,fill=red] (0.3,-0.6) circle [radius=0.8pt];
\draw [red,decorate glaubr] (0.3,0.6) to [bend left=0] (0.3,-0.6);
\draw [red,fill=red] (0.5,0.6) circle [radius=0.8pt];
\draw [red,fill=red] (0.5,-0.6) circle [radius=0.8pt];
\draw [red,decorate glaubr] (0.5,0.6) to [bend left=0] (0.5,-0.6);
\draw [red,fill=red] (0.85,0.6) circle [radius=0.8pt];
\draw [red,fill=red] (0.85,-0.6) circle [radius=0.8pt];
\draw [red,decorate glaubr] (0.85,0.6) to [bend left=0] (0.85,-0.6);
\node[scale=0.5] at (0.7,0) {$\cdots$};
\draw [-{Latex[length=3pt]}] (1,0.2)--(1,0.0)node[right, scale=0.5]{$\ell$};
 \draw [-{Latex[length=3pt]}] (-0.9,0.9) node[above,  scale=0.5]{$\ccO$}--(-0.7,0.7);
 \draw [-{Latex[length=3pt]}] (-0.9,-0.9) node[below,  scale=0.5]{$\ccT$}--(-0.7,-0.7);
 \draw [-{Latex[length=3pt]}] (0.6,0.7)--(0.9,0.7) node[above, scale=0.5]{$\ccTH$};
  \draw [-{Latex[length=3pt]}] (0.6,-0.7)--(0.9,- 0.7) node[below, scale=0.5]{$\ccF$};
\end{tikzpicture}
\end{align}
The diagram with $n$ Glauber gluons is related to the one with $n-1$ Glauber gluons through the following
recurrence relation
\be
\cM_G^{\n} ( q_T, p_{\ccTH, \perp} ) =    
 \frac{ -i \alpha_s(\mu) }{2 }  \, (\TT_\ccTH  \cdot \TT_\ccF) \,    \frac{1}{n}  \int \frac{d^{d-2} \ell_\perp }{ \pi^{d/2-1} }   \, 
  \frac{1}{\vec \ell_\perp^2 } \, \cM^{\nmo}_G 
  (q_T,  \, p_{\ccTH,\perp}+ \ell_\perp )  \label{MGRecur}
\ee
where $\alpha_s (\mu) \equiv 4\pi g_s^2 (4\pi \mu^2)^\ep $.  

The integral in Eq.~\eqref{MGRecur} can be divergent due to the region around $\ell_\perp = 0$. To remove this divergence, let us define 
 an amplitude $\cM_{G,R} \equiv \sum_{n=0}^\infty \cM^n_{G, R} $ recursively through the following equations: 
\be
 \cM^\zero_{G,R} (q_T,  p_{\ccTH,\perp} )  \equiv \cM^{\zero}  (q_T,  p_{\ccTH,\perp} ) ,  
\label{MR0}
\ee
 and for $n\geq 1$, 
\be
  \cM^\n_{G, R} (q_T,  p_{\ccTH,\perp} ) \equiv  \frac{ -i \alpha_s}{ 2 }  (\TT_\ccTH  \cdot \TT_\ccF)   \frac{1}{n} 
 \int \frac{d^{d-2} \ell_\perp }{ \pi^{d/2-1} }   
\bigg[   \frac{1}{\vec \ell_\perp^2 }   + \frac{\pi^{d/2-1}}{  \epsilon } \delta^{d-2} ( \ell_\perp )  \bigg]
 \cM^{\nmo}_{G, R}  (q_T, \,  p_{\ccTH,\perp}+ \ell_\perp  )    \label{MGRRecur}
 \ee
We claim that $\cM_{G, R}$ thus defined is IR finite at all orders in $\alpha_s$. The proof is by induction.
Assume that $\cM^{\nmo}_{G, R} (q_T,  p_{\ccTH,\perp} )$ is IR finite. 
Then the IR divergence in the integral in \Eq{MGRRecur} 
comes from the region where $\ell_\perp \rightarrow 0$.   
Since $\cM^{\nmo}_{G, R}  ( q_T, p_{\ccTH,\perp}+ \ell_\perp,   ) $ is smooth around $\ell_\perp = 0$, the integrand behaves like  $\sim \frac{1}{\vec \ell_\perp^2} $ in the singular limit.  Asymptotically, 
\begin{align} 
 \int \frac{d^{d-2} \ell_\perp }{  \vec \ell_\perp^2 }    \xrightarrow{\ell_\perp \rightarrow 0}  
  \Omega_{d-2} \int  \frac{d |\ell_\perp| }{ |\ell_\perp|^{1+2\ep}}  \sim -  \frac{\Omega_{d-2}}{2\ep} \int d |\ell_\perp|   \delta (|\ell_\perp|)  
\end{align} 
Thus the IR divergence that arises from
$\vec{\ell}_\perp \rightarrow 0$ limit is cancelled by the second term in brackets in \Eq{MGRRecur}, and therefore $\cM^n_{G, R}$ is IR finite.  

Next, we show that $\cM_{G, R}$ thus defined differs from $\cM_G$ only by a pure phase. Explicitly,
 \begin{align} \label{poleExp}
 \cM_{G} = e^{   \frac{i  \alpha_s}{ 2 \ep}   ( \TT_\ccTH \cdot \TT_\ccF )  } \cM_{G, R} 
 \end{align} 
This equation holds at each order in $\alpha_s$. It is consistent with observations made in~\cite{Rothstein:2016bsq} using a different regulator. 
Again we prove this result by induction, starting from $n=0$ where \Eq{poleExp}  holds by the definition, Eq.~\eqref{MR0}. 
Now we assume that the exponentiation holds at $(n-1)-$loop order, 
\begin{align} 
\cM_G^{\nmo} & = \sum_{m= 0}^{n-1} \frac{1}{ (n-1-m)!}  \left[ \frac{ i \alpha_s}{ 2\ep}  ( \TT_\ccTH \cdot \TT_\ccF )   \right]^{\nmom} \cM_{G, R}^{\m}  
\end{align} 
Plugging  $\cM_G^{\nmo}$  into \Eq{MGRecur} and using  \Eq{MGRRecur}, the $n-$loop amplitude becomes
\begin{align} 
\cM_G^\n 
&=   \sum_{m= 0}^{n-1} \frac{1}{ (n-1-m)!} \frac{1}{n}  \left[ \frac{  i \alpha_s }{ 2\ep }  ( \TT_\ccTH \cdot \TT_\ccF ) \right]^{\nmom} \left[ (m+1) \cM_{G, R}^{\mpo}  +  
   \frac{ i \alpha_s }{  2 \ep }  ( \TT_\ccTH \cdot \TT_\ccF )  \,  \cM_{G, R}^{\m}    \right]  \\
  &=\sum_{m= 0}^{n-1}  \left[ \frac{1}{ (n-m)!} \frac{m}{n} + \frac{1}{ (n-1-m)!} \frac{1}{n}  \right]    \left[ \frac{ i \alpha_s }{ 2 \ep }  ( \TT_\ccTH \cdot \TT_\ccF )    \right]^{\nmm} \cM_{G, R}^{\m} +  \cM_{G, R}^{\n}  \\
  &=  
  \sum_{m= 0}^{n} \frac{1}{ (n-m)!}  \left[ \frac{ i  \alpha_s }{ 2\ep }  ( \TT_\ccTH \cdot \TT_\ccF ) \right]^{\nmm} \cM_{G, R}^{\m} 
\end{align} 
So \Eq{poleExp}  holds at $n-$loop.  
Therefore the IR divergence of the Glauber ladder diagrams exponentiates. 
In particular, to all orders in $\alpha_s$,  the squared amplitudes are equal
\begin{align} 
\big| \cM_G \big|^2 =  \big| \cM_{G, R} \big|^2
\end{align}
and IR finite.

\section{Singular region to all-orders \label{app:pssing}}
In this appendix, we examine the behavior of amplitude in the phase-space-singular regime, when either  $p_\ccTH$  is aligned 
 with $p_\ccO$ or when $ p_\ccF$ is aligned with $p_\ccT$.
These singular regions correspond to  $u = 0, v=1$ and  $ u= 1, v=0$. 
Only the leading power terms in this expansion generate phase-space singularities,
so we denote the leading-power approximation to the amplitude as $\cM_{\text{sing}}$. The remainder is integrable, i.e. there exists some $\delta > 0$ such that
 \begin{align} 
  \cM_G - \cM_{\text{sing}} =    \cO((uv)^{\delta}) \; \cM^\zero , 
\end{align} 
 Thus we need to evaluate the singular part in $d$ dimensions and the remainder can be evaluated
 in 4 dimensions. 
 
 $\cM_G$ is determined by master integrals  in $d=2-2\ep$ Euclidean space with three mass terms: $\vec p_{\ccTH, \perp}^2, (\vec q_\perp-\vec p_{\ccTH, \perp})^2, q_T^2$. In the singular limit where one of the masses becomes small,  $\cM_\text{sing}$ is determined by single-scale integrals which we can compute to all-loop order. 
 For example, in the limit where $  u = \frac{\vec p_{\ccTH, \perp}^2}{q_T^2}  \rightarrow 0 $, 
 \begin{align} 
 C_0 (\vec p_{\ccTH, \perp}^2, (\vec q_\perp-\vec p_{\ccTH, \perp})^2, q_T^2) \rightarrow  \frac{1}{q_T^2 } B_0 (\vec{p}^{\,2}_{\ccTH, \perp} ) +  \frac{1}{q_T^{4 + 2\ep} } \cO(u )
 \end{align}
 where $C_0$ and $B_0$ are one-loop scalar triangle and bubble integrals $d=2-2\ep$ dimension. 
$\cM_\text{sing}$ in the first few orders of $\alpha_s$ are
\begin{align} 
 \cM_\text{sing}^\zero &=  \cM^\zero  \nn \\
\cM_\text{sing}^\one  &= - \frac{ i \alpha_s}{2}  (\TT_\ccTH \cdot \TT_\ccF)  \,
 e^{\ep \gamma_E   } (q_T^2)^{-\ep} (uv)^{-\ep} \Gamma (- \ep )     \frac{ \Gamma(1-\ep) \Gamma ( 1 + \ep)    }{ \Gamma (1-2 \ep ) }  \,   \cM^\zero , \\
 \cM_\text{sing}^\two  &= - \frac{ \alpha_s^2}{ 8 }  (\TT_\ccTH \cdot \TT_\ccF)^2  \,
 e^{ 2 \ep  \gamma_E  } (q_T^2)^{- 2\ep} (uv)^{-2\ep} [\Gamma (- \ep ) ]^2   \frac{ \Gamma(1- \ep) \Gamma ( 1 + 2\ep)    }{ \Gamma (1-3 \ep ) }  \,   \cM^\zero 
\end{align} 
The general formula of  $\cM_\text{sing}$  at  $n$-loop order is,
\be
 \cM_\text{sing}^{n}  =  \frac{1}{n!} \left( \frac{  -i \alpha_s}{ 2 } \right)^n  (\TT_\ccTH \cdot \TT_\ccF)^n  \,
 e^{  n  \ep   \gamma_E} (q_T^2)^{- n\ep} (uv)^{-n\ep} [\Gamma (- \ep ) ]^n   \frac{ \Gamma(1- \ep) \Gamma ( 1 + n\ep)    }{ \Gamma (1- (n+1) \ep ) }  \,   \cM^\zero 
\ee
which we have found by direct calculation. 

Expanding in $\ep$, $\cM_\text{sing}  = \sum_{n} \cM_\text{sing}^n$ can be written as
\be
 \label{AmpLPE}
\cM_\text{sing}  
  = e^{i  {\bf \phi_g} ( \widehat{\alpha_s} ( \TT_\ccTH \cdot \TT_\ccF )  )}  \bigg\{ 1+ \ep  \left(  3 \zeta_3 \, \widehat{\alpha_s}^2 (\TT_\ccTH \cdot \TT_\ccF)^2    + \cO(\widehat{\alpha_s}^3) \right)   +  \cO(\ep^2)    \bigg\}  \cM^\zero
\ee
where 
\be
\widehat{\alpha_s} ( \ep, q_T, u, v ) \equiv \frac{\alpha_s}{2}  e^{\ep \gamma_E} c_\Gamma (q_T^2 u v)^{-\ep} 
\ee
with $c_\Gamma= \frac{\Gamma(1-\ep)^2\Gamma(1+\ep) }{\Gamma (1-2 \ep) } $ and the all-orders phase factor $ \bf \phi_g$  is 
\be
 {\bf{ \phi}_g} ( \alpha ) 
   \equiv   \alpha  \left(  \frac{1}{ \ep} +  2 \gamma_E \right)  
 -i \ln \frac{ \Gamma (1+ i \alpha)   }{ \Gamma(1- i   \alpha  )}   \\ 
  =   
  \frac{    \alpha  }{ \ep}    +2 
  \sum_{k=1}^\infty   \frac{\zeta_{2k+1} }{2k+1 } \, \alpha^{2k+1}    
\ee
Note that the $\frac{1}{\ep}$ term in ${\bf \phi_g}$ agrees with \Eq{poleExp}.
The key point represented in the expression in Eq.~\ref{AmpLPE} is that once the phase is factored out the singular terms start at order $\ep$. 

\section{Regular terms to two loops \label{app:2loopreg}}
We can define an IR finite subtracted amplitude $\cM_\text{reg}$ as follows, 
\begin{align} 
\cM_{\text{reg}}  \equiv e^{-i \bf \phi_g} \Big( \cM_G - \cM_{\text{sing}}  \Big) 
\end{align} 
which is regular at all points of phase-space. 
Expanding in orders of $\alpha_s$, 
\begin{align} 
\cM_\text{reg} \equiv  \frac{i \alpha_s}{2} \cM_\text{reg}^\one - \frac{\alpha_s^2}{8} \cM_\text{reg}^\two + \cdots 
\end{align}
these terms can be obtained at fixed order in $\ep$ by matching both sides of the following equation:  
\begin{multline}
\left( 1- i  {\bf \phi_g} (\widehat{\alpha_s}) -  
 \frac{{\bf \phi_g^2}(\widehat{\alpha_s}) }{2} + \cdots \right)  \Big( \cM^0  + \cM_G^\one + \cM_G^\two +\cdots   \Big)   \\
 =  \cM^\zero \bigg\{ 1+ \ep  \left(  3 \zeta_3 \, \widehat{\alpha_s}^2    + \cdots   \right)   +  \cO(\ep^2)   \bigg\}  +\frac{i \alpha_s }{2 } \cM_\text{reg}^\one  -\frac{\alpha_s^2}{8} \cM_\text{reg}^\two  + \cdots 
\end{multline} 
Due to the absence of phase-space singularities, we can take $d=4$ and
 write $\cM_\text{reg}$ in the form of helicity amplitude.  
Each helicity amplitude $\cM_\text{reg} (\{ h_\cci\})$  is a function of $z$ and $\bar z$, where $h_\cci$ are the helicities of external partons. 
The precise definition of complex variables  $z, \bar z$ is 
\begin{align} 
\frac{z }{  1-z } \equiv         
 \frac{  \l 3 1 \r [ 2 3 ] }{  \l 41 \r [2 4]  }  ,  \quad \frac{\bar z }{ 1- \bar z } \equiv   \frac{  \l 3 2 \r [1 3 ] }{  \l 42 \r [1 4]  } 
\end{align} 
$z,\bar z$ thus defined satisfy \Eq{conformvar}.

We find that the form of $\cM_\text{reg} (\{ h_\cci\})$  only depends on the relative sign between quark and gluon helicities in each collinear sectors. Thus without loss of generality, we take $h_\ccO= +, h_\ccT = -$. 
 The one and two loop results are the following,   
\begin{align} 
&\cM_{\text{reg}}^\one (+, -, h_\ccTH, h_\ccF)  =    \cM^\zero  (+, -, h_\ccTH,  h_\ccF  )  \nn \\
&~~~~~~~ \times \bigg\{   \delta_{h_\ccTH +}  \delta_{ h_\ccF - }  
\left[  z \ln u + (1-z) \ln v \right]   +  \delta_{h_\ccTH -}  \delta_{ h_\ccF +}    \left[  \bar z \ln u + (1- \bar z) \ln v \right]      \nn \\
&~~~~~~~   +   \ep \bigg(  \delta_{h_\ccTH +}  \delta_{ h_\ccF - }  
\left[  z  \ln u ( \frac12 \ln u + \ln q_T^2 ) + (1-z) \ln v ( \frac12 \ln v +  \ln q_T^2 )   -   z ( 1-  z )   \frac{ 2i P_2 (z)}{ z - \bar z }    \right]    \nn \\
&~~~~~~~ + 
 \delta_{h_\ccTH -}  \delta_{ h_\ccF +}  
   \left[  \bar z   \ln u (\frac12 \ln u  + \ln q_T^2 )  + (1- \bar z)   \ln v  ( \frac12 \ln v +  \ln q_T^2 )  -  \bar z ( 1-  \bar z )   \frac{2  i  P_2 (z)}{ z - \bar z }   \right]        
  \nn \\
&~~~~~~~ + \delta_{ h_\ccTH + } \delta_{h_\ccF+}    \Big[  \ln u \ln v  +   z ( 1- \bar z )  \frac{ 2i P_2(z) }{z - \bar z }\Big]
    + \delta_{ h_\ccTH -} \delta_{h_\ccF - }   \Big[  \ln u \ln v  +  \bar z ( 1-  z )  \frac{2 i P_2(z) }{z - \bar z }\Big]    \bigg) 
  \bigg\}
\end{align} 
where $P_2(z) \equiv \text{Im}  \Big\{ \text{Li}_2(z) - \text{Li}_1(z) \ln |z| \Big\}  $ is Bloch-Wigner dilogarithm and
\begin{multline}
 \cM_{\text{reg}}^\two    (+, -, h_\ccTH, h_\ccF)    =       \cM^\zero  (+, -, h_\ccTH,  h_\ccF  )    \\ 
 \times \bigg\{   \delta_{h_\ccTH +}  \delta_{ h_\ccF - }   
\left[  z \ln^2 u + (1-z) \ln^2 v \right]   +  \delta_{h_\ccTH -}  \delta_{ h_\ccF +}    \left[  \bar z \ln^2 u + (1-\bar z) \ln^2 v \right] \\
\quad  +  \delta_{h_\ccTH,  h_\ccF} \, \Big[ -  \ln u \ln v -    h_\ccTH \, 2i P_2(z) \Big]  + \cO(\ep)  \bigg\} 
\end{multline} 
Note that we need the $\cO(\ep)$ terms in $\cM_{\text{reg}}^\one$ in order to extract the $\cO(\ep^0)$ part of the 2-loop result. 
 
Putting things together, keeping only the finite terms in $\cM_{\text{reg}}$, the full amplitude is 
\begin{multline} 
\cM_G ( + ,  -,    h_\ccTH,  h_\ccF )  =  
e^{i  {\bf \phi_g} ( \widehat{\alpha_s} (\TT_\ccTH \cdot \TT_\ccF )  )  } \times \bigg\{  1 
 \\
%\times \bigg\{  1 +
+ \frac{i\alpha_s}{2}   (\TT_\ccTH \cdot \TT_\ccF )  \,  \bigg(  \delta_{h_\ccTH +}  \delta_{ h_\ccF - }  
\left[  z \ln u + (1-z) \ln v \right]   +  \delta_{h_\ccTH -}  \delta_{ h_\ccF +}    \left[  \bar z \ln u + (1- \bar z) \ln v \right]     
    \bigg) \\ 
 - \frac{ \alpha_s^2}{8}   (\TT_\ccTH \cdot \TT_\ccF )^2\, 
  \bigg(  \delta_{h_\ccTH +}  \delta_{ h_\ccF - }  
\left[  z \ln^2 u + (1-z) \ln^2 v \right]   +  \delta_{h_\ccTH -}  \delta_{ h_\ccF +}    \left[  \bar z \ln^2 u + (1-\bar z) \ln^2 v \right]       \\
+  \delta_{h_\ccTH,  h_\ccF} \, \Big[ -  \ln u \ln v +  h_\ccTH \, 2i P_2(z) \Big]     %\\
+    e^{2 \ep \gamma_E }  c_\Gamma^2  (q_T^2 u v)^{-2\ep} \,  (-6 \zeta_3 ) \, \ep   \bigg) 
  \\
  + \cO(\alpha_s^3)    \bigg\}  \cM_\zero  (+, -, h_\ccTH,  h_\ccF  )
\end{multline} 

The singularities in the Glauber phase cancel the squared amplitude and the result for $|\cM_G|^2$ to 2-loops has the relatively simple form
\begin{multline} 
\big| \cM_G ( + ,  -,    h_\ccTH,  h_\ccF )  \big|^2  =  
\big| \cM_\zero  (+, -, h_\ccTH,  h_\ccF  ) \big|^2 
 \\
\times \bigg\{  1  
  + \frac{   {\rm c}_{\rm 2G} }{ {\rm c}_\zero } \left[  \frac{ \alpha_s^2}{4} 
  \bigg(  \delta_{h_\ccTH  h_\ccF }    \ln u \ln v -  \delta_{h_\ccTH, -h_\ccF }  \frac{1}{2} (1-u -v ) \ln^2 \frac{u}{v} + \cO(\ep)  \bigg) +     \widehat{\alpha}_s^2 \bigg( 6 \zeta_3 \, \ep   + \cO(\ep^2) \bigg)  \right]   \\
  + \cO(\alpha_s^3)    \bigg\}
\end{multline} 
This result was quoted in Eqs.~\eqref{hel1} and \eqref{hel2}. 
Here $  {\rm c}_{\rm 2G} = \frac18 C_A^2 (C_A^2 +2) C_F, \, {\rm c}_\zero=  C_A C_F^2$, and 
\begin{align}
\big| \cM_\zero  (+, -, h_\ccTH,  h_\ccF  ) \big|^2  
&=  {\rm c}_\zero \, \frac{1}{s_{\ccO \ccTH} s_{\ccT \ccF} } 
  \Big| \text{Split}_{-, h_\ccTH} \Big(  \frac{1}{ z_{\ccO \ccTH } }  \Big) \Big|^2  \Big|\text{Split}_{+, h_\ccF} \Big(  \frac{1}{  z_{ \ccT\ccF }} \Big)  \Big|^2 \nn \\
  & = {\rm c}_\zero  \, \frac{1}{q_T^4 \, u v } \,
 \frac{ s_{ \ccT \ccTH } s_{\ccO \ccF } }{ s_{\ccO \ccT}^2  } \, \Big| \text{Split}_{-, h_\ccTH} \Big( \frac{1}{ z_{\ccO \ccTH } }  \Big) \Big|^2  \Big|\text{Split}_{+, h_\ccF} \Big( \frac{1}{  z_{ \ccT\ccF }}\Big)  \Big|^2 , 
 \label{M02}
 \end{align}
% with
% \be
%   z_{\ccO \ccTH }  \equiv  \frac{ \bar n \cdot (p_\ccO + p_\ccTH ) }{ \bar n \cdot p_\ccO }< 1  , 
%  \quad \text{and} \quad   z_{\ccT \ccF}  \equiv  \frac{  n \cdot (p_\ccT+ p_\ccF) }{  n \cdot p_\ccT }< 1  .
%\ee
$\text{Split} (z_{\cci \ccj})$ encodes the energy dependence of the splitting amplitude, where $z_{\cci \ccj}$ is defined as the energy fraction  of the daughter fermion in $q \rightarrow qg$ time-like splitting, see Eqs.~\eqref{z12} and \eqref{splitdef}.
% (this $z_{\cci \ccj}$ is not to be confused with the conformal coordinate $z$)
%\be 
%\Big|\text{Split}_{++} ( z ) \Big|^2=  \frac{1}{1-z}, \quad  \Big|\text{Split}_{+-} ( z ) \Big|^2=  \frac{z^2}{1-z},
%\ee  
Except for an overall factor of $q_T^{-4}$ in  $|\cM^\zero|^2$,  the squared amplitude depends only on dimensionless parameters $u$ and $v$, up to terms suppressed by $\ep$.  This is a consequence of the scale symmetry of ladder integrals in $d=2$ transverse plane, as well as the cancellation of IR singularities.

\section{Two-loop phase-space integrals \label{app:ContourPhase}}
In this appendix, we give some details of computing the integrals in Eqs.~\eqref{intpp} and \eqref{intpm}.  
The two-dimensional phase space can be parametrized by radial distance $|z|$ and azimuthal angle $\text{arg } z$. 
Define
\be 
\sigma \equiv  |z|, \quad  \omega \equiv e^{i \text{arg} \, z} = \frac{z}{|z|} , 
\ee
such that
\be
u = \sigma^2  , \quad  v =  (\sigma -\omega ) (\sigma - \frac{1}{\omega} ) 
\ee
The phase-space integrals in Eqs.~\eqref{intpp} and \eqref{intpm}  now become
\begin{align} 
I^{+ - }_{\text{reg}}  & \equiv   \int  d^2 z  \frac{1}{ u v}  \ln u \ln v   \nn \\
& = \oint_{|\omega|=1} \frac{d \omega}{  i \omega}  \int_0^\infty  \frac{d \sigma}{\sigma}    \frac{1 }{  (\sigma -\omega ) (\sigma - \frac{1}{\omega} )  }  \ln \sigma^2  \ln \left[ (1- \omega \sigma) (1- \frac{\sigma}{ \omega} )   \right]
\end{align} 
\begin{align} 
I^{+ + }_\text{reg}  & \equiv   - \int  d^2 z  \frac{1}{ u v} (1-u -v)  \frac{1 }{2}  \ln^2 \frac{u}{v}   \nn \\
& = \oint_{|\omega|=1} \frac{d \omega}{  i \omega}  \int_0^\infty d \sigma   \frac{ (\sigma -\omega ) + (\sigma - \frac{1}{\omega} )  }{  (\sigma -\omega ) (\sigma - \frac{1}{\omega} )  }  \frac12 \ln^2 \left[ \frac{ (1- \omega \sigma) (1- \frac{\sigma}{ \omega} ) }{ \sigma^2 } \right] 
\end{align} 
where the azimuthal angle integrals can be done by contour integration in the $\omega-$complex plane. 

First we compute $I^{+-}_\text{reg}$. The integral over the radial distance $\sigma$ can be  converted into integrals from $0$ to $1$, so that
\begin{align} 
I^{+ - }_{\text{reg}} 
& = \oint_{|\omega|=1} \frac{d \omega}{  i \omega}  \int_0^1  \frac{d \sigma}{\sigma}    \frac{1 -\sigma^2 }{  (\sigma -\omega ) (\sigma - \frac{1}{\omega} )  }  \ln \sigma^2  \ln \left[ (1- \omega \sigma) (1- \frac{\sigma}{ \omega} )     \right]  \nn \\
& +  \oint_{|\omega|=1} \frac{d \omega}{  i}  \int_0^1  d \sigma     \frac{ 2 \sigma  }{  ( \sigma - \omega ) (\sigma \omega- 1 )  }   
2\ln^2 \sigma    \label{IPLUSPLUS}
\end{align} 
The  integrand on the second line of \Eq{IPLUSPLUS} has two poles in the $\omega-$complex plane. 
Thus the azimuthal angle integral is given by the residue at $\omega = \sigma$, which is 
\begin{align} 
 2\pi \int_0^1  d \sigma    \frac{ 2 \sigma }{  1- \sigma^2 }  \, 2 \ln^2 \sigma  = 2\pi \zeta_3    \label{SecondLine}
\end{align}  
To compute the first line of \Eq{IPLUSPLUS}, we first carry out  the integral over radial coordinate $\sigma$,  which can be writen as
\begin{align} 
 \cI_\omega^{+-} \equiv & - \int_{0}^1 d \ln \left[ \frac{(1- \omega \sigma) (1- \frac{\sigma}{ \omega} ) }{\sigma}   \right] \,  \ln \sigma^2 \ln 
\left[  (1- \omega \sigma) (1- \frac{\sigma}{ \omega} )    \right]   \\ 
 & = 2 \left[  G (  0, \omega , \omega ; 1) +  G (  0, \omega , \frac1\omega ; 1) + G (  0, \frac1\omega , \omega ; 1)  +G (  0, \frac1\omega ,  \frac1\omega ; 1)    \right.  \nn \\
 & \qquad   \left.-G (  0, \omega , 0 ; 1)  - G (  0, \frac1\omega , 0 ; 1)  - G (  0,0, \omega  ; 1)  - G (  0, 0, \frac1\omega ; 1)    \right] \\
 & \equiv  2 \left[ \cG_{\omega} +  \cG_{\frac1\omega}  + \cF_{ \{ \omega , \frac1\omega \} }   \right] 
\end{align}  
where 
\begin{align} 
\cG_{\omega}  & \equiv  G( 0,1,1 ; \omega ) - G (0,1,0  ; \omega)   - G( 0,0,1; \omega)  \\
\cF_{ \{ \omega, \frac1\omega \} }  & \equiv  G( 0,\omega,\frac1\omega ; 1 ) +  G ( 0, \frac1\omega, \omega; 1) 
\end{align} 
Since $\cG_\omega$ is analytic in region $|\omega| < 1$,  $\cG_\frac1\omega$ is analytic in region $|\omega| > 1$, and $\cG_0 =0$,  then by Cauchy theorem, 
\begin{align} 
 \oint_{|\omega|=1} \frac{d \omega}{  i \omega} \,  \cG_\omega =  \oint_{|\omega|=1} \frac{d \omega}{  i \omega}  \, \cG_\frac1\omega = 0 .
\end{align} 
$\cF_{ \{ \omega, \frac1\omega \} }$ has branch cuts along the real axis of $\omega$, starting from $\omega =0$. 
So we pick a contour that wraps around the branch cut from $\omega =0$ to $\omega =1$, 
\begin{align} 
 \oint_{|\omega|=1} \frac{d \omega}{  i \omega} \,  \cF_{ \{  \omega , \frac1\omega \} } =   -\int  d \ln \omega  \, \theta_{|\omega| <1} \left[  \frac{1}{i} \text{Disc}_\omega\, 
    \cF_{ \{ \omega, \frac1\omega \} }    \right]  
\end{align}  
To figure out the discontinuity of  $\cF_{ \{ \omega, \frac1\omega \} }$, consider the following  function 
\begin{align} 
F_z (\omega) \equiv G(0, \omega, \frac1\omega; z) + G(0, \frac1\omega, \omega; z), \quad  z \leq 1.
\end{align} 
By computing its iterated coproduct, 
\begin{align} 
\Delta_{1,1,1} [F_z (\omega) ] 
&  =G( \omega;  z)  \otimes (1- \omega z )   \otimes \omega z  
  -  G( \omega ; z) \otimes ( 1- \omega^2 )\otimes \omega^2 \nn \\
 & + G\Big( \frac1\omega;  z \Big)  \otimes \Big( 1- \frac{z}{\omega}  \Big) \otimes \frac{z}{\omega} 
  -  G\Big( \frac{1}{\omega}; z \Big) \otimes  \Big( 1- \frac{1}{\omega^2} \Big)  \otimes \frac{1}{\omega^2} 
 \end{align} 
 we know that  $F_z (\omega)$  has  branch cuts  on real axis : $\omega  \in [ 0, z ] $  and  $\omega \in [  \frac{1}{z} ,  \infty)$, 
 and
\begin{align} 
\cS \left[  \text{Disc}_{\omega} \, F_z (\omega)  \right]  & =  
  \theta\Big(  \frac{z}{\omega} -1 \Big)( 2\pi i  )  \, \left[  (1- \omega z )   \otimes \omega z    -     ( 1- \omega^2 )\otimes \omega^2  \right]  \nn \\
& +  \theta ( \omega z - 1 )    
  (- 2\pi i ) \, \left[   \Big( 1- \frac{z}{\omega}  \Big) \otimes \frac{z}{\omega}  -  \Big( 1- \frac{1}{\omega^2} \Big)  \otimes \frac{1}{\omega^2} 
  \right]
\end{align}  
 Matching on to boundary values at $\omega =z$ and $\omega =\frac1z$ , 
 \begin{align}  
 & \text{Disc}_{\omega =z } \, F_z (\omega)
   =- \text{Disc}_{\omega = \frac1z } \, F_z (\omega) \\
 &  =  \text{Disc}  \Big\{   G(0, 1, \frac{1}{z^2}; 1) + G(0, \frac{1}{z^2}, 1 ; 1)   \Big\}   =0, \quad \forall z \leq 1, \nn 
 \end{align} 
 we now conclude that 
 \begin{align} 
 \text{Disc}_{\omega} \, F_z (\omega)  & = 
 \theta\Big(  \frac{z}{\omega} -1 \Big)( 2\pi i  )  \left[ -\text{Li}_2 ( \omega z ) + \text{Li}_2  ( \omega^2 ) \right]  \nn \\
& +  \theta ( \omega z - 1 )    
  (- 2\pi i ) \left[ - \text{Li}_2 \Big(  \frac{z}{\omega}  \Big) + \text{Li}_2  \Big( \frac{1}{\omega^2} \Big)  \right]
 \end{align} 
 Setting $z=1$, we obtain
\begin{align} 
 \text{Disc}_{\omega} \,  \cF_{\{ \omega, \frac1\omega \}}  & = 
 \theta\Big(  \frac{1}{\omega} -1 \Big)( 2\pi i  )  \left[ -\text{Li}_2 ( \omega  ) + \text{Li}_2  ( \omega^2 ) \right]  - \Big(  \omega \leftrightarrow \frac{1}{\omega}  \Big) 
\end{align} 
Thus the first line of \Eq{IPLUSPLUS} evaluates to
\begin{align} 
 \oint_{|\omega|=1}   \frac{d\omega}{i \omega} \, 2 \left[ \cG_{\omega} +  \cG_{\frac1\omega}  + \cF_{ \{ \omega , \frac1\omega \} }   \right]   & = 
   -\int  d \ln \omega  \, \theta_{|\omega| <1} \left[  \frac{2}{i} \text{Disc}_\omega\, 
    \cF_{\{\omega, \frac1\omega\}}    \right]  
    \\
   & =  - (4\pi) \int_0^1  d \ln \omega  \,   \left[ -\text{Li}_2 ( \omega  ) + \text{Li}_2  ( \omega^2 ) \right]  \\
    & = 2 \pi  \, \zeta_3  \label{FirstLine}
\end{align}   
Collecting results for the first and second lines of \Eq{IPLUSPLUS}, 
\begin{align}
I^{+-}_\text{reg} =  \text{Eq. }\eqref{FirstLine} + \text{Eq. }\eqref{SecondLine}  =  4\pi \zeta_3
\end{align} 

Now we move onto the second integral $ I^{+ + }_\text{reg}  $. 
\begin{align} 
I^{+ + }_\text{reg}
& = \oint_{|\omega|=1} \frac{d \omega}{  i \omega}  \int_0^\infty d \sigma   \frac{ (\sigma -\omega ) + (\sigma - \frac{1}{\omega} )  }{  (\sigma -\omega ) (\sigma - \frac{1}{\omega} )  }  \frac12 \ln^2 \left[ \frac{ (1- \omega \sigma) (1- \frac{\sigma}{ \omega} ) }{ \sigma^2 } \right] 
\end{align} 
The integral over the radial distance can be written as 
\be
\cI^{++}_\omega  \equiv    \int _0^\infty  d \ln \left[ ( 1- \omega \sigma)  (1- \frac{\sigma }{ \omega} ) \right]   \circ \int  
d \ln  \left[ \frac{ (1- \omega \sigma) (1- \frac{\sigma}{ \omega} ) }{ \sigma^2 } \right]    \circ  \ln \left[ \frac{ (1- \omega \sigma) (1- \frac{\sigma}{ \omega} ) }{ \sigma^2 } \right]
\ee
The integral over $\sigma$ from $1$ to $\infty$ can be inverted into an integral from $0$ to $1$, so that 
\begin{align} 
\cI_\omega^{++}  & = 
   4 \int_{0}^1  d \ln \sigma  \circ \int d \ln \left[   \frac{ (1- \omega \sigma) (1- \frac{\sigma}{ \omega} ) }{ \sigma}  \right]  \circ  \int d \ln \left[   \frac{ (1- \omega \sigma) (1- \frac{\sigma}{ \omega} ) }{ \sigma}  \right]   \nn \\   
 & ~~~~~~~~~~  + \frac{1}{3}  \ln^3 \sigma  \bigg|_{0}^1     - 
   \ln \sigma  \ln^2    \left[   \frac{ (1- \omega \sigma) (1- \frac{\sigma}{ \omega} ) }{ \sigma}  \right] \bigg|_{0}^1     \\
   & = 4 \left[  G (  0, \omega , \omega ; 1) +  G (  0, \omega , \frac1\omega ; 1) + G (  0, \frac1\omega , \omega ; 1)  +G (  0, \frac1\omega ,  \frac1\omega ; 1)    \right.  \nn \\
 & ~~~~~~~~~~   \left.-G (  0, \omega , 0 ; 1)  - G (  0, \frac1\omega , 0 ; 1)  - G (  0,0, \omega  ; 1)  - G (  0, 0, \frac1\omega ; 1)    \right]    
 \\
 &= 4  \left[ \cG_{\omega} +  \cG_{\frac1\omega}  + \cF_{ \{ \omega , \frac1\omega \} }   \right] 
\end{align} 
Therefore  
\begin{align}
 I^{+ +}_\text{reg}   =  \oint_{|\omega|=1}   \frac{d\omega}{i \omega} \, 4\left[ \cG_{\omega} +  \cG_{\frac1\omega}  + \cF_{ \{ \omega , \frac1\omega \} }   \right]  = 4\pi \zeta_3  
 \end{align}

 \end{fmffile}
 
 \bibliography{Glauber-cancel}
 
\bibliographystyle{utphys}

\end{document}